\documentclass[twocolumn,showpacs,preprintnumbers,amsmath,amssymb,pre]{revtex4}

\usepackage{graphicx}
\usepackage{dcolumn}
\usepackage{bm}
\usepackage{epsfig} 

\renewcommand{\d}{{\rm d}}
\renewcommand{\pl}{\partial}
 
\newcommand{\beq}{\begin{equation}} 
\newcommand{\eeq}{\end{equation}} 
\newcommand{\beqa}{\begin{eqnarray}} 
\newcommand{\eeqa}{\end{eqnarray}} 
\newcommand{\bea}{\begin{array}} 
\newcommand{\ea}{\end{array}} 
 
\newcommand{\rhob}{\overline{\rho}} 
\newcommand{\lag}{\langle} 
\newcommand{\rag}{\rangle}
\newcommand{\varphib}{\overline{\varphi}}
\newcommand{\Phib}{\overline{\Phi}}
\newcommand{\Vb}{\overline{V}}

\begin{document}

\preprint{APS/123-QED}

\title{Relaxation of a 1-D gravitational system}

\author{P.~Valageas}
\affiliation{Service de Physique Th\'eorique, CEA Saclay, 91191 Gif-sur-Yvette, France}

\date{\today}

\begin{abstract}
We study the relaxation towards thermodynamical equilibrium of a 1-D 
gravitational system. This model shows a series of critical energies
$E_{cn}$ where new equilibria appear and we focus on the homogeneous ($n=0$),
one-peak ($n=\pm 1$) and two-peak ($n=2$) states. Using numerical simulations
we investigate the relaxation to the stable equilibrium $n=\pm 1$ of this 
$N-$body system starting from initial conditions defined by equilibria $n=0$
and $n=2$.
We find that in a fashion similar to other long-range systems the relaxation
involves a fast violent relaxation phase followed by a slow collisional phase 
as the system goes through a series of quasi-stationary states. Moreover, in
cases where this slow second stage leads to a dynamically unstable 
configuration (two peaks with a high mass ratio) it is followed by a new 
sequence ``violent relaxation/slow collisional relaxation''. We obtain
an analytical estimate of the relaxation time $t_{2\rightarrow \pm 1}$
through the mean escape time of a particle from its potential well in a
bistable system. We find that the diffusion and dissipation coefficients
satisfy Einstein's relation and that the relaxation time scales
as $N e^{1/T}$ at low temperature, in agreement with numerical simulations.
\end{abstract}

\pacs{Valid PACS appear here}

\maketitle

\section{\label{Intro}Introduction}

The thermodynamics and dynamics of systems with long-range interactions 
have been the focus of many studies in recent years \cite{Dauxois2002,HMF2005}.
Indeed, such systems exhibit many peculiar features due to their long-range
nature and to their non-additivity, such as ensemble inequivalence between 
micro-canonical, canonical and grand-canonical ensembles and regions of 
negative specific heat \cite{Padmanabhan1990}.
On the other hand, their dynamics presents many interesting phenomena 
\cite{Chavanis2006II}.
In particular, the relaxation to thermodynamical equilibrium can be very slow
and diverge with the number $N$ of particles 
\cite{Luwel1985,Tsuchiya1998,Yawn2003}.
Moreover, this relaxation often proceeds in 
two steps with very different time-scales. First, there is a 
collisionless relaxation over a few dynamical times which involves collective 
dynamical instabilities (this step is also called violent relaxation 
in astrophysics \cite{Lynden-Bell1967}). 
Secondly, after the system has reached a mean-field 
equilibrium a collisional relaxation associated with two-body encounters,
or more generally due to the discrete nature of the matter distribution
which gives rise to fluctuations with respect to the smooth mean-field 
potential (finite $N$ effects), leads to a slow relaxation towards statistical
equilibrium over a time-scale which diverges with $N$ \cite{Yamaguchi2004}.

A prototype of such systems with long-range interactions is the Hamiltonian
Mean Field (HMF) model defined by a cosine interaction for particles moving
on a circle \cite{HMF2005}. It has been shown \cite{Yamaguchi2004} that 
this system first converges to a stable stationary solution of the 
mean-field Vlasov equation. 
Next, the relaxation to thermodynamical equilibrium proceeds over a much 
longer time-scale through a slowly varying sequence of stable stationary states
of the Vlasov dynamics.
This process implies that the dynamics of the system strongly depends on the 
initial conditions since there are an infinite number of stable stationary 
solutions of the Vlasov equation \cite{Yamaguchi2004,Barre2005}.

In this article we study the relaxation of the 1-D gravitational system
described in details in \cite{Valageas2006}. This One-dimensional Static 
Cosmology (OSC) model consists of particles moving between two reflecting 
walls within an external potential $V$ which balances the 1-D gravitational
self-interaction $\Phi$ so that the homogeneous state (i.e. constant density)
is an equilibrium solution. This model also corresponds to the evolution of
1-D density fluctuations in a 3-D cosmological background over time-scales
much smaller than the Hubble time so that the expansion of the universe can
be neglected. As shown in \cite{Valageas2006}, this system exhibits a series
of critical energies $E_{cn}$ ($E_{c1}>E_{c2}>..$). At high energies the
only stable thermodynamical equilibrium is the homogeneous state (called
``$n=0$'') and at each transition $E_{cn}$ two new thermodynamical equilibria 
($\pm n$) appear. Moreover, equilibria $\pm 1$ are stable below $E_{c1}$
(while the homogeneous state turns unstable) whereas other equilibria $\pm n$
with $n\geq 2$ are unstable (both from a thermodynamical and a mean-field
dynamical analysis) except for equilibrium $n=2$ which becomes stable for
the Vlasov dynamics at low energy but remains thermodynamically
unstable. In this article we study the relaxation of the OSC model starting
from either the $n=0$ or $n=2$ equilibria, which allows us to investigate
both collisionless and collisional processes. 
We first recall in sect.~\ref{OSC_model} the thermodynamical properties of
the OSC model. Next, we study the relaxation of the system, starting from the 
homogeneous state in sect.~\ref{Homogeneous} and starting from equilibrium
$n=2$ in sect.~\ref{n=2}. Finally we conclude in sect.~\ref{Conclusion}.

\section{\label{OSC_model}The OSC model}

\subsection{\label{Description}Description of the model}

The OSC model \cite{Valageas2006} consists of $N$ particles of mass $m$ which 
move along the
$x$-axis in the interval $0<x<L$ (with reflecting walls) within an external 
concave quadratic potential $V(x)$ and which interact through 1-D gravity. 
Thus, the Hamiltonian ${\cal H}_N$ of the system is:
\beq
{\cal H}_N = m \sum_{i=1}^N \frac{v_i^2}{2} + g m^2 \sum_{i>j} |x_i-x_j| 
+ m \sum_{i=1}^N V(x_i) ,
\label{HN}
\eeq
where $v_i=\dot{x}_i$ is the velocity of particle $i$ (we note by a dot the
derivative with respect to time $t$), $g$ is the coupling constant of the
1-D gravitational interaction and the external potential $V(x)$ is:
\beq
V(x)= - g \rhob [ (x-L/2)^2 + L^2/4 ] ,
\label{V}
\eeq
where $\rhob=M/L$ is the mean density of the system ($M=N m$ is the total 
mass). In \cite{Valageas2006} the thermodynamics and stability properties of
the OSC system were studied in the mean-field limit (i.e. continuum limit)
where the mass $m$ goes to zero at fixed density $\rhob$. Then, the 
gravitational self-interaction is described by the potential $\Phi(x)$ with:
\beq
\Phi(x)= g \int_0^L\d x'\rho(x') \; |x-x'| ,
\label{Phi}
\eeq
and the dynamics of the system is governed by the Vlasov equation for the
phase-space density $f(x,v,t)$ within the total potential $\phi$:
\beq
\frac{\pl f}{\pl t} + v . \frac{\pl f}{\pl x} - \frac{\pl \phi}{\pl x} . 
\frac{\pl f}{\pl v} = 0  , \;\;\; \phi= \Phi+V .
\label{Vlasov}
\eeq
The main feature of the OSC model which comes from its cosmological context is 
the appearance of the background potential $V(x)$ which ensures that the 
homogeneous configuration $\rho=\rhob$ is a solution of the equations of 
motion (this is the counterpart of the Hubble flow for the expanding universe).
Alternatively, the constant density $\rhob$ defined from the potential $V$
can be interpreted as a cosmological constant (in which case it is not
necessarily equal to the mean matter density) if we work in physical 
coordinates. The statistical mechanics of such a system has been studied in 
the 3-D case in \cite{deVega2005a,deVega2005b}.
Here, following \cite{Valageas2006} we interpret the OSC model as derived from
a simple cosmological framework without cosmological constant. Then, the
external potential $V$ appears through the change to comoving coordinates
\cite{Peebles1980} and the Hamiltonian (\ref{HN}) applies on time-scales 
which are shorter than the Hubble time (so that the expansion of the universe 
can be neglected and there is no explicit time dependence in the 
Hamiltonian (\ref{HN})).
However, for our purposes the OSC model can also be studied for its own sake
as a simple model of systems with scale-free long-range interaction
which exhibits an interesting behavior (in particular the series of critical
energies discussed in \cite{Valageas2006} and recalled below in 
Eq.(\ref{Tcn})). In the following we shall decompose the total energy $E$ 
from (\ref{HN}) into its kinetic ($K$), self-gravity ($\Phib$) and 
potential ($\Vb$) components as $E=K+\Phib+\Vb$ with:
\beqa
K= m \sum_{i=1}^N \frac{v_i^2}{2} , \;\;\;\;\; \Vb= m \sum_{i=1}^N V(x_i) , 
\label{EN1} \\
\Phib= g m^2 \sum_{i>j} |x_i-x_j| = m \sum_{i=1}^N \frac{\Phi(x_i)}{2} .
\label{EN2}
\eeqa

\subsection{\label{Thermodynamics}Thermodynamical equilibria}

At thermodynamical equilibrium the phase-space density is the usual 
Maxwell-Boltzmann distribution:
\beq
f(x,v) = \rho(x) \sqrt{\frac{\beta}{2\pi}} \; e^{-\beta v^2/2} 
\propto e^{-\beta[v^2/2+\phi(x)]}  ,
\label{Maxwell}
\eeq
where we introduced the inverse temperature $\beta=1/T$. The density is
related to the potential by the Poisson equation modified by a constant
term due to the background $V$:
\beq
\frac{\d^2\varphi}{\d x^2} = 2 g (\rho-\rhob) \;\; \mbox{with} \;\;
\phi=\varphi+\varphib \;\; \mbox{and} \;\; \rho = \rhob e^{-\beta \varphi} .
\label{offset}
\eeq
Here we introduced for convenience the offset $\varphib$ defined by the 
condition $\rho=\rhob$ for $\varphi=0$. Therefore, the thermodynamical 
equilibrium is set by the equations:
\beq
\frac{\d^2\varphi}{\d x^2} = 2 g \rhob \left( e^{-\beta\varphi}-1 \right) 
\;\;\; \mbox{and} \;\;\; \varphi'(0)=\varphi'(L)=0 .
\label{LaneEmden}
\eeq
The homogeneous state $\rho=\rhob$ (i.e. $\varphi=0$) is a solution of 
Eqs.(\ref{LaneEmden}) which we label as ``equilibrium $n=0$''. Its total energy
is:
\beq
E_0= \frac{MT}{2} - \frac{gM^2L}{6} .
\label{E0}
\eeq
Note that the energy $E_0$ is bounded from below by $E_{\rm min}(0)=-gM^2L/6$.
Besides, as shown in \cite{Valageas2006}, Eqs.(\ref{LaneEmden}) yield a series 
of critical temperatures $T_{cn}$ and energies $E_{cn}$ ($n=1,2,..$) where new 
thermodynamical equilibria appear, with:
\beq
T_{cn} = \frac{2gML}{n^2\pi^2} , \hspace{0.5cm} 
E_{cn} = - \frac{n^2\pi^2-6}{6n^2\pi^2} gM^2L .
\label{Tcn}
\eeq
Thus, at high temperatures above $T_{c1}$ the only thermodynamical equilibrium 
is the homogeneous solution $\rho=\rhob$ ($n=0$). Below $T_{c1}$ this state 
becomes unstable and two new stable equilibria ``$n=\pm 1$'' appear. The state 
$n=1$ corresponds to a density peak at $x=0$ and a density minimum at $x=L$, 
whereas the state $n=-1$ is its reflection through $x=L/2$. Similarly, at each 
critical temperature $T_{cn}$ two new equilibria $\pm n$ appear, which 
consist of $n$ half-oscillations (i.e. from a density peak to a density
 minimum). The state $n>0$ shows a peak at $x=0$ whereas the state $n<0$ shows 
a minimum at $x=0$. As described in \cite{Valageas2006} the OSC model can also
be extended to the whole real line by symmetry with respect to $x=0$ and
periodicity of $2L$. Then, the state $n<0$ is a mere translation of the 
state $|n|$ by 
$L/|n|$. Thus, the equilibrium $n=2$ displays a density peak at each boundary 
$x=0,L$ and a minimum at $x=L/2$ (whereas state $n=-2$ shows one density peak
at $x=L/2$ and two density minima at $x=0,L$). Moreover, all thermodynamical 
equilibria $\pm n$ can be obtained from state $n=1$ at a rescaled temperature:
\beq
\rho_n(\zeta;\beta) = \rho_1(\zeta;\frac{\beta}{n^2}) , \;\;
\rho_{-n}(\zeta;\beta) = \rho_1(\zeta-\frac{\zeta_L}{n};\frac{\beta}{n^2}) ,
\label{rhon}
\eeq
where we wrote explicitly the dependence on inverse temperature $\beta$ and
we defined the dimensionless coordinate $\zeta$ by:
\beq
\zeta = \frac{x}{L_J} = \zeta_L \frac{x}{L} 
\;\;\; \mbox{with} \;\;\; L_J=\frac{1}{\sqrt{2g\rhob\beta}} , \;\; 
\zeta_L = \frac{L}{L_J} ,
\label{zeta}
\eeq
where we introduced the ratio $\zeta_L$ of the system size $L$ to the 
Jeans length $L_J$.
All equilibria $\pm n$ with $n\geq 2$ are thermodynamically unstable 
(in the three micro-canonical, canonical and grand-canonical ensembles, 
see \cite{Valageas2006}). They also exhibit a dynamical linear instability
in the continuum limit where the motion follows the Vlasov equation, except 
for state $n=2$ which becomes linearly stable very close to $T_{c2}$. 
Therefore, above $T_{c1}$ the stable thermodynamical equilibrium of the
discrete system (\ref{HN}) is the state $n=0$ with $\rho=\rhob$ whereas
below $T_{c1}$ it is the state $n=1$. However, as discussed in 
\cite{Yamaguchi2004} quasi-stationary long-lived states can also exist and
below $T_{c2}$ we can expect the equilibrium $n=2$ to be such a long-lived
configuration.

At low temperature 
($\zeta_L\rightarrow\infty$) the density profile of the peak in the 
equilibrium $n=1$ obeys \cite{Valageas2006}:
\beq
\zeta \ll \frac{\ln\zeta_L}{\zeta_L} : \;\;\; \frac{\rho_1(\zeta)}{\rhob} 
\simeq \frac{\zeta_L^2}{2\cosh^2(\zeta_L \zeta/2)} ,
\label{rhocosh}
\eeq
whereas the density minimum at $x=L$ scales as:
\beq
\rho_1(\zeta_L)\sim \rhob e^{-\zeta_L^2/2}= \rhob e^{-\pi^2 T_{c1}/2T} .
\label{rhovoid}
\eeq
For other states $\pm n$ we can obtain the asymptotics from (\ref{rhon}).
Note that the minimum energy which can be achieved by the OSC system is
$E_{\rm min}(1)=-gM^2L/2$ when all particles are at rest at the same boundary,
either $x=0$ or $x=L$. It can only be reached by the equilibrium $n=\pm 1$
at zero temperature whereas other equilibria $\pm n$ with $n\geq 2$ have
energies above $E_{\rm min}(n=2)=-gM^2L/4$.

\subsection{\label{Numerical} Numerical simulations}

From Eq.(\ref{HN}) we obtain the equations of motion of the $N$ particles as:
\beq
\ddot{x}_i - 2g\rhob x_i = g \rhob L \left( \frac{N_i^+ - N_i^-}N - 1\right) ,
\label{xi}
\eeq
where $N_i^+$ (resp. $N_i^-$) is the number of particles located to the right
(resp. to the left) of particle $i$, that is with $x>x_i$. In practice, 
following \cite{Noullez2003,Fanelli2002} we rank the particles in increasing
order of $x_i$ and when two particles ``collide'' we exchange their velocities
so that the ordering remains valid at all times: $x_i \leq x_{i+1}$. 
Therefore, the numbers of particles to the left and to the right of particle 
$i$ are constant: $N_i^-=i-1$ and $N_i^+=N-i$. Thus between two collisions
the dynamics of particle $i$ is given by:
\beqa
x_i(t) & = & x_i^{\rm eq} + (x_i^0-x_i^{\rm eq}) \cosh\left( 
\frac{t-t_i^0}{t_{\rm dyn}}\right) \nonumber \\
&& + v_i^0 t_{\rm dyn} \sinh\left(\frac{t-t_i^0}{t_{\rm dyn}}\right) ,
\label{chsh}
\eeqa
where we defined $x_i^{\rm eq}$ as the (unstable) equilibrium position of 
particle $i$ and $t_{\rm dyn}$ the typical dynamical time (for the homogeneous
configuration):
\beq
x_i^{\rm eq} = \frac{2i-1}{2N} L \;\;\; \mbox{and} \;\;\; 
t_{\rm dyn} = \frac{1}{\sqrt{2g\rhob}} . 
\label{xieq}
\eeq
In Eq.(\ref{chsh}) $x_i^0$ and $v_i^0$ are the particle coordinate and 
velocity at time $t_i^0$ (which is taken as the time of the last collision).
We use the event-driven scheme of \cite{Noullez2003} to follow the dynamics
of the system. We store the position and velocity $(x_i^0,v_i^0)$ of all
particles as well as the time $t_i^0$ of their last collision (initially we 
set $t_i^0=0$). Then, we compute the collision time of each particle with its
neighbors and we store the results in a heap structure, so that the next
collision is at the root of this heap. Next, we advance to this first
crossing by evolving the two colliding particles with Eq.(\ref{chsh}),
exchanging their velocities at collision and updating their last collision time
$t_i^0$. Finally, we compute the new three collision times associated with
these two particles and their neighbors which we store in the heap.
At the next step we take care of the next collision. Therefore, we advance
from one collision to the next and at each crossing we only need to update
the trajectories of the colliding particles and their next crossing times.
The collision time between particles $i$ and $i+1$ can be obtained 
analytically from Eq.(\ref{chsh}). Indeed, this yields for the distance
between neighbors an expression of the form 
$x_{i+1}-x_i = A + B e^{t/t_{\rm dyn}} +  C e^{-t/t_{\rm dyn}}$ where
$A,B$ and $C$ are constants. Then the condition $x_{i+1}=x_i$ leads to a 
quadratic equation for the variable $y=e^{t/t_{\rm dyn}}$ which is easily
solved. Therefore, both the trajectories and the collision times are computed
``exactly'' from analytical formulae. Their accuracy only depends on the
numerical accuracy of the computer and does not involve a discretization
procedure to compute integrals or differential equations.

\section{\label{Relaxation}Relaxation to thermodynamical equilibrium}

\subsection{\label{Homogeneous}Homogeneous initial state $n=0$}

We study in this section the dynamics of the discrete OSC model (\ref{HN}) 
starting from initial conditions defined by the homogeneous thermodynamical 
equilibrium $n=0$, i.e. Eq.(\ref{Maxwell}) with $\rho=\rhob$.
At high energies $E>E_{c1}$ the equilibrium $n=0$ is stable (and there are no
other thermodynamical equilibria) and we checked numerically that the system
remains in this configuration. Therefore, we focus here on systems with total 
energies $E$ below the critical energy $E_{c1}$ where the state $n=0$ becomes
unstable (both from the thermodynamical analysis and the mean-field Vlasov
dynamics) and two new stable equilibria $n=\pm 1$ appear, characterized by a 
density peak at $x=0$ ($n=1$) or at $x=L$ ($n=-1$). States $\pm 1$ are 
symmetric with respect to $x=L/2$ and are essentially the same configuration. 
Hence we study here the relaxation of the OSC system from the homogeneous 
state $n=0$ to the thermodynamical equilibria $\pm 1$.

\subsubsection{\label{time01}Transition times}

\begin{figure}
\begin{center}
\epsfxsize=4.15 cm \epsfysize=4.4 cm {\epsfbox{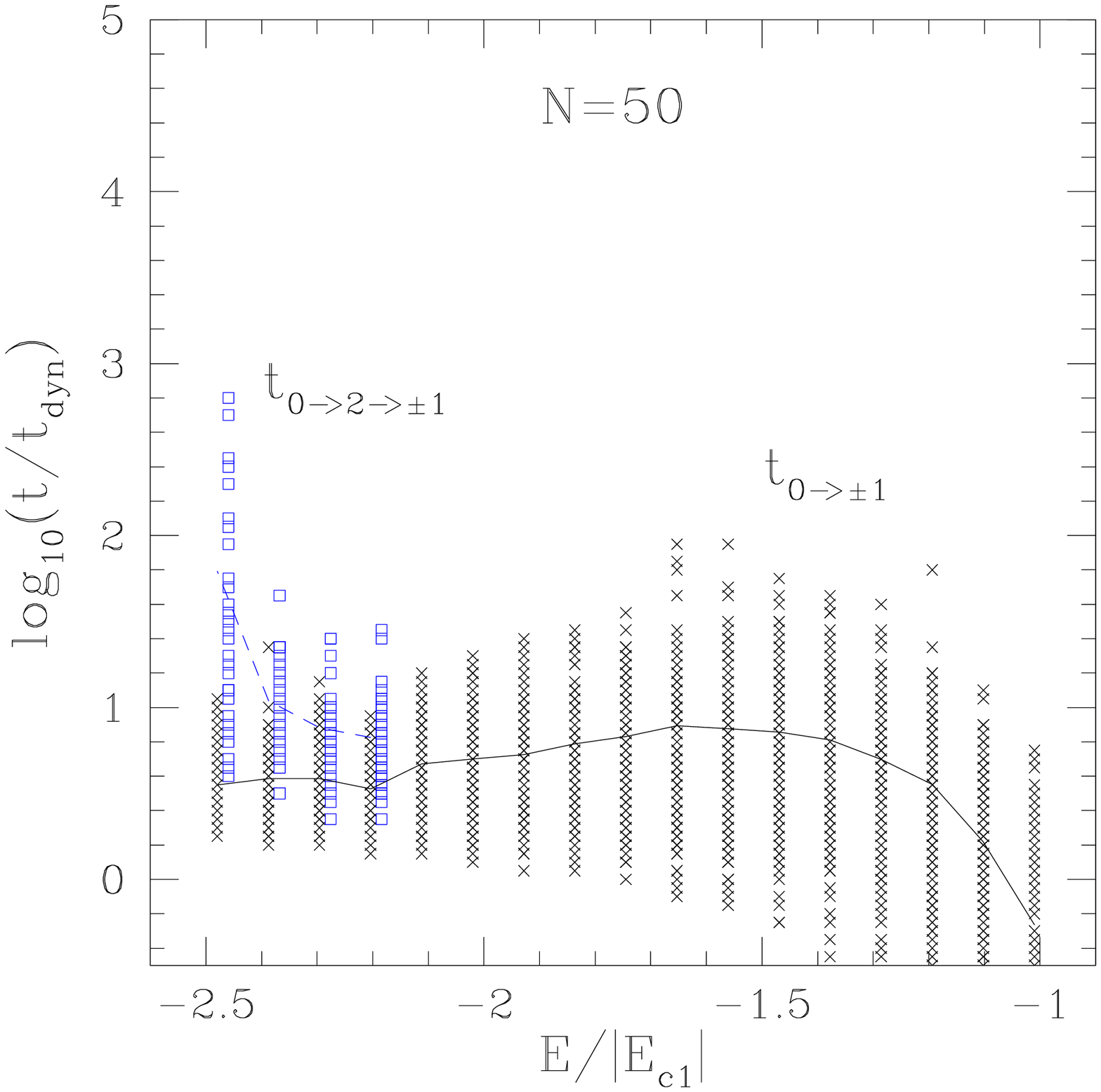}} 
\epsfxsize=4.15 cm \epsfysize=4.4 cm {\epsfbox{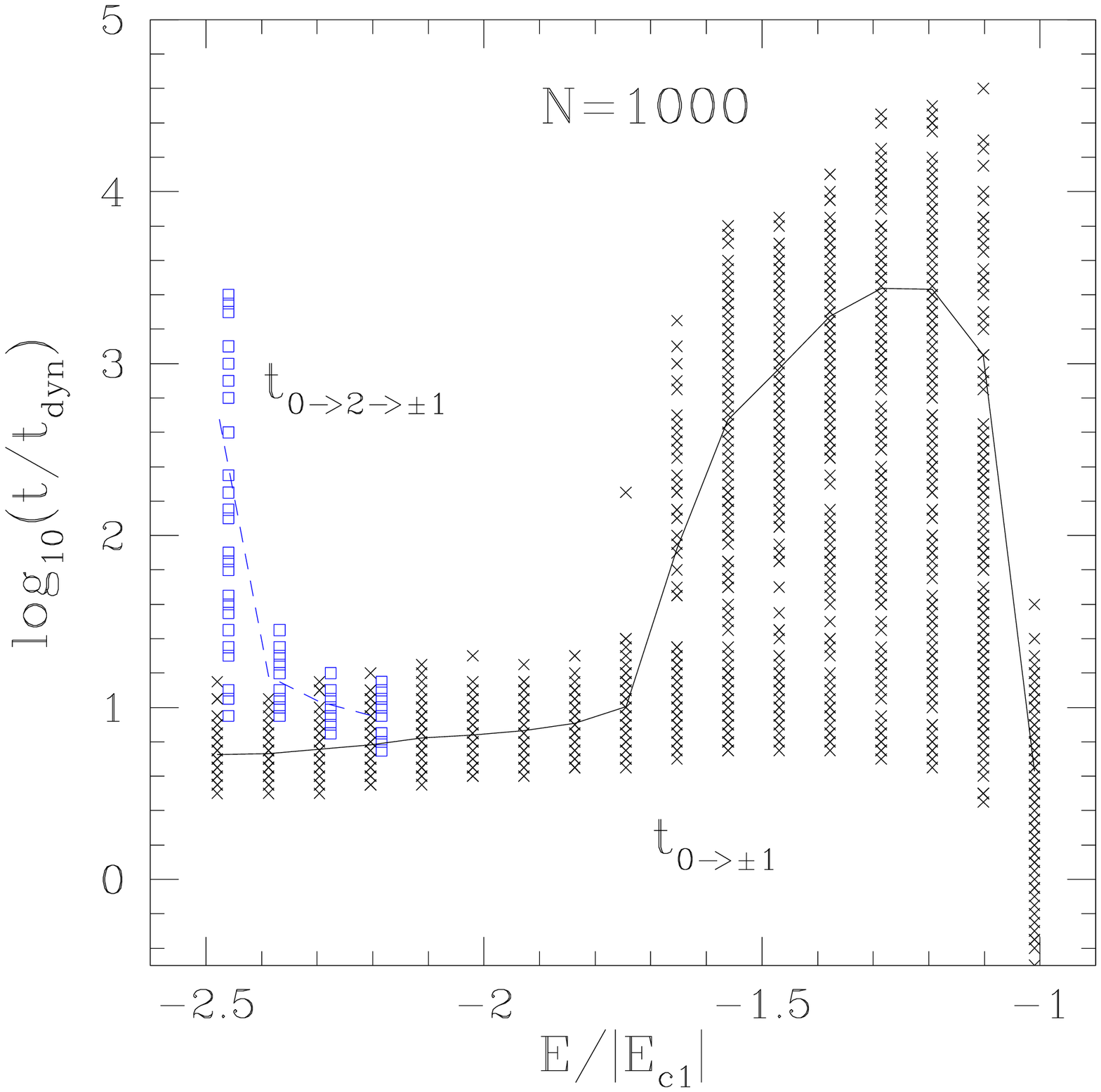}}
\end{center}
\caption{\label{figtime0to1}(Color online) The transition time 
$t_{0\rightarrow\pm 1}$ from 
the unstable homogeneous state $n=0$ to the stable equilibria $n=\pm 1$ as
a function of total energy $E$, below the critical energy $E_{c1}$. 
At each energy the crosses are the numerical results obtained for $200$ 
realizations of the initial condition $n=0$, for a system of $N=50$ 
(left panel) and $N=1000$ particles (right panel). 
The solid line is the mean transition time obtained by averaging over these 
numerical results.
The squares which are slightly shifted to the right show the transition times 
to states $n=\pm 1$ for realizations which happen to first relax to state 
$n=2$ over a few dynamical times. The dashed curve is the average obtained for 
these systems.}
\end{figure}

We first investigate the dependence on the energy $E$ and on the
number of particles $N$ of the transition time $t_{0\rightarrow\pm 1}$ from 
state $n=0$ to either state $n=\pm1$. Here we simply define the transition as 
the first time where $|K-K_1|<|K-K_0|$ and $|\Phib-\Phib_1|<|\Phib-\Phib_0|$ 
(i.e. the kinetic and self-gravity energies are closer to the levels of states 
$n=\pm 1$ than those of state $n=0$). We display in Fig.~\ref{figtime0to1}
the transition time $t_{0\rightarrow\pm 1}$ as a function of the total 
energy $E$, obtained for $200$ realizations (crosses) of the initial 
condition $n=0$ at each energy.
Left (resp. right) panel corresponds to systems of $N=50$ (resp. $N=1000$)
particles. The solid line is the average of these numerical results.

We can distinguish three regimes in Fig.~\ref{figtime0to1}. First, 
at low energies below $-1.7|E_{c1}|$ the transition proceeds
over a few dynamical times (crosses and solid line). Moreover, the comparison
of both panels shows that this time-scale does not display a significant 
dependence on the number of particles. Indeed, it is set by the mean-field 
dynamical instability growth rate associated with the fact that the homogeneous
equilibrium $n=0$ is linearly unstable for the Vlasov dynamics.

Secondly, at energies above $-1.7|E_{c1}|$, closer to the 
critical energy $E_{c1}$, we find that the transition times 
$t_{0\rightarrow\pm 1}$ can be quite large and increase 
with the number of particles. We can also note that they exhibit a very broad 
dispersion for different random realizations.

Thirdly, at low energies below $E_{c2}\simeq -2.2 |E_{c1}|$ we note that for 
some realizations of the homogeneous initial state the system does not directly
relax to the equilibria $n=\pm 1$. It first exhibits a transition to the 
equilibrium $n=2$ (which appears below the second critical energy $E_{c2}$) 
over a few $t_{\rm dyn}$ and remains trapped in this 
two-peak configuration over a long time-scale until it eventually evolves to 
a one-peak state $n=\pm 1$. This leads to the very long transition times
labeled $t_{0\rightarrow 2\rightarrow\pm 1}$ shown by the squares in 
Fig.~\ref{figtime0to1} (and the dotted curve for their average).

We shall discuss these three regimes in more details below.

\subsubsection{\label{dynamical01}Dynamical relaxation to $n=\pm 1$}

\begin{figure}
\begin{center}
\epsfxsize=4.4 cm \epsfysize=4.8 cm {\epsfbox{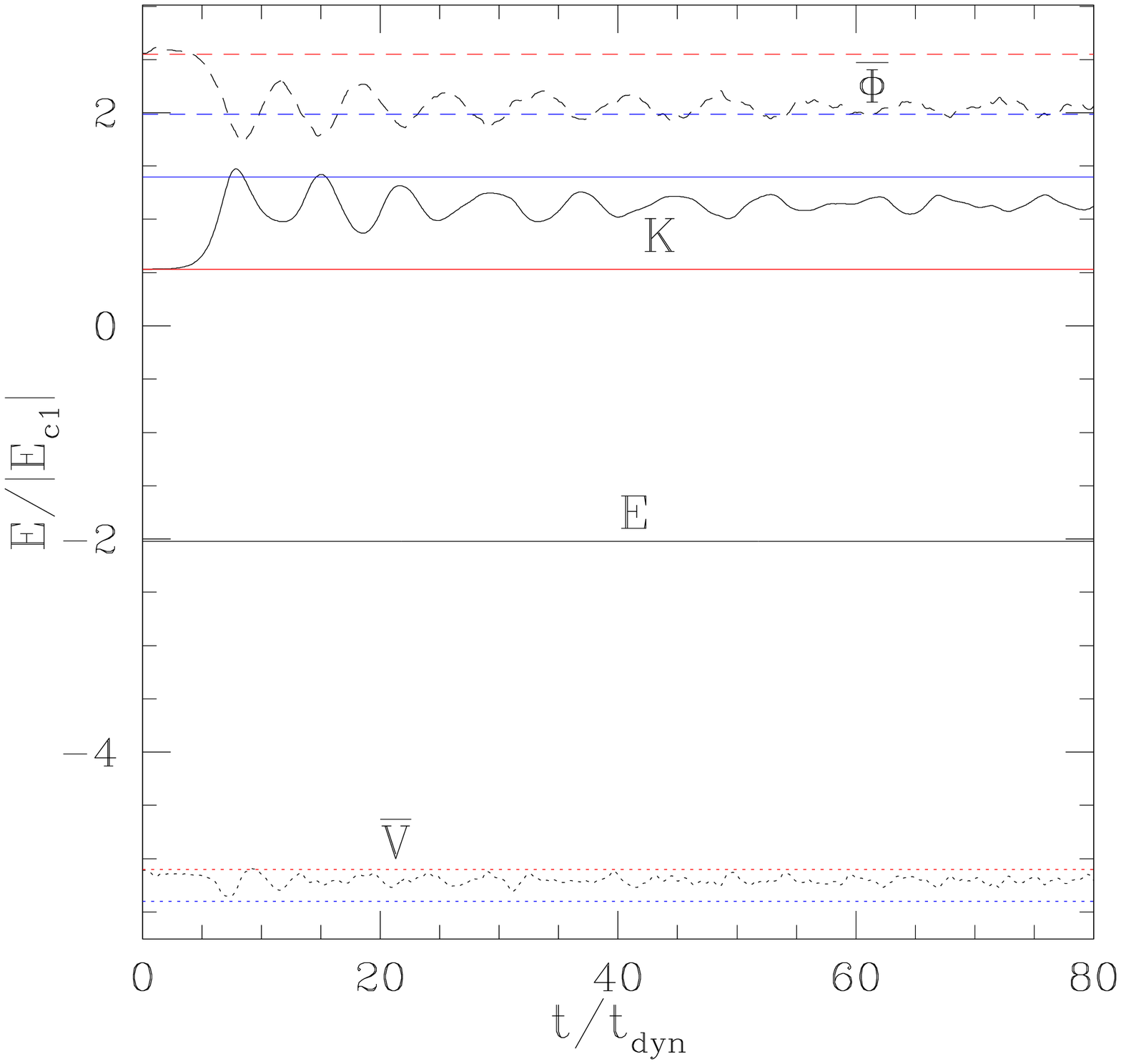}}
\epsfxsize=3.9 cm \epsfysize=4.8 cm {\epsfbox{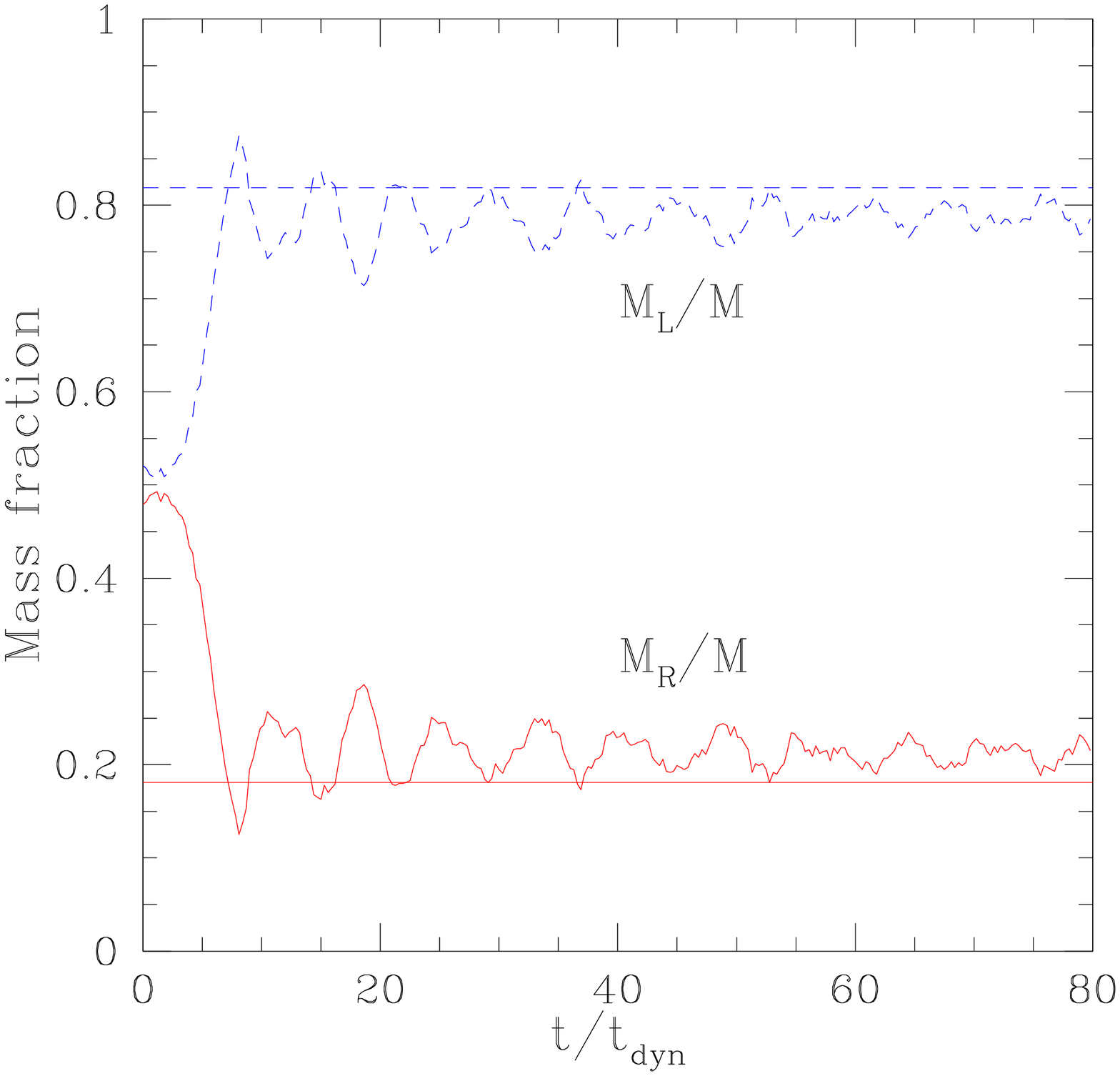}}
\end{center}
\caption{\label{figEM0to1}(Color online) {\it Left panel:} The evolution 
with time $t$ of 
the various contributions $K,\Phib$ and $\Vb$ to the total energy $E$ of a 
system of $N=1000$ particles. We display the curves obtained for a particular 
realization of the initial condition defined by the unstable equilibrium 
$n=0$ at energy $E = -2|E_{c1}|$. The constant curves are the 
mean-field energy levels of $K$ (solid lines), $\Phib$ (dashed lines) and 
$\Vb$ (dotted lines) for the equilibria $n=0$ and $n=\pm 1$. The system starts 
at levels $n=0$ at $t=0$ and undergoes a transition to levels $n=1$ at 
$t \simeq 7 t_{\rm dyn}$. {\it Right panel:}  The evolution with time of the
masses $M_L$ (dashed lines) and $M_R$ (solid lines) located in the left and 
right parts of the system ($x<L/2$ and $x>L/2$). The constant lines show 
the values $M_L,M_R$ of equilibrium $n=1$.}
\end{figure}

We first consider a typical configuration of the first regime which exhibits
a relaxation to equilibrium $\pm 1$ driven by the collective dynamical
instability. Thus we choose a particular realization with $N=1000$ particles 
of the initial condition defined by the unstable equilibrium $n=0$ at 
$E=-2|E_{c1}|$. We show in left panel of Fig.~\ref{figEM0to1} the evolution 
of the various contributions (\ref{EN1})-(\ref{EN2}) to the total energy $E$. 
The fluctuating curves are the
gravitational self-energy $\Phib$, the kinetic energy $K$ and the external
potential energy $\Vb$ of the $N-$body system (from top to bottom). 
The constant curve labeled $E$ is the total energy which is conserved 
(as verified in Fig.~\ref{figEM0to1}). 
The other constant curves show the mean-field energy 
levels of $K,\Phib$ and $\Vb$ for the equilibria $n=0$ and $n=\pm 1$. Thus, we
can see that the system starts at levels $n=0$ and displays a transition 
to levels $n=\pm 1$ over a few dynamical times $t_{\rm dyn}$. The fact that 
the transition time-scale is of order of a few $t_{\rm dyn}$ and does not 
depend on the number of particles, as was checked in Fig.~\ref{figtime0to1} 
for $E<-1.7|E_{c1}|$, is due to the collective character of the
instability. Indeed, as shown in 
\cite{Valageas2006} the thermodynamical equilibrium $n=0$ becomes linearly 
unstable below the critical energy $E_{c1}$ for the mean-field Vlasov dynamics.
The growth rate of this instability saturates to $e^{t/t_{\rm dyn}}$ at
low energies \cite{Valageas2006} which sets the time-scale for the transition
to equilibrium $n=\pm 1$. The right panel in Fig.~\ref{figEM0to1} shows the
evolution of the masses $M_L$ and $M_R$ located in the left and right parts 
of the system. It also clearly shows the dynamical instability which amplifies
the small initial random imbalance $M_L-M_R$ to build a left-peak state after
a few $t_{\rm dyn}$. We can note that the relaxation is not fully complete
as the system still exhibits some mean deviation and oscillations from the 
levels $n=1$ for $K$ and $\Vb$ (and for masses $M_L$ 
and $M_R$), although the largest energy component $\Phib$ 
relaxes to its equilibrium level after $\sim 60 t_{\rm dyn}$. We found that
these small departures persist up to $10^4 t_{\rm dyn}$.

\begin{figure}
\begin{center}
\epsfxsize=4.15 cm \epsfysize=4.5 cm {\epsfbox{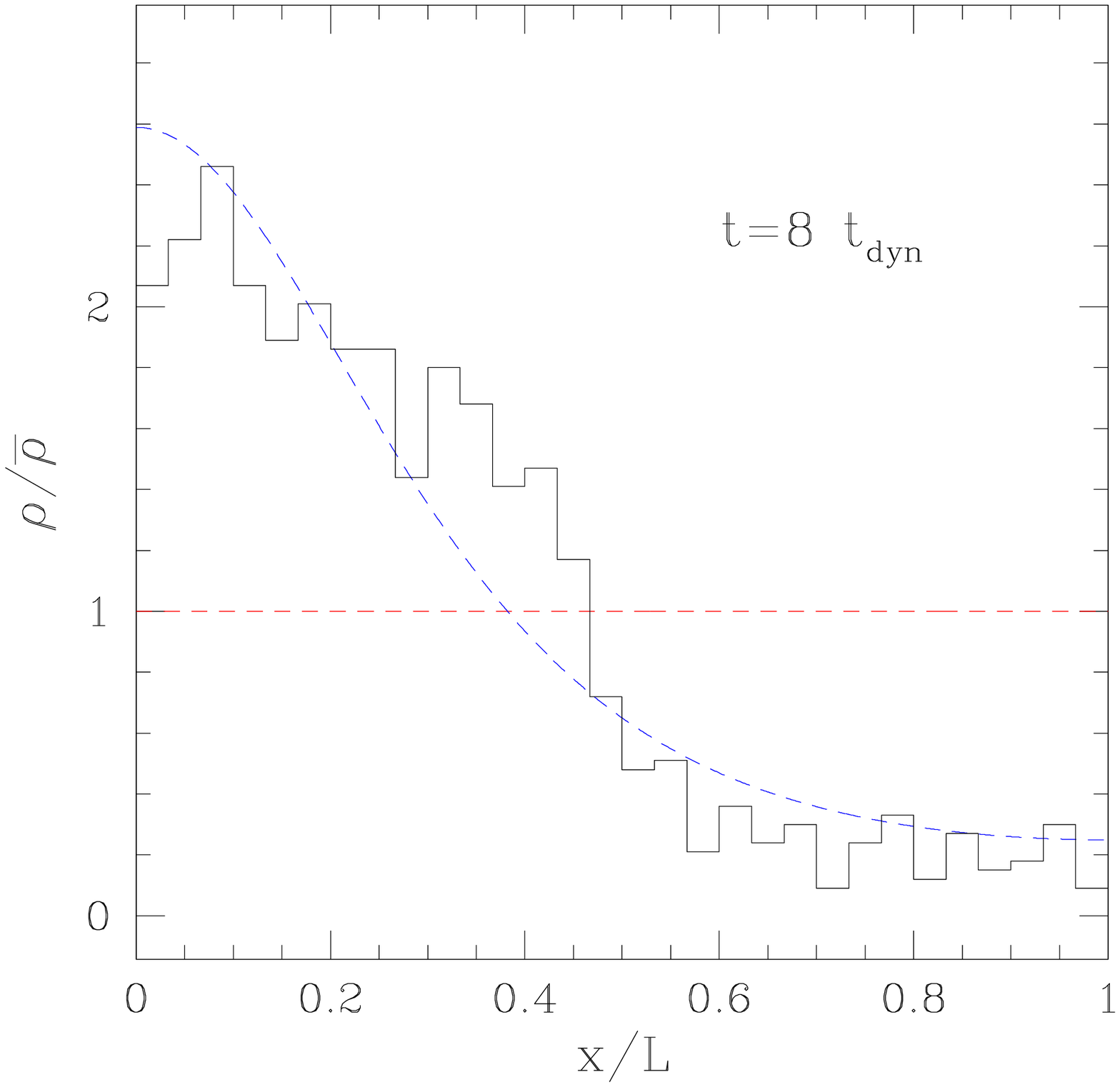}} 
\epsfxsize=4.15 cm \epsfysize=4.5 cm {\epsfbox{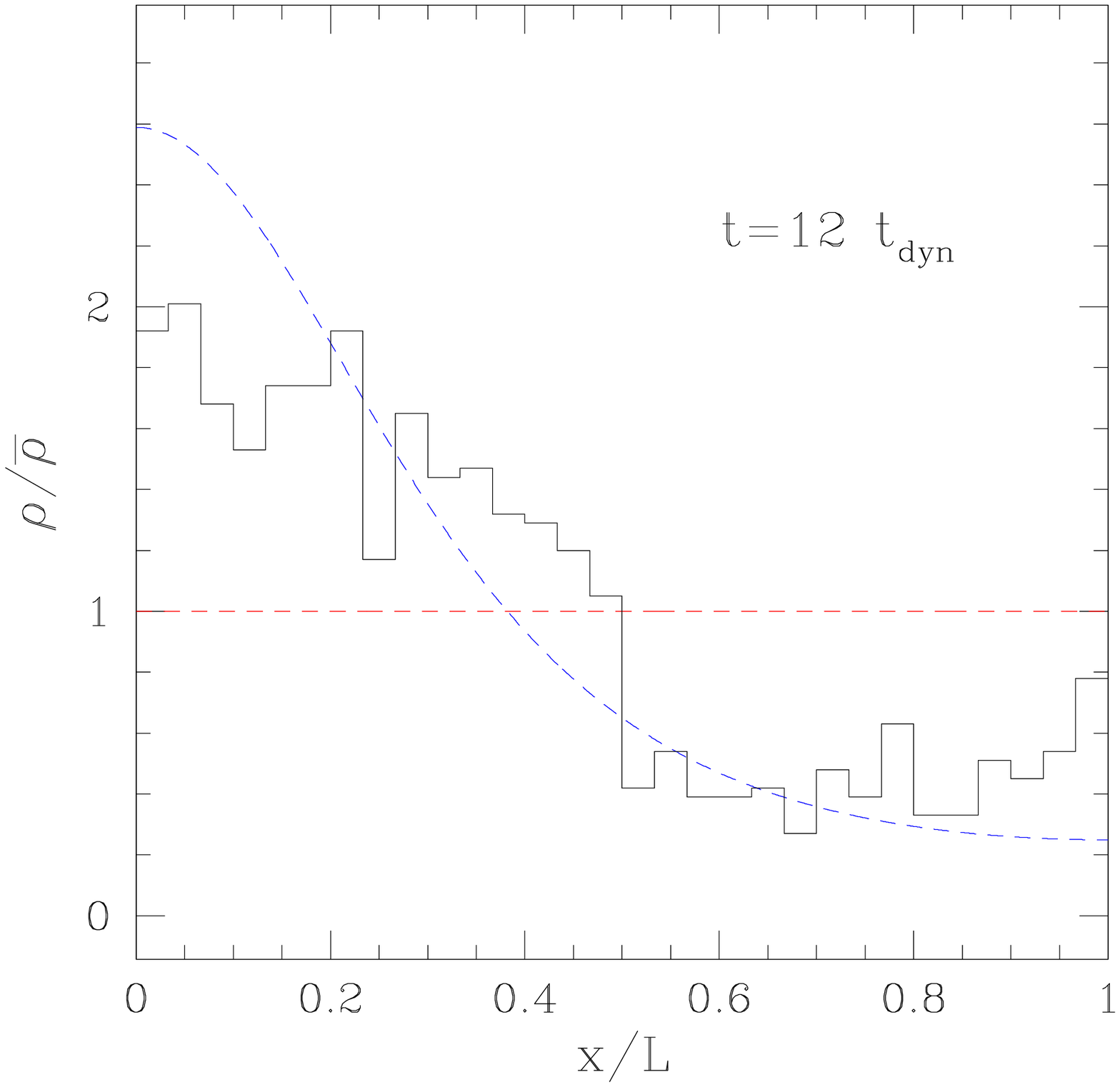}}
\end{center}
\caption{\label{figrho0to1}(Color online) Two snapshots of the density 
distribution $\rho(x)$
at times $t=8$ and $12 t_{\rm dyn}$. The histogram shows the
matter distribution of the $N-$body system over $30$ bins. The dashed curves
are equilibria $n=0$ (constant density) and $n=1$ (peak at $x=0$).}
\end{figure}

We present in Fig.~\ref{figrho0to1} two snapshots of the density distribution
for the realization used in Fig.~\ref{figEM0to1}, at times $t=8$ and 
$12 t_{\rm dyn}$. We can see that the matter distribution has indeed evolved
from the homogeneous state to a one-peak configuration close to equilibrium 
$n=1$ at $t=8 t_{\rm dyn}$. However, it has not yet fully relaxed and some
collective oscillations persist as already seen in Fig.~\ref{figEM0to1}, which
also lead to oscillations of the density and width of the peak.

\begin{figure}
\begin{center}
\epsfxsize=4.15 cm \epsfysize=4.5 cm {\epsfbox{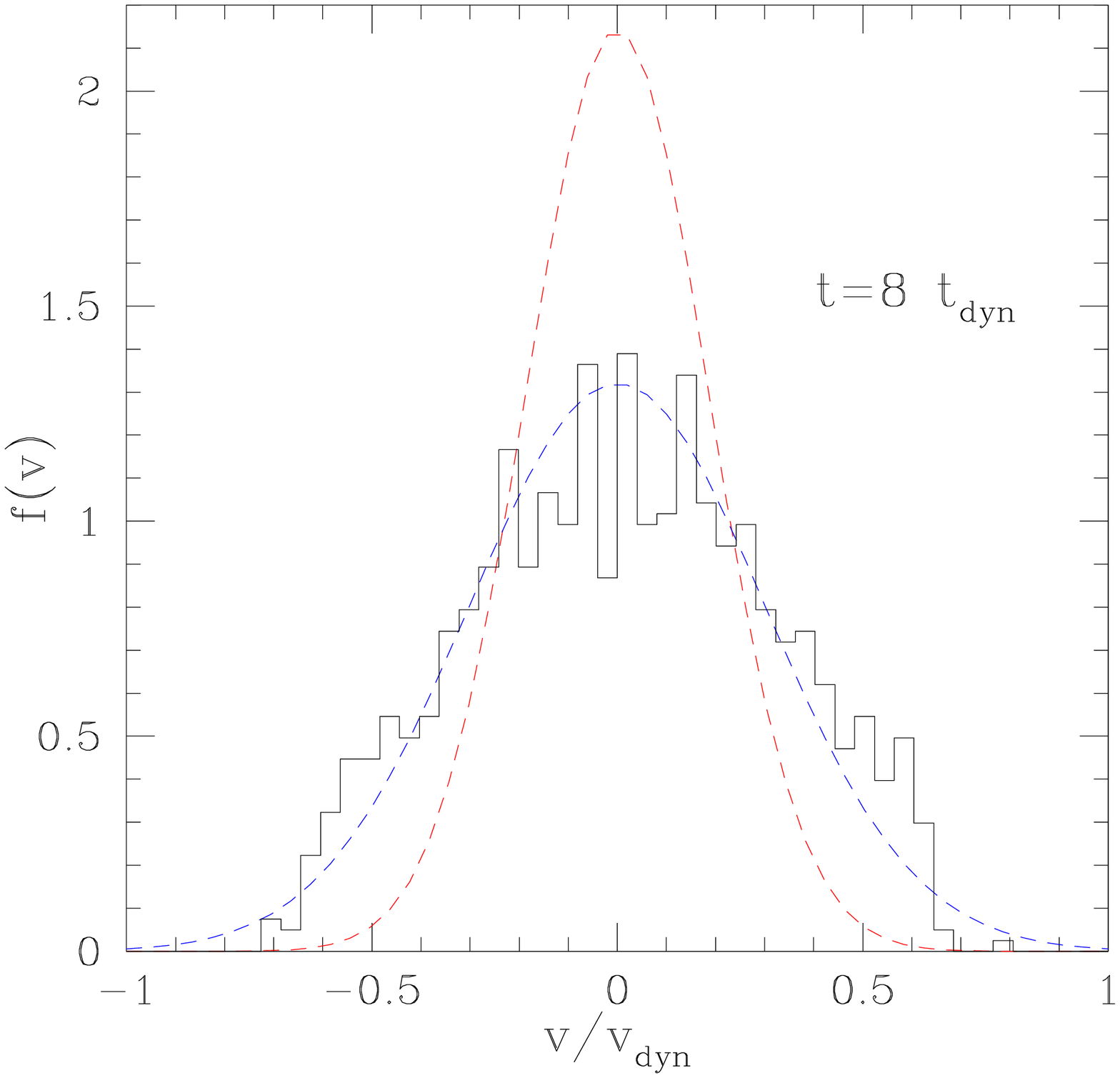}} 
\epsfxsize=4.15 cm \epsfysize=4.5 cm {\epsfbox{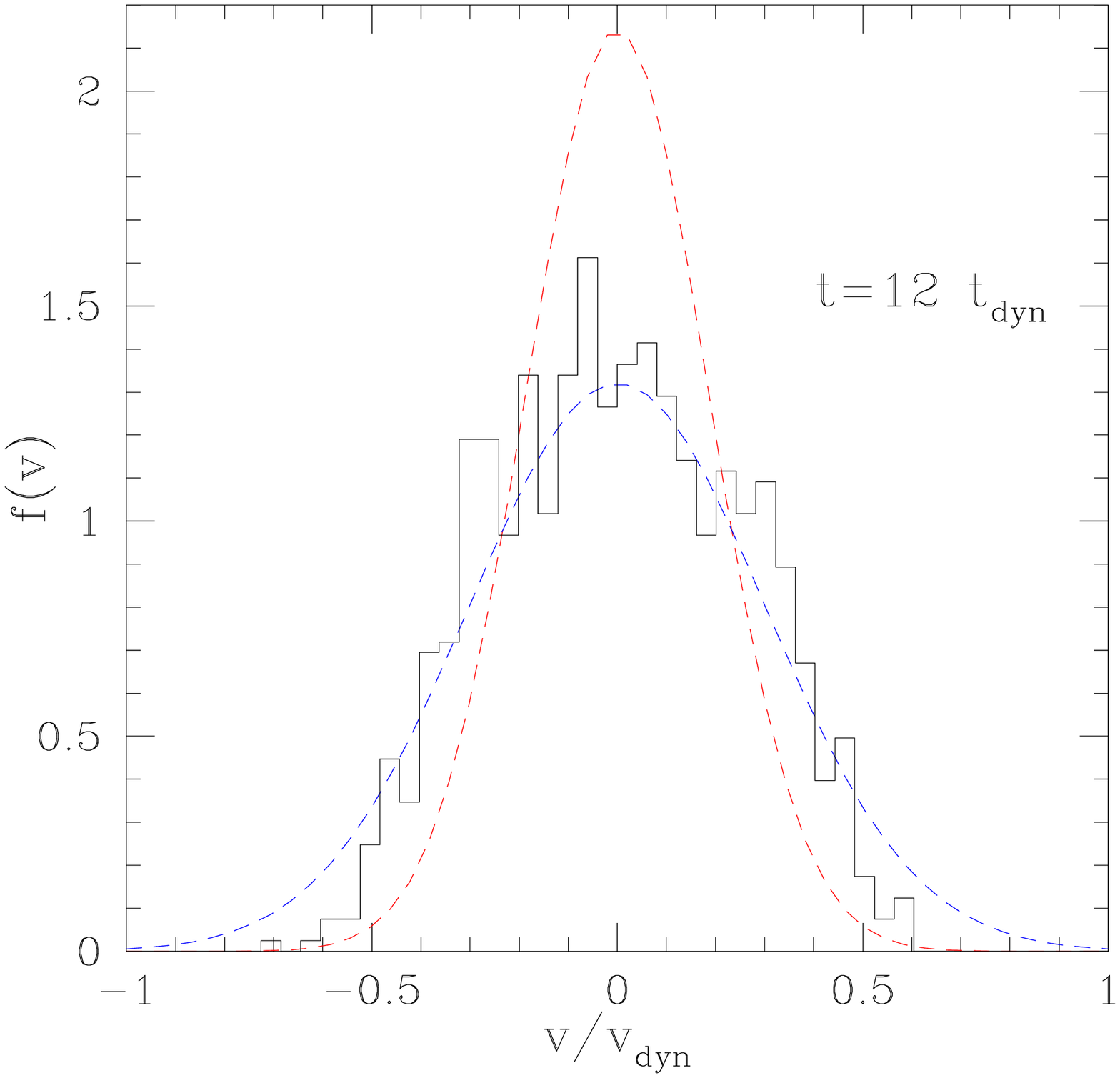}}
\end{center}
\caption{\label{figfv0to1}(Color online) Two snapshots of the velocity 
distribution $f(v)$
at times $t=8$ and $12 t_{\rm dyn}$, in units of $v/v_{\rm dyn}$ where
we defined $v_{\rm dyn}=L/t_{\rm dyn}$. The 
histogram corresponds to the $N-$body system whereas the dashed curves
are equilibria $n=0$ (narrow Gaussian) and $n=1$ (broad Gaussian).}
\end{figure}

We show in Fig.~\ref{figfv0to1} two snapshots of the velocity distribution
at the same times $t=8$ and $12 t_{\rm dyn}$. It has again evolved from the 
distribution
$n=0$ to the distribution $n=1$ at time $t=8 t_{\rm dyn}$. Although there are
some oscillations in the tails (as seen from the right panel at time 
$12 t_{\rm dyn}$ and the fluctuations of $K$ in Fig.~\ref{figEM0to1}) we can 
see that the relaxation of the velocity distribution is rather efficient
and always remains close to the equilibrium $n=1$.

We checked that for a small number of particles $N=50$ the behavior of the
system is the same in this energy range. The states $n=\pm 1$ and $n=0$ are 
well separated and the system exhibits a transition to equilibrium $\pm 1$ 
over a few dynamical times.

\subsubsection{\label{Close_to_critical01}Close to the critical energy $E_{c1}$}

\begin{figure}
\begin{center}
\epsfxsize=4.4 cm \epsfysize=4.8 cm {\epsfbox{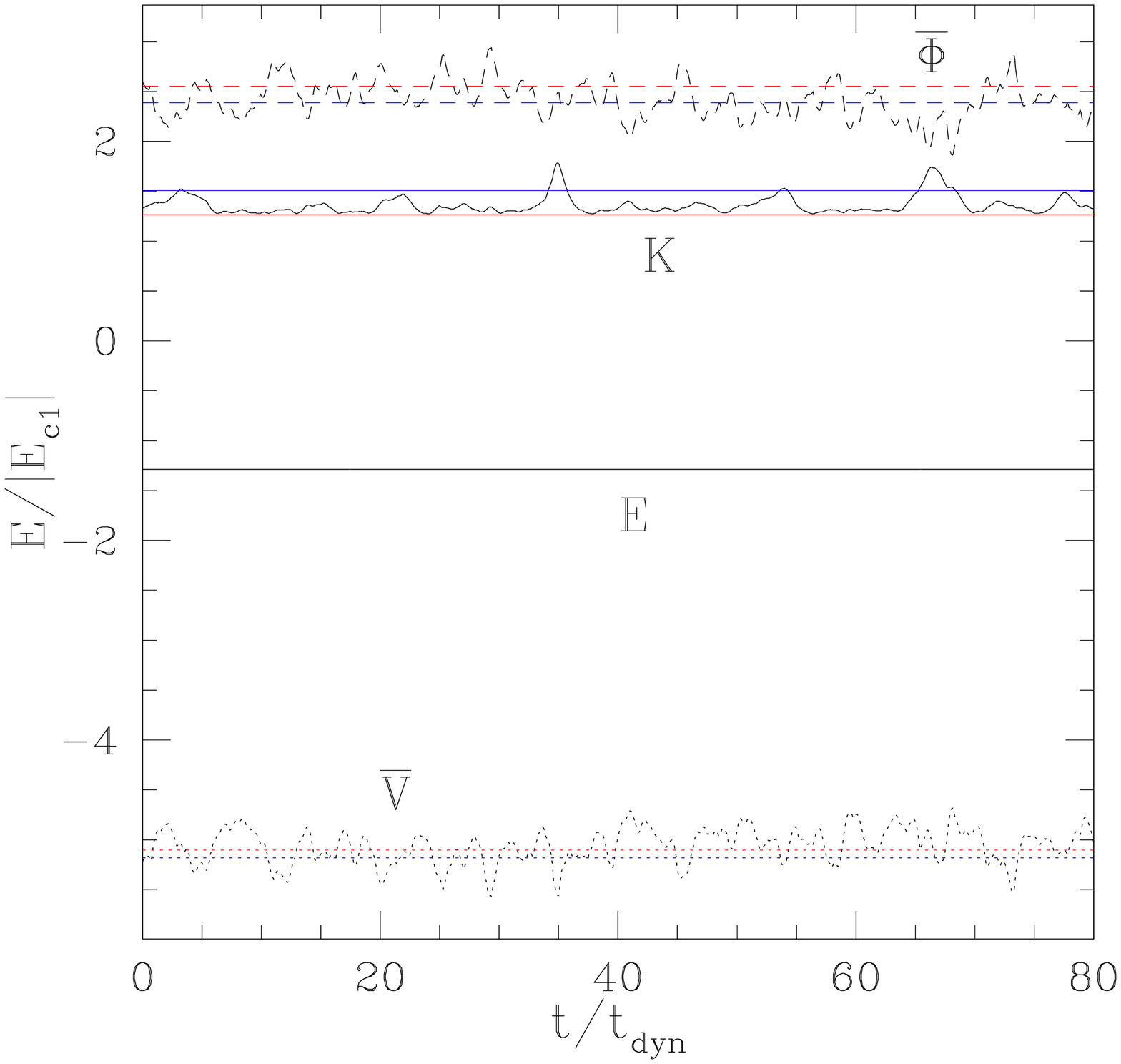}}
\epsfxsize=3.9 cm \epsfysize=4.8 cm {\epsfbox{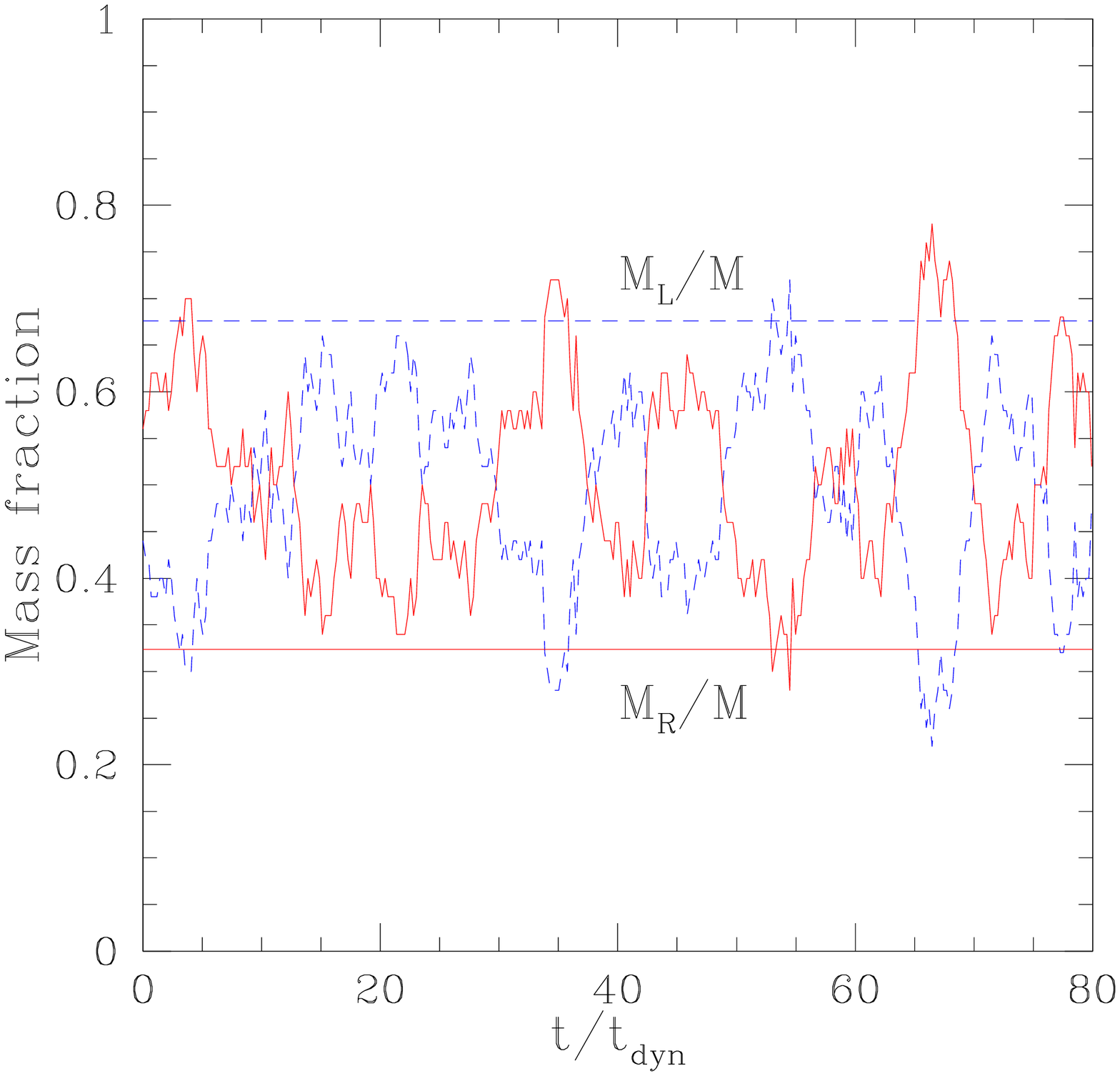}}
\end{center}
\caption{\label{figEM0to1_long50}(Color online) The evolution of the energy 
components 
$K,\Phib,\Vb$ and $E$ (left panel) and of the masses $M_L,M_R$ (right panel)
as in Fig.~\ref{figEM0to1}. We show the curves obtained for a particular 
realization with $N=50$ particles of the initial homogeneous state at
$E=-1.3|E_{c1}|$. The system keeps wandering over states $n=0, \pm 1$ and does
not settle into a stable equilibrium.}
\end{figure}

\begin{figure}
\begin{center}
\epsfxsize=4.4 cm \epsfysize=4.8 cm {\epsfbox{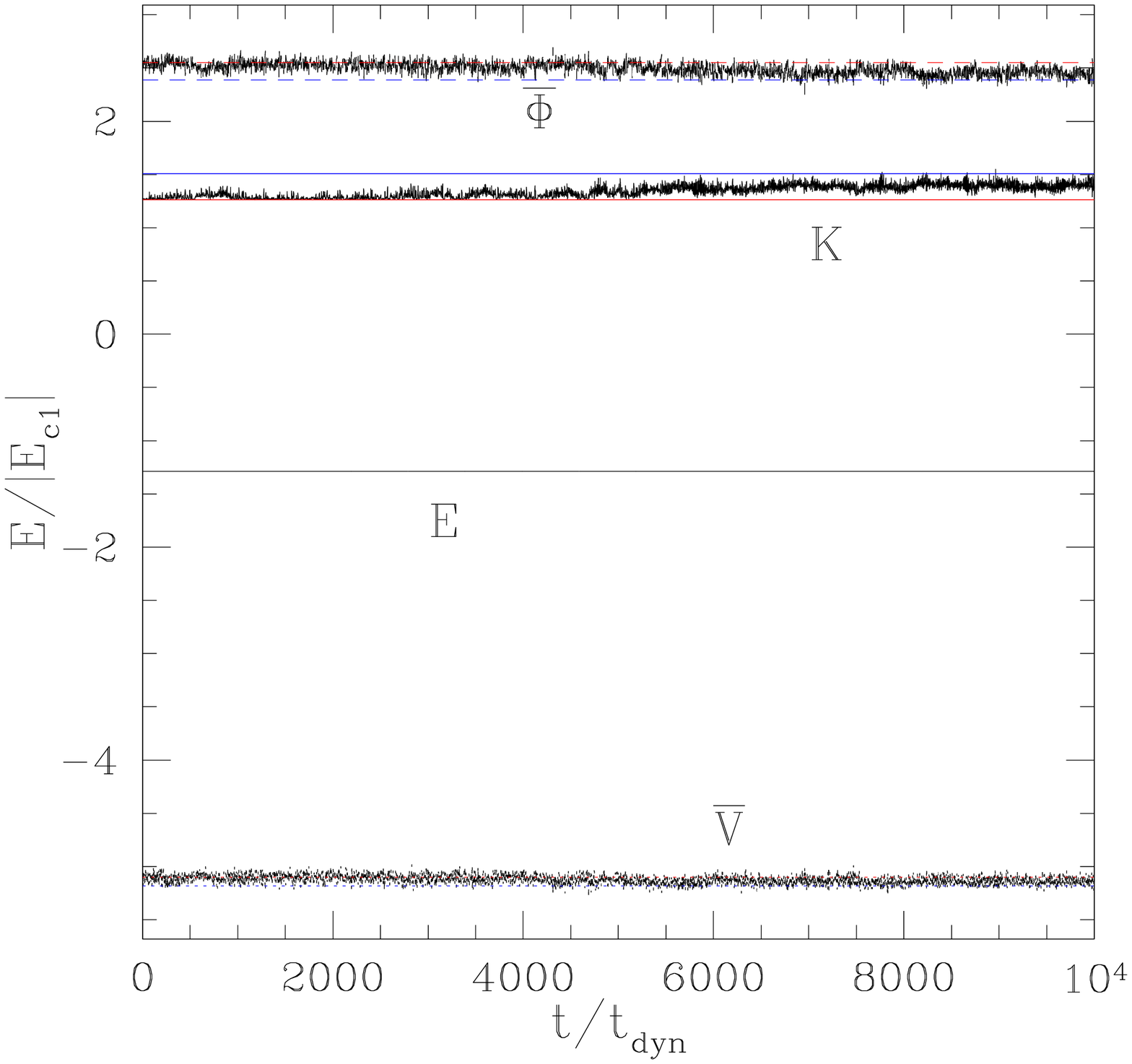}}
\epsfxsize=3.9 cm \epsfysize=4.8 cm {\epsfbox{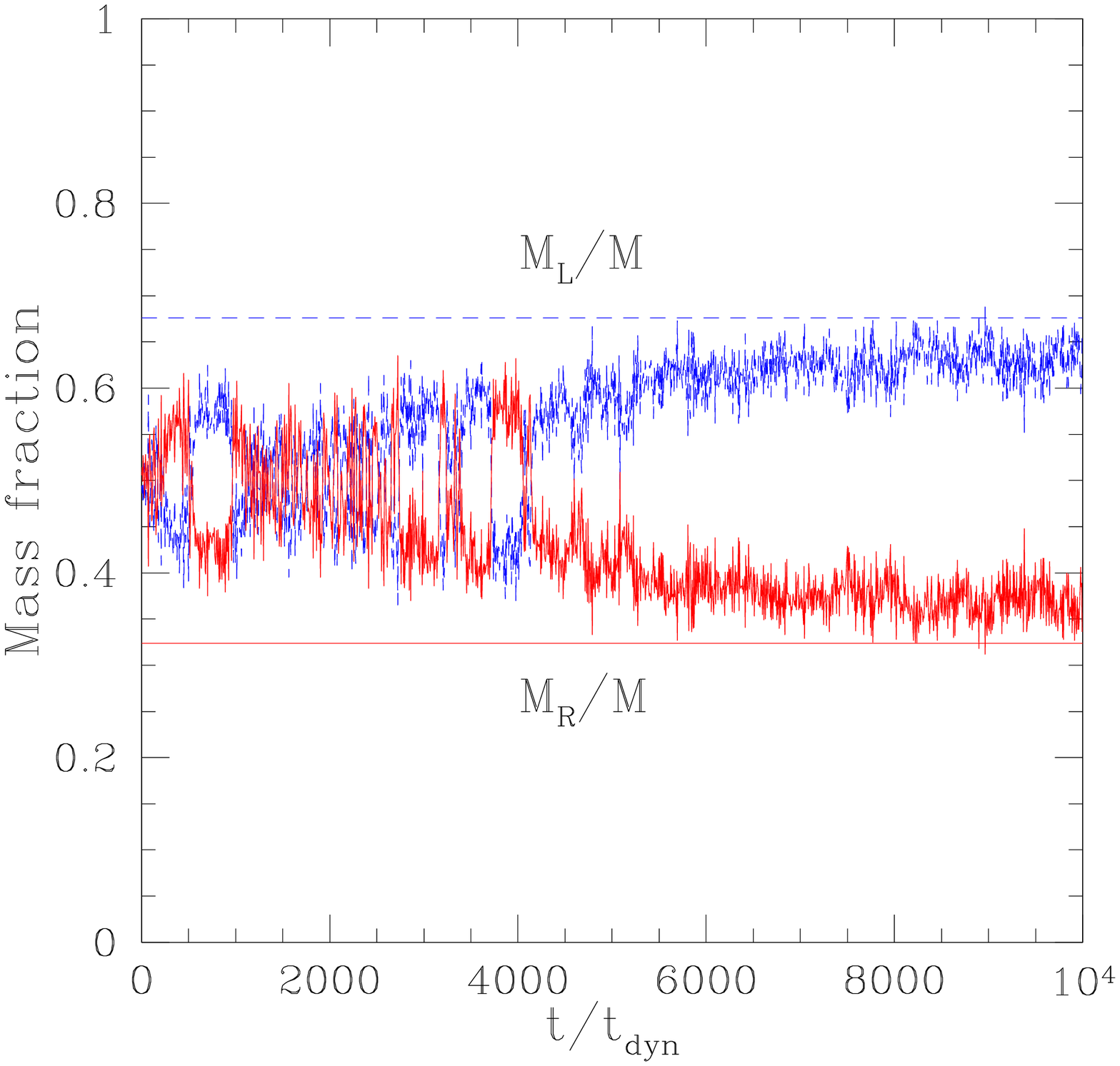}}
\end{center}
\caption{\label{figEM0to1_long1000}(Color online) The evolution of the 
energy components 
$K,\Phib,\Vb$ and $E$ (left panel) and of the masses $M_L,M_R$ (right panel)
as in Fig.~\ref{figEM0to1}. We show the curves obtained for a particular 
realization with $N=1000$ particles of the initial homogeneous state at
$E=-1.3|E_{c1}|$ which only relaxes to equilibrium $n=1$ over 
$\sim 8000 t_{\rm dyn}$.}
\end{figure}

We now investigate the evolution for higher energies 
$E>-1.7|E_{c1}|$, closer to the critical energy $E_{c1}$. We first display in
Fig.~\ref{figEM0to1_long50} the evolution of the energy and mass components 
for a system of $N=50$ particles at energy $E=-1.3|E_{c1}|$. The right
panel shows that the system evolves over a time scale set by the dynamical
time $t_{\rm dyn}$, in agreement with Fig.~\ref{figtime0to1}, but it does not
settle into a stable equilibrium. Indeed, it keeps fluctuating indefinitely
from left-peak to right-peak configurations and wanders over states 
$n=0, \pm 1$; this can also be seen from the fluctuating energy levels in 
left panel of Fig.~\ref{figEM0to1_long50} (we checked numerically that this
fluctuating behavior remains unchanged up to $10^4 t_{\rm dyn}$ at least).
Therefore, in this regime the transition time $t_{0\rightarrow\pm 1}$
which was shown in Fig.~\ref{figtime0to1} only corresponds to the first time
where the system gets close to a state characterized by a density peak close
to a boundary, as equilibria $\pm 1$, but the fluctuations are too large to
let the system settle down in such a stable configuration. Hence it keeps 
exploring various states with both left or right overdensities.

Next, we show in Fig.~\ref{figEM0to1_long1000} the evolution of a system
of $N=1000$ particles with the same energy $E=-1.3|E_{c1}|$. Because of the
larger number of particles the system starts with a more balanced state
$(M_L-M_R)/M \sim 1/\sqrt{N}$ and the fluctuations with time of various
quantities are smaller. Therefore, we find that the system again wanders for 
a long time over states with $M_L\sim M_R$ and even exhibits several 
oscillations from left-peak to right peak configurations but it now manages
to eventually settle down into a stable equilibrium $n=\pm 1$ after 
$\sim 8000 t_{\rm dyn}$.
Indeed, as the system starts close to $M_L=M_R$ small fluctuations are
initially sufficient to evolve from $M_L>M_R$ to $M_L<M_R$ but once it has
converged close to either one of equilibria $\pm 1$ the fluctuations are no
longer sufficient to escape to the symmetric state.

\subsubsection{\label{Below_second_critical}Below the second critical energy $E_{c2}$}

\begin{figure}
\begin{center}
\epsfxsize=4.15 cm \epsfysize=4.5 cm {\epsfbox{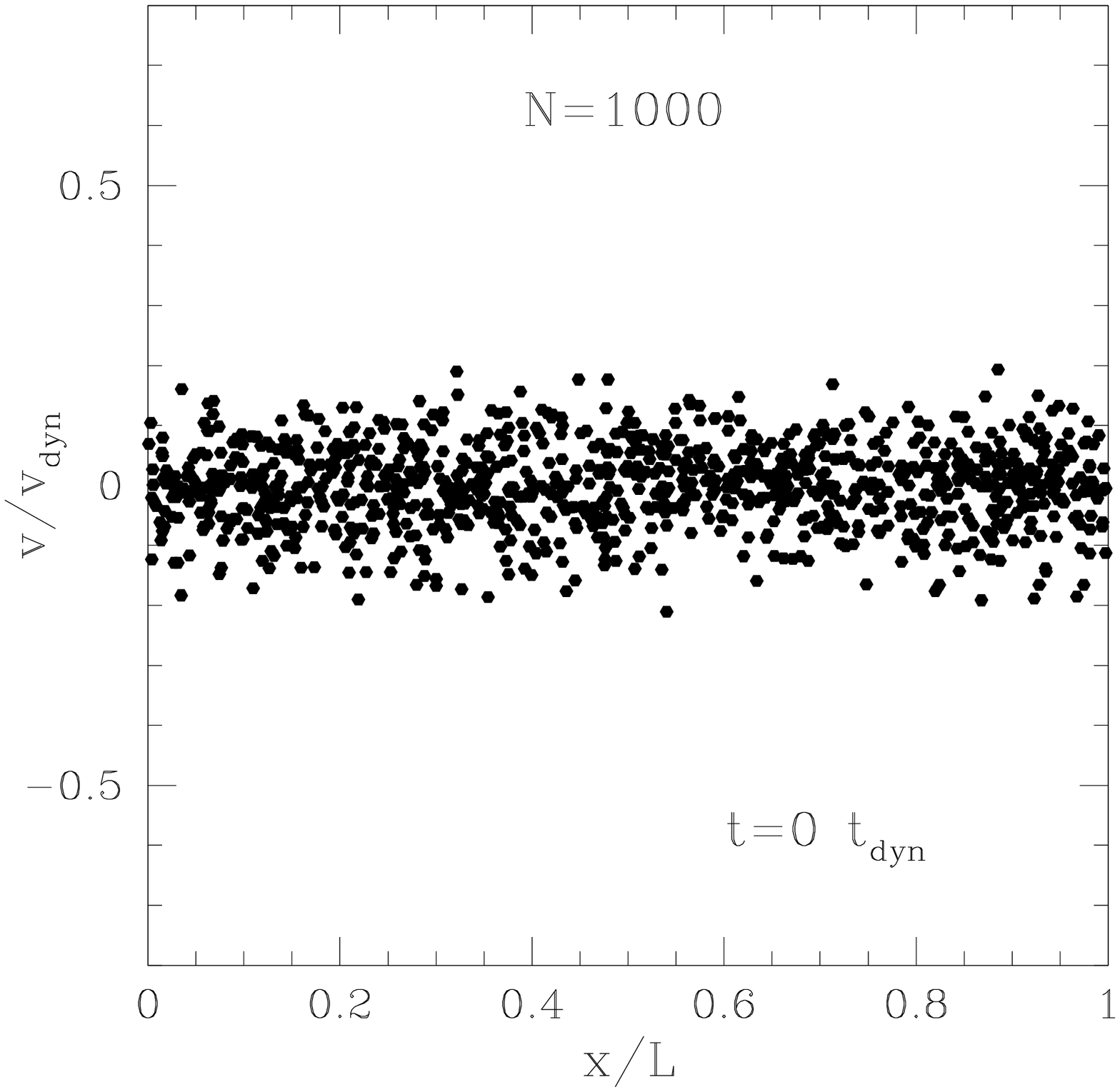}} 
\epsfxsize=4.15 cm \epsfysize=4.5 cm {\epsfbox{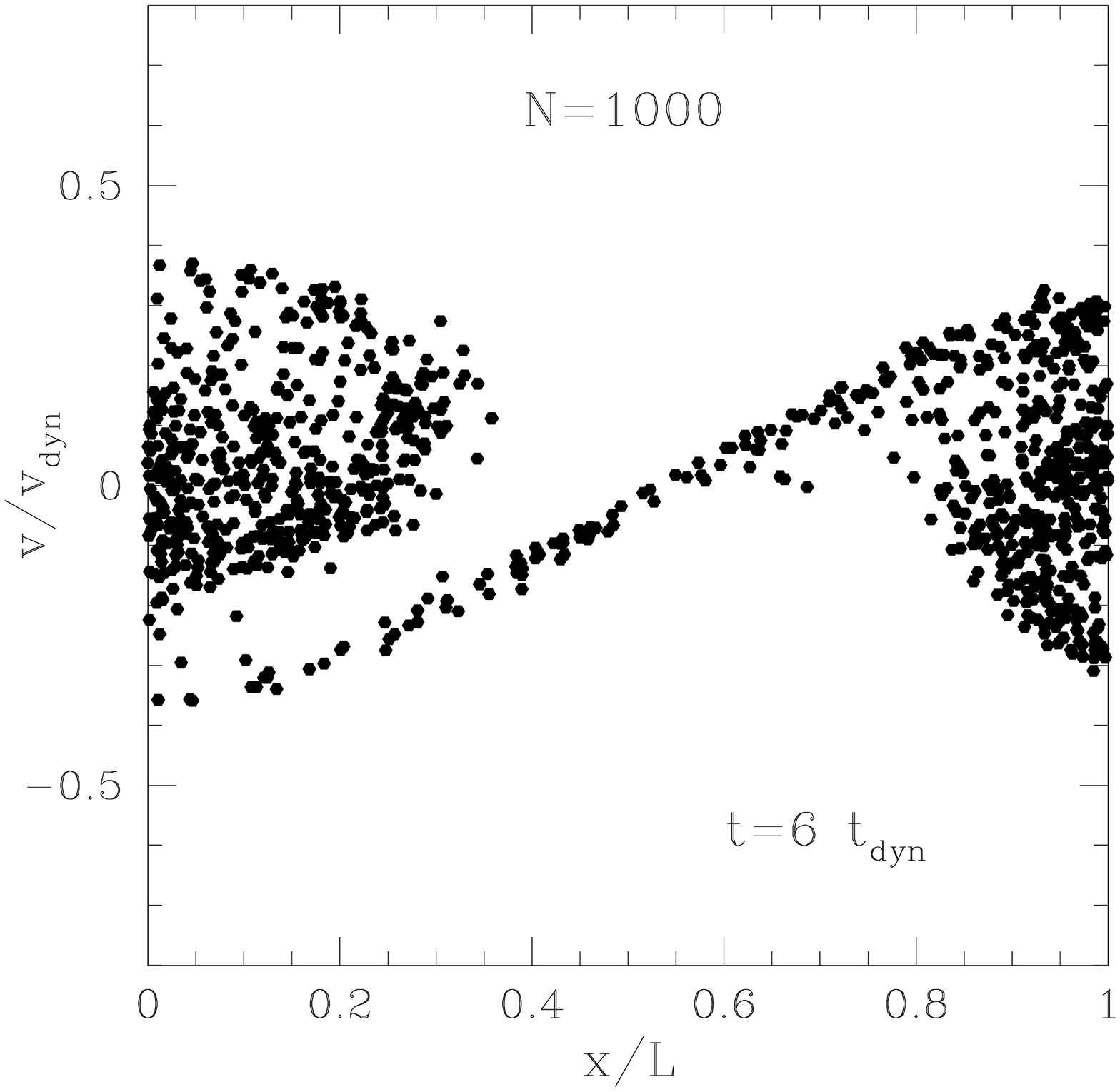}}\\
\epsfxsize=4.15 cm \epsfysize=4.5 cm {\epsfbox{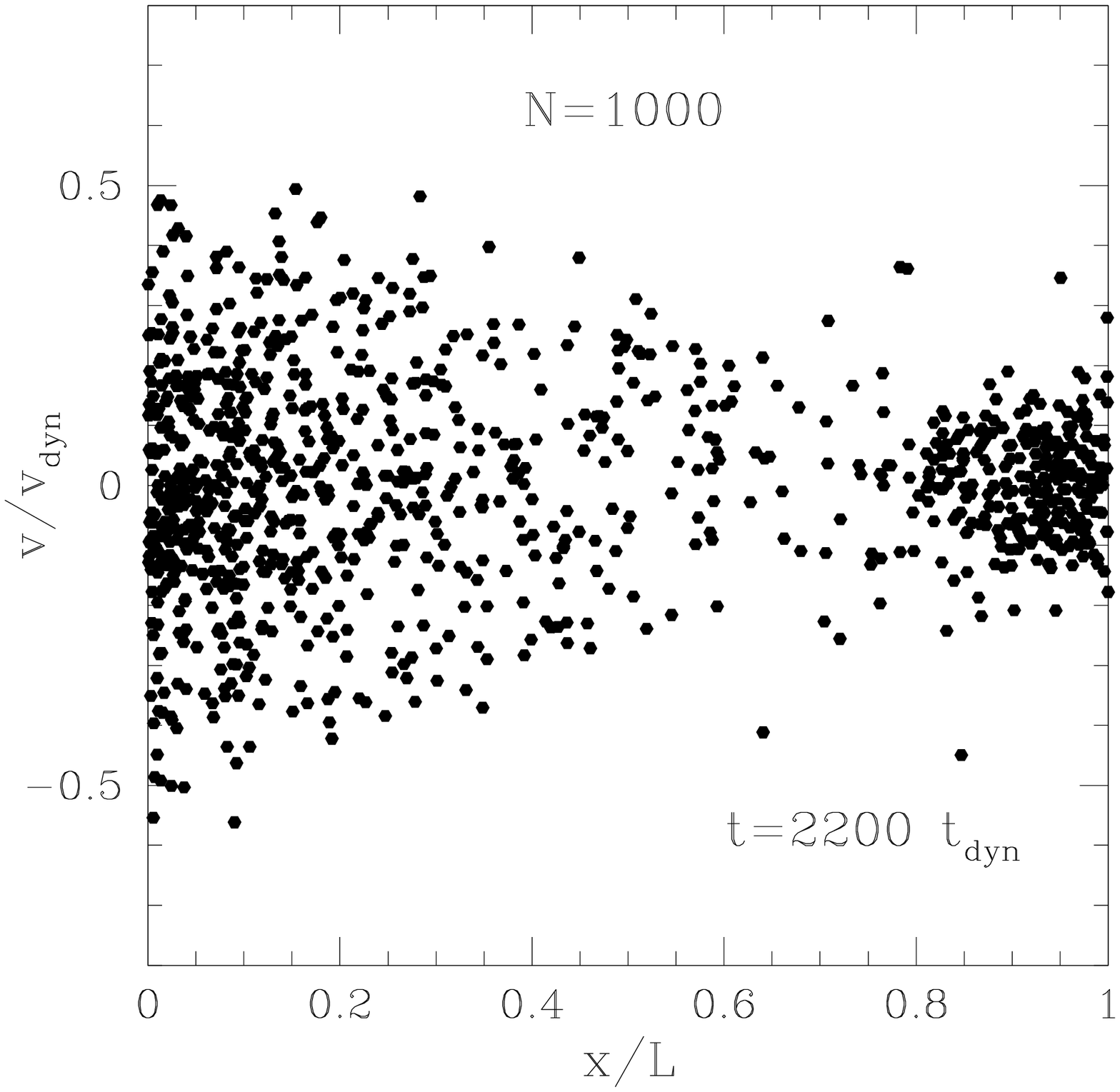}} 
\epsfxsize=4.15 cm \epsfysize=4.5 cm {\epsfbox{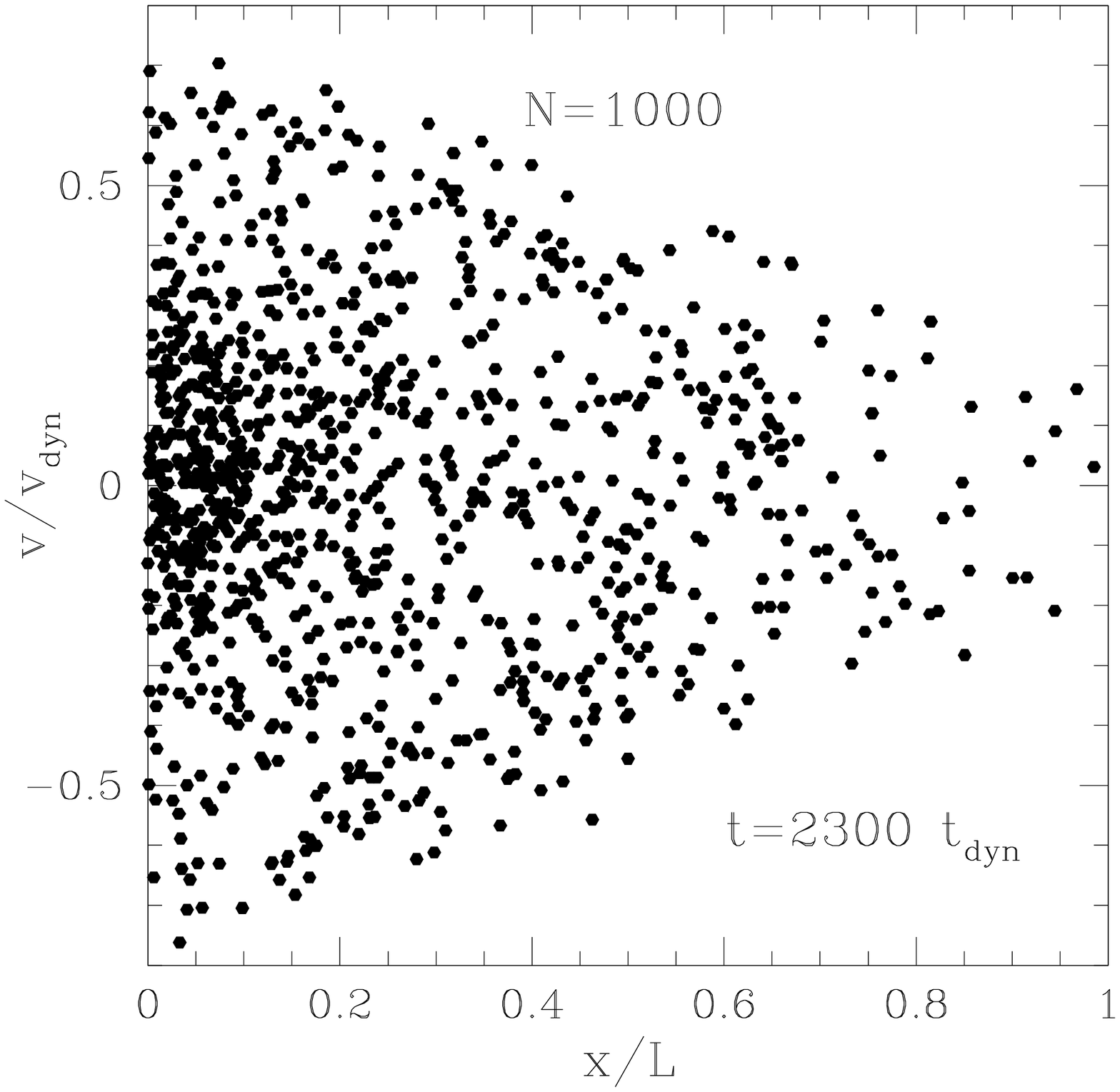}}
\end{center}
\caption{\label{figphase0to1}(Color online) Four snapshots of the 
phase-space distribution 
$f(x,v)$ at times $t=0, 6, 2200$ and $2300 t_{\rm dyn}$ for a system of 
$N=1000$ particles. For this particular realization of the initial homogeneous
state $n=0$ the system first undergoes a transition to the equilibrium $n=2$
over $6 t_{\rm dyn}$ and only relaxes to the one-peak state $n=1$ at 
$t \sim 2250 t_{\rm dyn}$.}
\end{figure}

Finally, we noticed in Fig.~\ref{figtime0to1} that at low energies below 
$E_{c2}$ there are cases where the system does not exhibit a direct transition
to stable equilibria $\pm 1$. It first evolves to the two-peak equilibrium
$n=2$ over a few dynamical times and next relaxes to the stable 
equilibrium $n=\pm 1$ over a much longer time-scale.
We show in Fig.~\ref{figphase0to1} four snapshots of the phase-space 
distribution $f(x,v)$ for such a realization at $E=-2.5|E_{c1}|$, with
$N=1000$ particles, which first relaxes to state $n=2$. Thus, we can see
that the system has already evolved from an homogeneous configuration to
a two-peak state at $t=6 t_{\rm dyn}$. It remains in such a configuration 
until $\sim 2250 t_{\rm dyn}$ while particles slowly diffuse out of the right
density peak and at $t=2300 t_{\rm dyn}$ it has relaxed to the stable 
equilibrium $n=1$. Of course the energy levels show the same transitions
from $n=0$ to $n=2$ and finally to $n=1$. In agreement with the calorific 
curve shown in Fig.~4 of \cite{Valageas2006} we note that for a fixed total 
energy $E$ the temperature grows (the velocity distribution becomes broader) 
as we go from state $n=0$ to $n=2$ and finally to $n=1$.
We shall investigate in more details 
the transition from equilibrium $n=2$ to state $n=1$ in sect.~\ref{n=2} below.
For systems with $N=50$ particles we found that $58$ out of $200$ realizations
exhibit this two-stage behavior at $E=-2.5|E_{c1}|$ whereas for $N=1000$ 
particles this only occurs for $30$ out of $200$ realizations.
Indeed, as seen in \cite{Valageas2006} the linear growth rate of the two-peak
instability obtained for the mean-field Vlasov dynamics is smaller than the
growth rate associated with a one-peak perturbation (higher wavenumbers are
less unstable thanks to the finite temperature, following the usual Jeans
instability). Therefore, systems with a larger number of particles which
follow more closely the mean-field dynamics should more frequently evolve
directly towards a one-peak state.

\subsection{\label{n=2}Initial equilibrium $n=2$}

We now study in this section the relaxation to the stable equilibrium 
$n=\pm 1$ starting from the equilibrium $n=2$.

\subsubsection{\label{time21}Transition times}

\begin{figure}
\begin{center}
\epsfxsize=4.15 cm \epsfysize=4.5 cm {\epsfbox{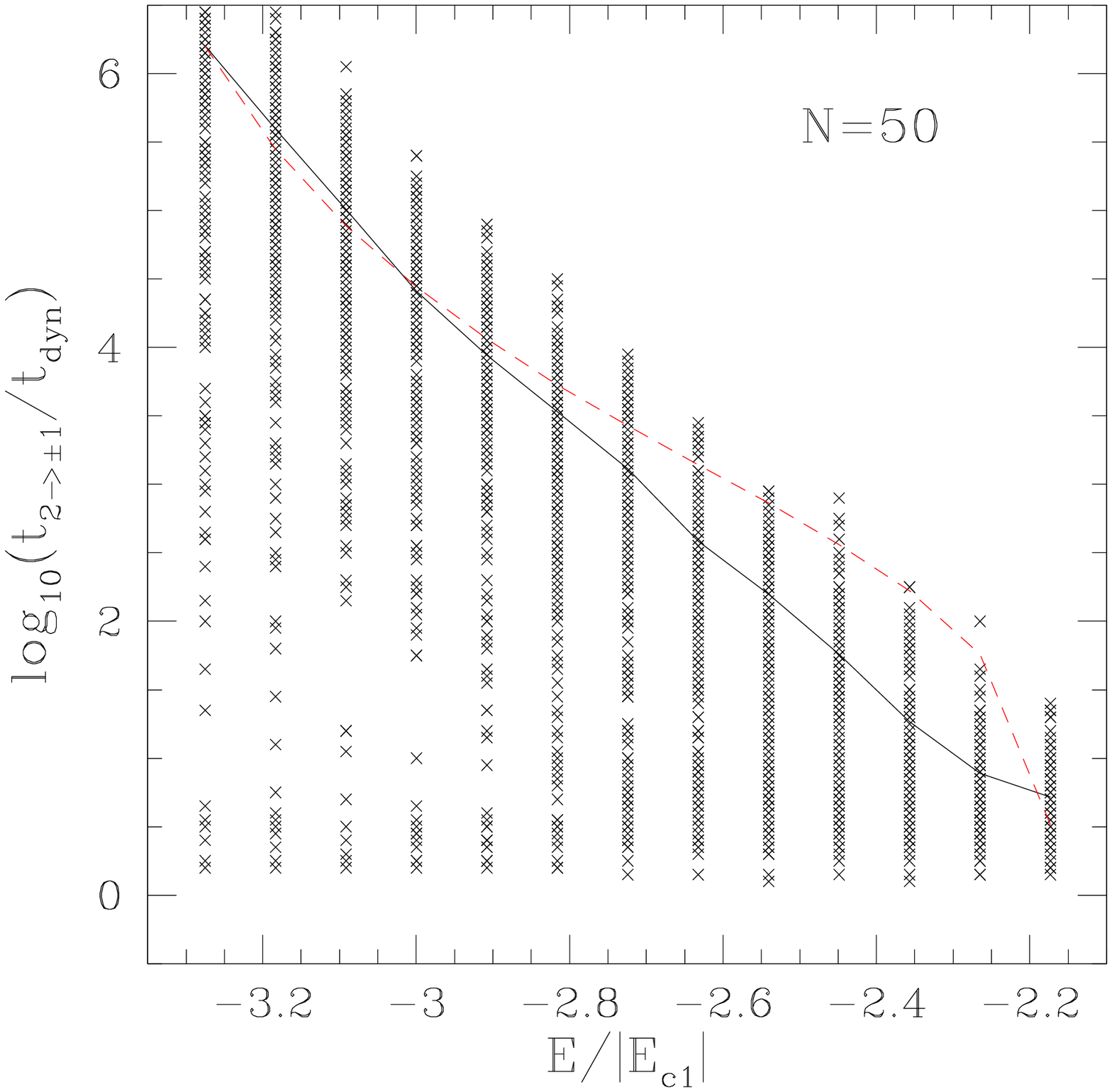}} 
\epsfxsize=4.15 cm \epsfysize=4.5 cm {\epsfbox{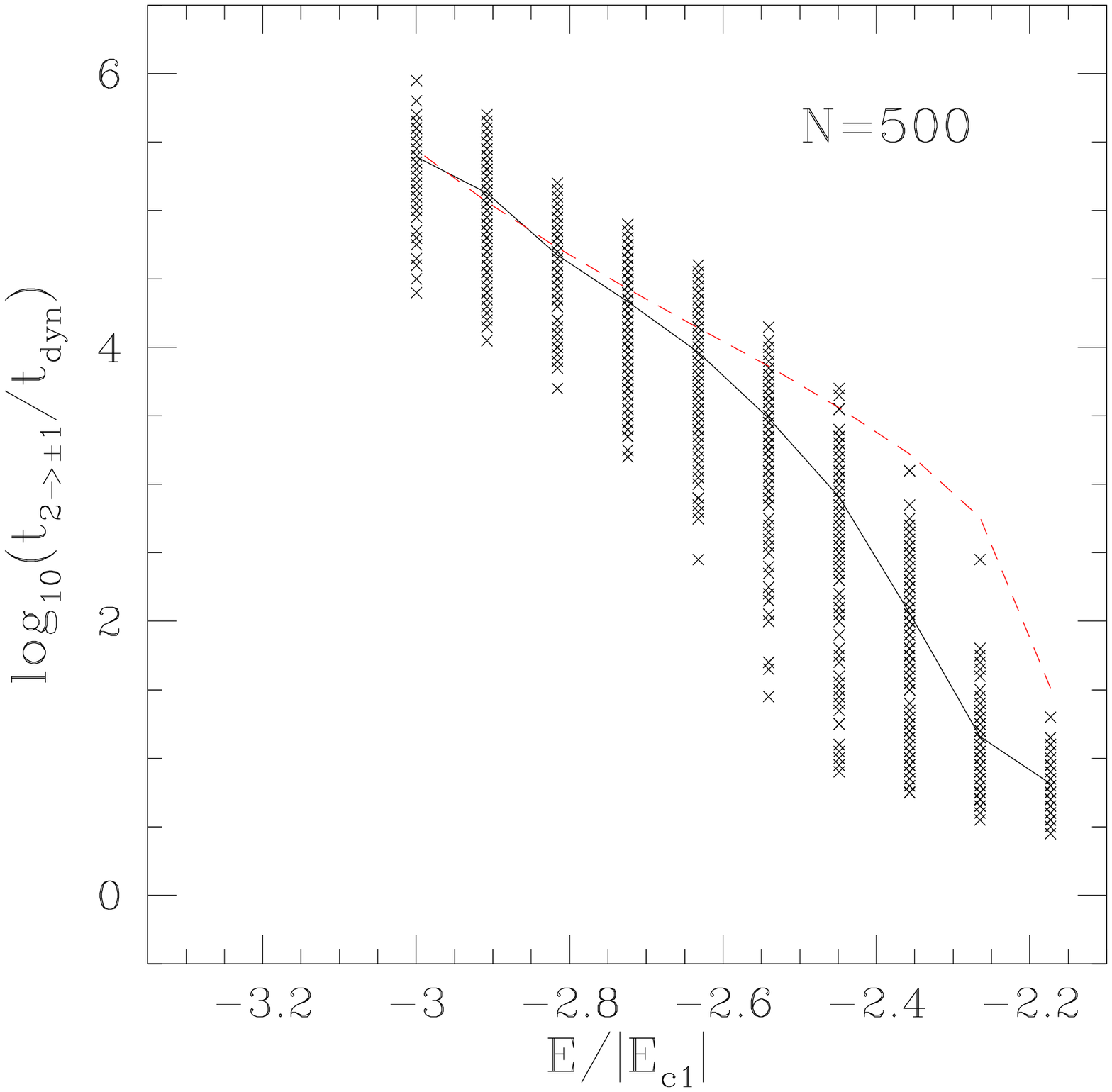}}
\end{center}
\caption{\label{figtime2to1}(Color online) The transition time 
$t_{2\rightarrow\pm 1}$ from 
the two-peak equilibrium $n=2$ to the one-peak stable equilibria $n=\pm 1$
as a function of total energy $E$. The crosses correspond to various 
realizations of the initial equilibrium $n=2$ at a given energy, for systems
of $N=50$ (left panel) and $N=500$ particles (right panel). 
The solid line is the mean transition time obtained from these realizations
whereas the dashed-line is the theoretical prediction (\ref{t21}).}
\end{figure}

We first show in Fig.~\ref{figtime2to1} the transition times 
$t_{2\rightarrow\pm 1}$ obtained for various realizations of the initial 
equilibrium $n=2$, as a function of total energy, for systems with 
$N=50$ (left panel) and $N=500$ particles (right panel). We also
plot the mean transition times derived from these realizations 
(solid line) and the theoretical prediction (dashed-line) of eq.(\ref{t21}).
As in sect.~\ref {time01} we defined the transition as the first time where 
$|K-K_1|<|K-K_2|$ and $|\Phib-\Phib_1|<|\Phib-\Phib_2|$ (i.e. the kinetic and 
self-gravity energies are closer to the levels of states $n=\pm 1$ than those 
of state $n=2$).
We can see that there is a large dispersion from one realization to another
for small systems ($N=50$) but the mean transition time agrees reasonably 
well with Eq.(\ref{t21}). For larger $N$ where discrete effects are less
violent ($N=500$) the dispersion is much smaller and we recover the 
theoretical prediction (\ref{t21}) which is derived in the limit of large $N$
(which allows a perturbative analysis). In particular, note the steep increase 
at low energies when the two narrow density peaks are separated by an almost 
void region and there is a very slow diffusion of particles out of the 
smallest peak until the system reaches the abrupt transition point discussed 
below in Fig.~\ref{figEM2to1} where a collective instability merges the 
smallest peak into the largest one. We further discuss the dependence on energy
of $t_{2\rightarrow\pm 1}$ in Sect.~\ref{Diffusion_time} below where we detail
the theoretical calculation of the transition time.

\begin{figure}
\begin{center}
\epsfxsize=7 cm \epsfysize=6.5 cm {\epsfbox{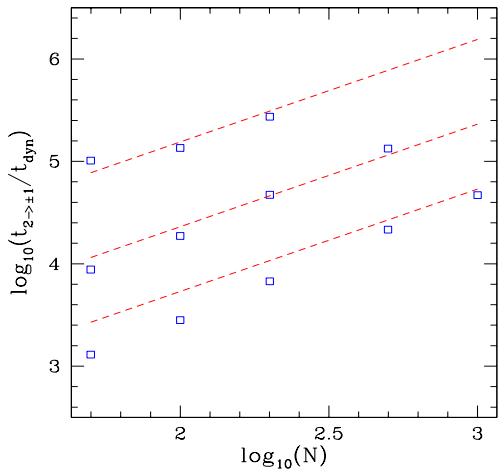}} 
\end{center}
\caption{\label{figtN2to1}(Color online) The mean transition 
time $t_{2\rightarrow\pm 1}$ 
as a function of the number $N$ of particles. The various lines correspond 
to total energies $E=-2.73|E_{c1}|, -2.9|E_{c1}|$ and $-3.1|E_{c1}|$ from
bottom to top. 
The squares are the numerical results (averaged over many realizations, solid
line in Fig.~\ref{figtime2to1}) whereas the dashed-lines are the theoretical 
prediction (\ref{t21}).}
\end{figure}

Next, we present in Fig.~\ref{figtN2to1} the mean transition time 
$t_{2\rightarrow\pm 1}$ as a function of the number of particles $N$, for 
various total energies.
We compare the numerical results (squares) with the theoretical 
predictions (dashed lines) of Eq.(\ref{t21}). We can check that the
agreement is reasonably good and that the transition time scales linearly
as $N$. In the limit $N\rightarrow\infty$ the equilibrium $n=2$ becomes stable,
in agreement with the mean-field analysis of \cite{Valageas2006} where
it was shown that this state is linearly stable for the Vlasov dynamics.

\subsubsection{\label{typical21}Diffusive relaxation to $n=\pm 1$}

\begin{figure}
\begin{center}
\epsfxsize=4.4 cm \epsfysize=4.8 cm {\epsfbox{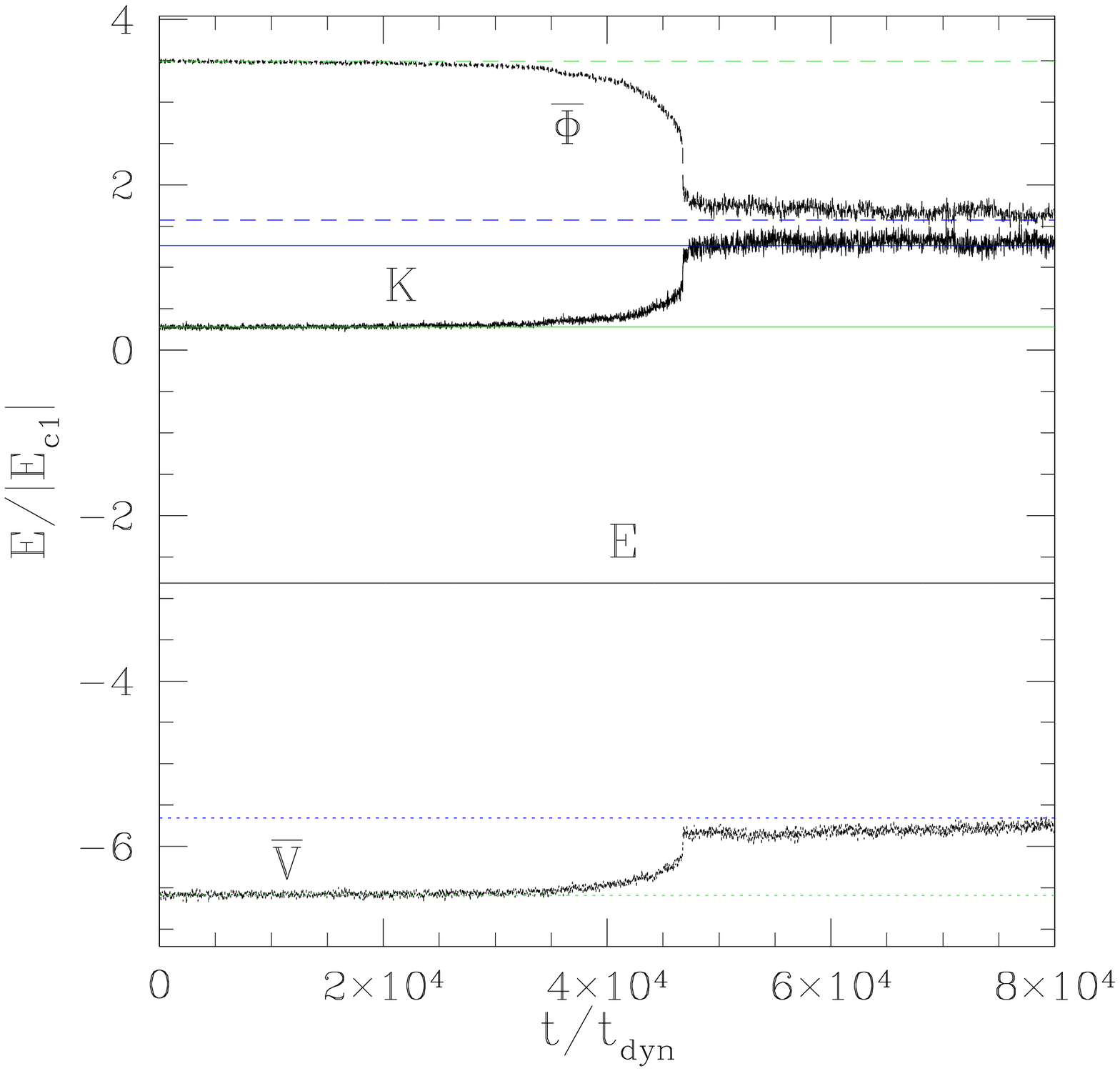}}
\epsfxsize=3.9 cm \epsfysize=4.8 cm {\epsfbox{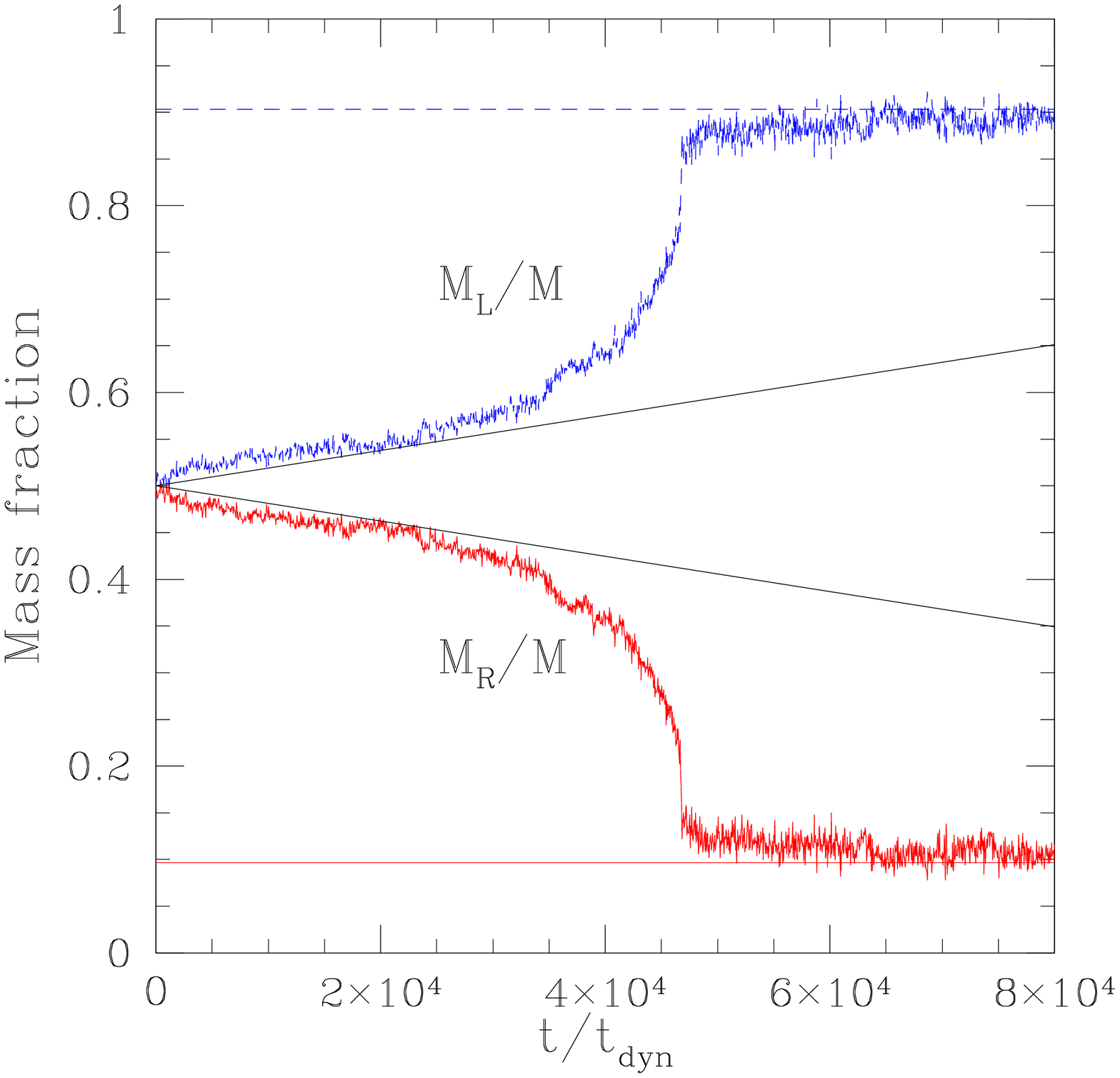}}
\end{center}
\caption{\label{figEM2to1}(Color online) {\it Left panel:} The evolution 
with time $t$ of 
the various contributions $K,\Phib$ and $\Vb$ to the total energy $E$ of a 
system of $N=500$ particles.
We display the curves obtained for a particular realization of the initial 
condition defined by the unstable equilibrium $n=2$ at energy 
$E=-2.8|E_{c1}|$. The constant curves are 
the mean-field energy levels of $K$ (solid lines), $\Phib$ (dashed lines) and 
$\Vb$ (dotted lines) for the equilibria $n=1$ and $n=2$. The system undergoes 
a transition from levels $n=2$ to levels $n=1$ at 
$t \sim 4.7 \times 10^4 t_{\rm dyn}$.
{\it Right panel:}  The evolution with time of the
masses $M_L$ (dashed lines) and $M_R$ (solid lines) located in the left and 
right parts of the system ($x<L/2$ and $x>L/2$). The constant lines show 
the values $M_L,M_R$ of equilibrium $n=1$. The two linear solid lines starting
from $M_L=M_R=M/2$ at $t=0$ are the theoretical estimate (\ref{dMdt}) for the 
mass transfer between both density peaks.}
\end{figure}

\begin{figure}
\begin{center}
\epsfxsize=4.15 cm \epsfysize=4.5 cm {\epsfbox{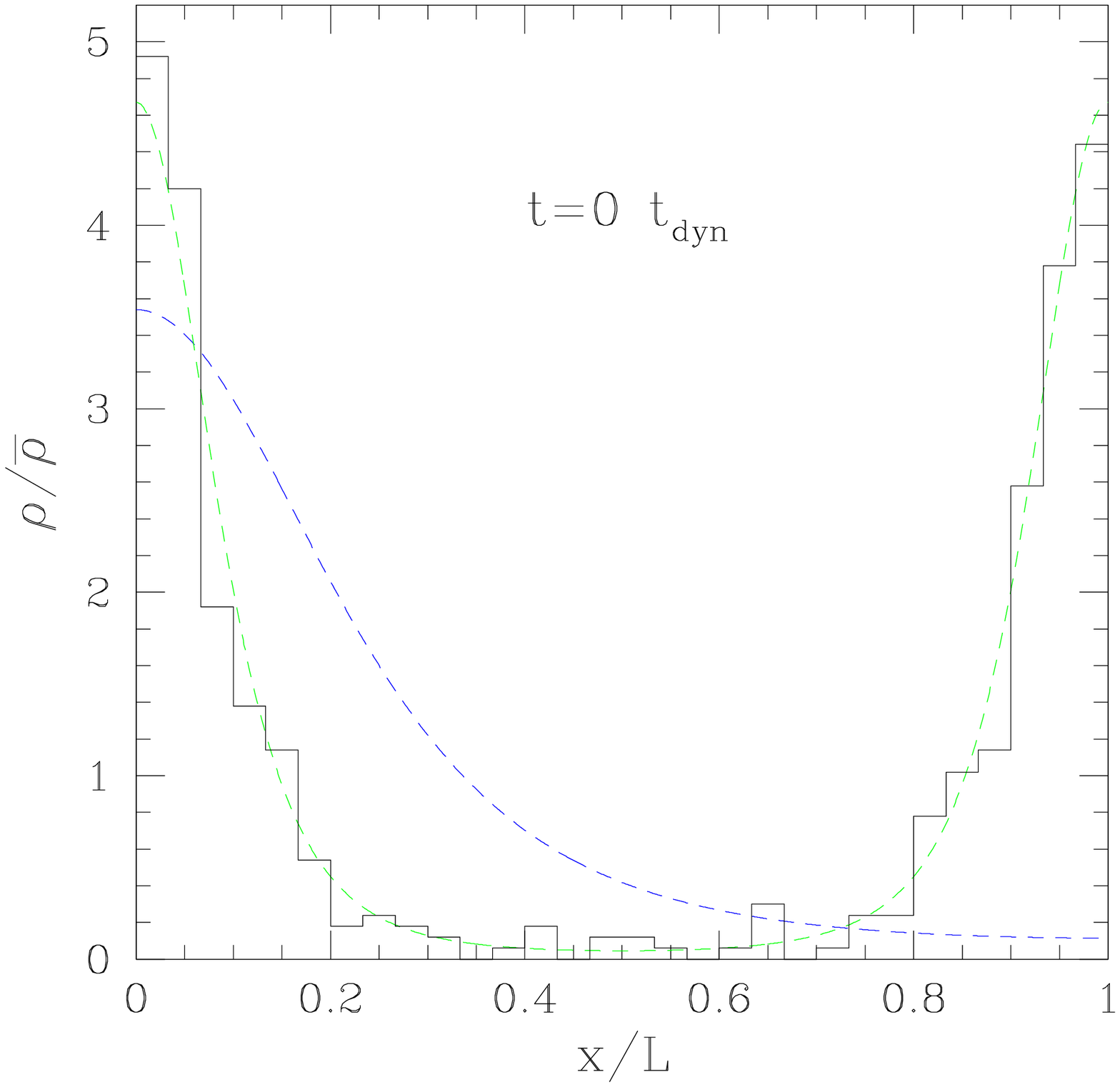}} 
\epsfxsize=4.15 cm \epsfysize=4.5 cm {\epsfbox{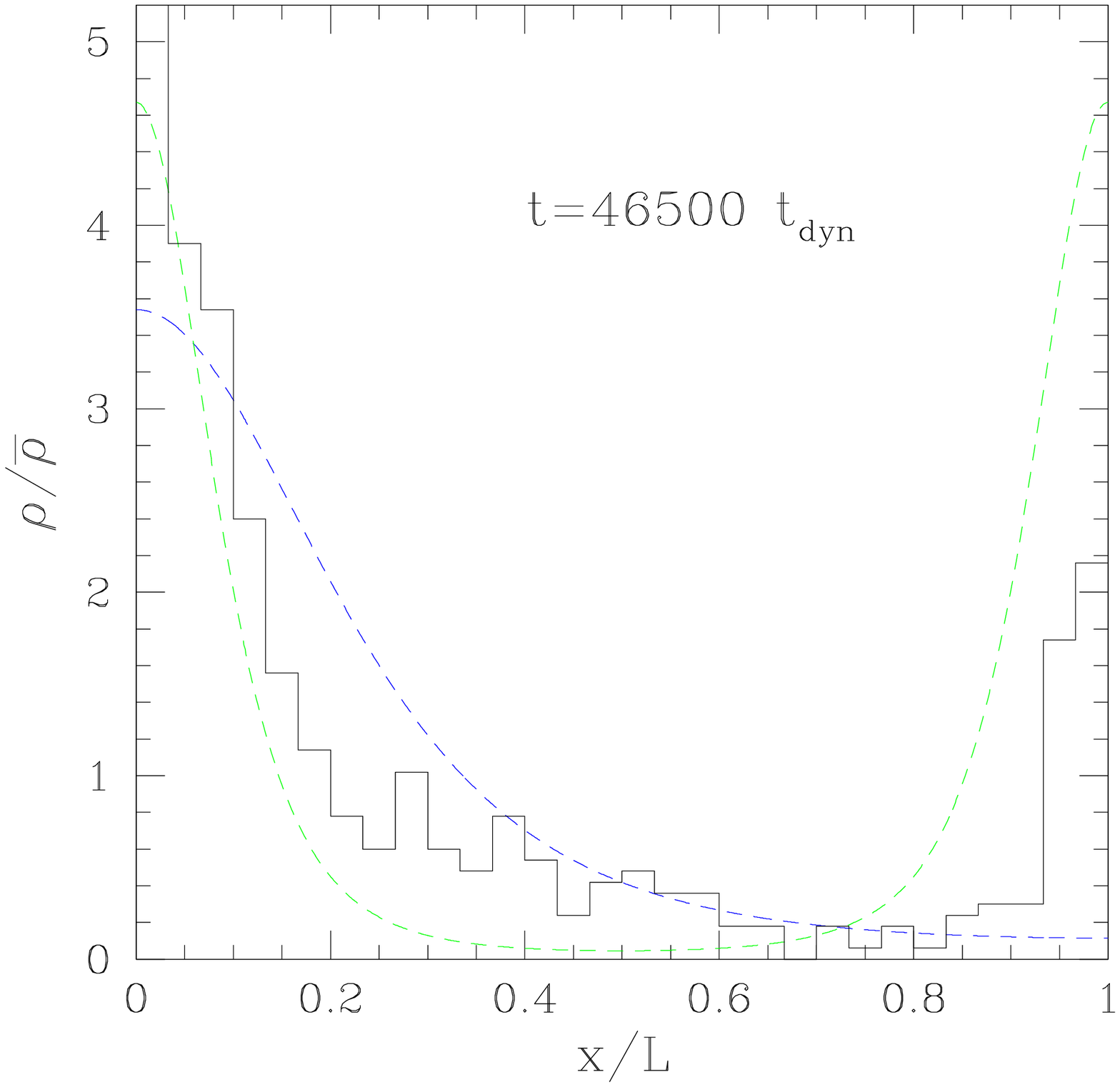}}\\
\epsfxsize=4.15 cm \epsfysize=4.5 cm {\epsfbox{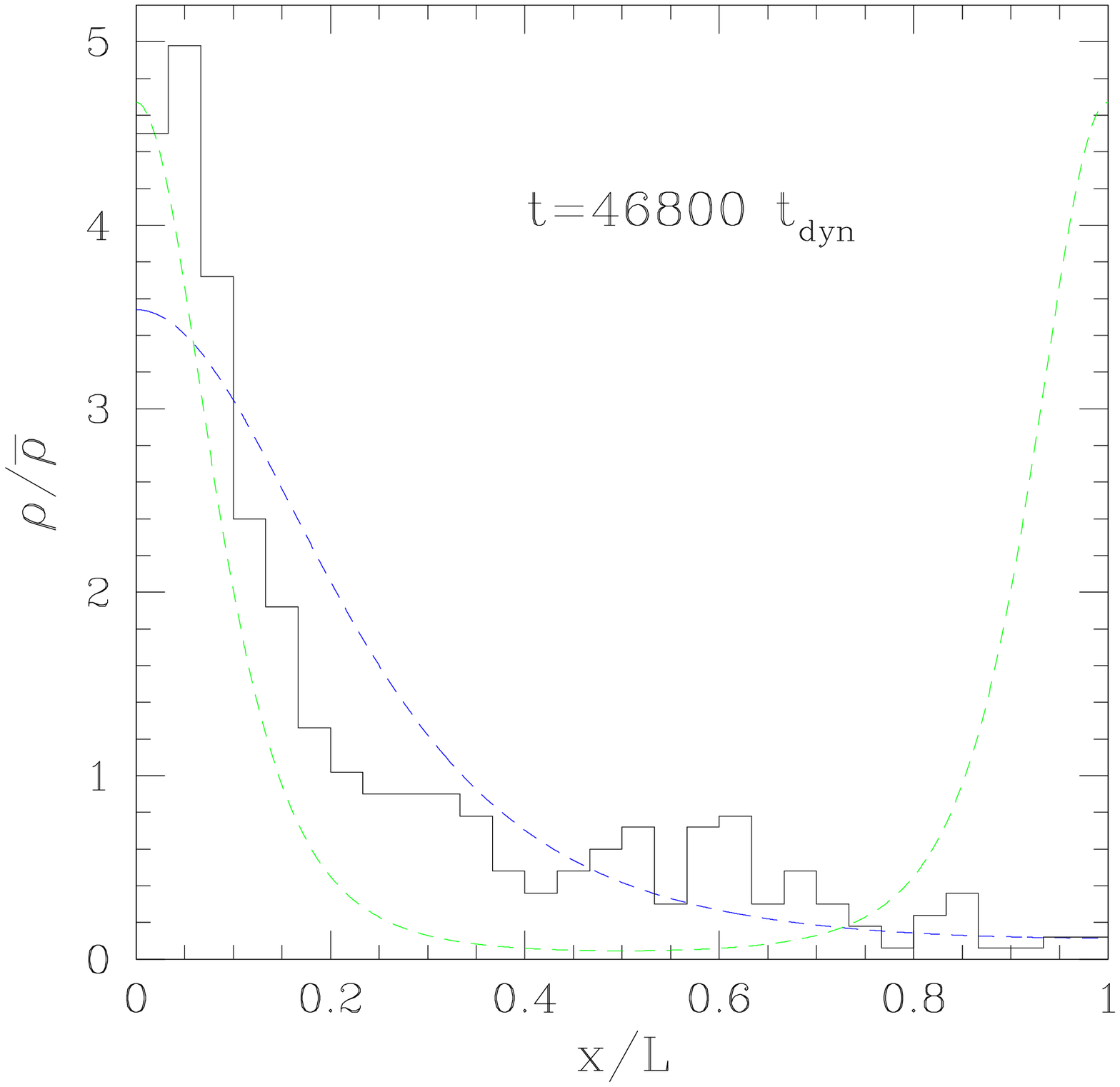}} 
\epsfxsize=4.15 cm \epsfysize=4.5 cm {\epsfbox{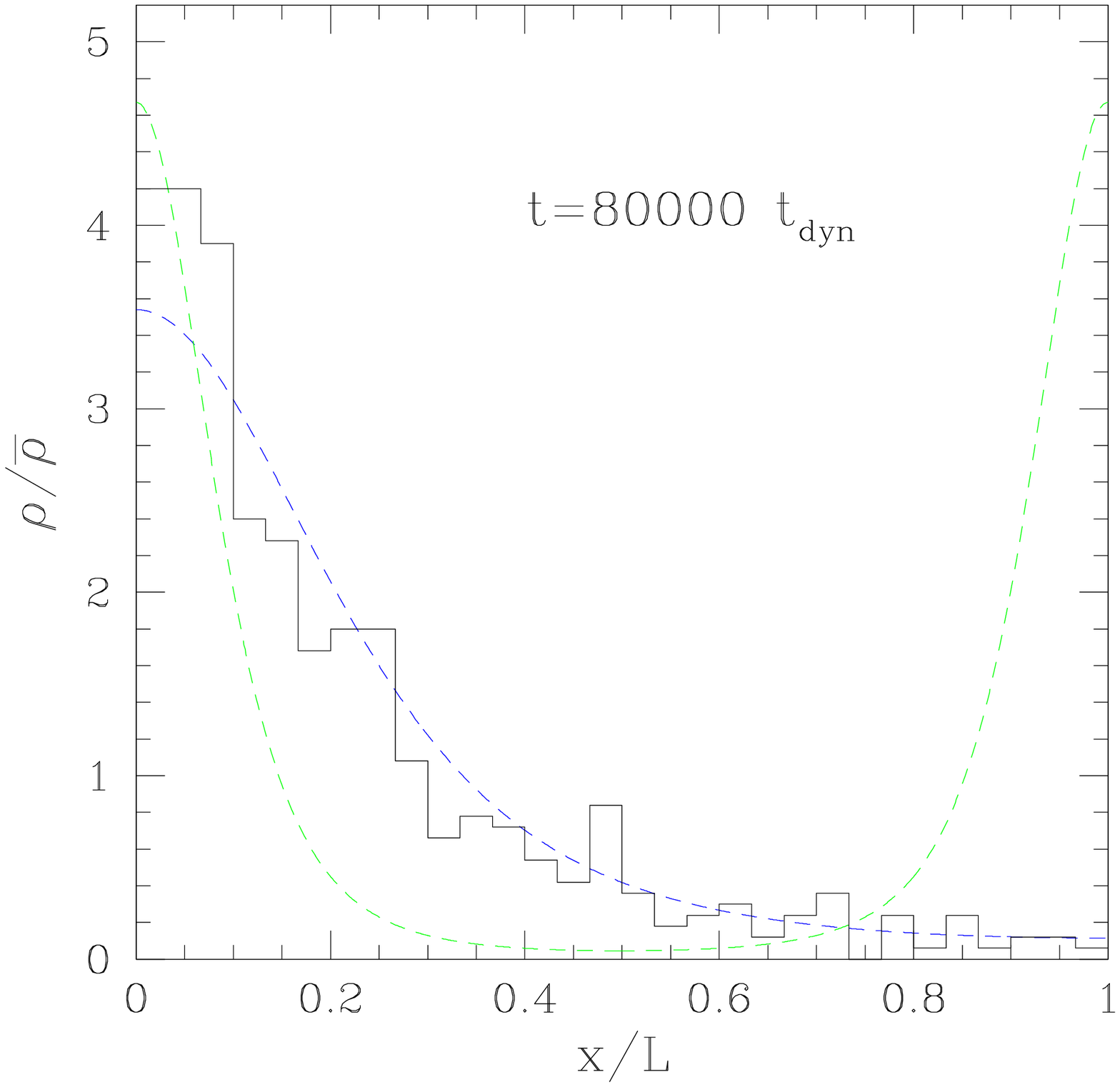}}
\end{center}
\caption{\label{figrho2to1}(Color online) Four snapshots of the density 
distribution $\rho(x)$
at times $t=0, 46500, 46800$ and $80000 t_{\rm dyn}$. The histogram shows the
matter distribution of the $N-$body system over $30$ bins. The dashed curves
are equilibria $n=1$ and $n=2$.}
\end{figure}

We first show in Fig.~\ref{figEM2to1} the evolution of the various 
contributions to the total energy, for a particular realization with $N=500$
particles of the initial condition defined by the unstable equilibrium $n=2$
at energy $E=-2.8|E_{c1}|$. 
We can see that the system displays a transition from levels $n=2$ to levels
$n=1$ at $t \sim 4.7 \times 10^4 t_{\rm dyn}$. We can note that there is 
first a slow drift up to $\sim 4.7 \times 10^4 t_{\rm dyn}$ 
where there is a sudden transition towards energy levels of $n=1$.
This is more clearly seen in the right panel which displays the evolution of 
the masses located to the left ($M_L$ at $x<L/2$) and to the right 
($M_R$ at $x>L/2$) of the system. We can see that there is a slow transfer 
of matter from the right peak to the left peak until 
$\sim 4.7 \times 10^4 t_{\rm dyn}$. Then, there is a sudden jump
where the mass ratio reaches its equilibrium value after a few dynamical times.
This suggests that there is first a slow evolution over a few thousand 
$t_{\rm dyn}$ with a quasi-static exchange of matter between both density 
peaks which remain close to local thermodynamical equilibrium states for 
separate clouds until the ratio of both peak masses becomes so large that 
the system turns dynamically unstable and relaxes towards equilibrium $n=1$ 
over a few dynamical times $t_{\rm dyn}$.
Indeed, as shown in \cite{Valageas2006} both statistical equilibria $n=1$ and 
$n=2$ are stable at low temperature for the mean-field Vlasov dynamics 
(although only state $n=1$ is thermodynamically stable). By continuity, this 
means that there is a threshold where the system leaves the basin of attraction
of two-peaks configurations to enter the basin of attraction of the one-peak
equilibrium. Indeed, two-peak states close to the equilibrium $n=2$ (i.e. with
a mass ratio of both clouds close to unity) have negative stability eigenvalues
and remain stable with respect to the mean-field Vlasov dynamics. On the other
hand, two-peak states with a high mass ratio can be seen as small perturbations
of the one-peak equilibrium $n=1$ which is itself stable. Therefore, such
two-peak states undergo a collective instability which leads to relaxation
towards the stable equilibrium $n=1$. For the system shown in 
Fig.~\ref{figEM2to1} this transition between two different stability regions
of the mean-field dynamics occurs at $t \sim  4.7 \times 10^4 t_{\rm dyn}$.

The solid lines starting from $M_L=M_R=M/2$ at $t=0$ are the theoretical 
estimate (\ref{dMdt}) which has been normalized by a factor $0.2$ of order
unity to match the slope at early times of the mass transfer. This 
was then used to predict the transition times shown in 
Figs.~\ref{figtime2to1}-\ref{figtN2to1} for a broad range of total energies and
particle numbers.

We present in Fig.~\ref{figrho2to1} four snapshots of the density distribution 
$\rho(x)$ for the $N-$body system as compared with the mean-field equilibria
$n=1$ and $n=2$. In agreement with Fig.~\ref{figEM2to1} we recover the sudden 
transition at $\sim 47000 t_{\rm dyn}$ from a two-peak state to a one-peak 
state, after the right peak has slowly lost about half its mass to the left 
peak. We can also check that the density distribution relaxes to the 
statistical equilibrium $n=1$.

\begin{figure}
\begin{center}
\epsfxsize=4.15 cm \epsfysize=4.5 cm {\epsfbox{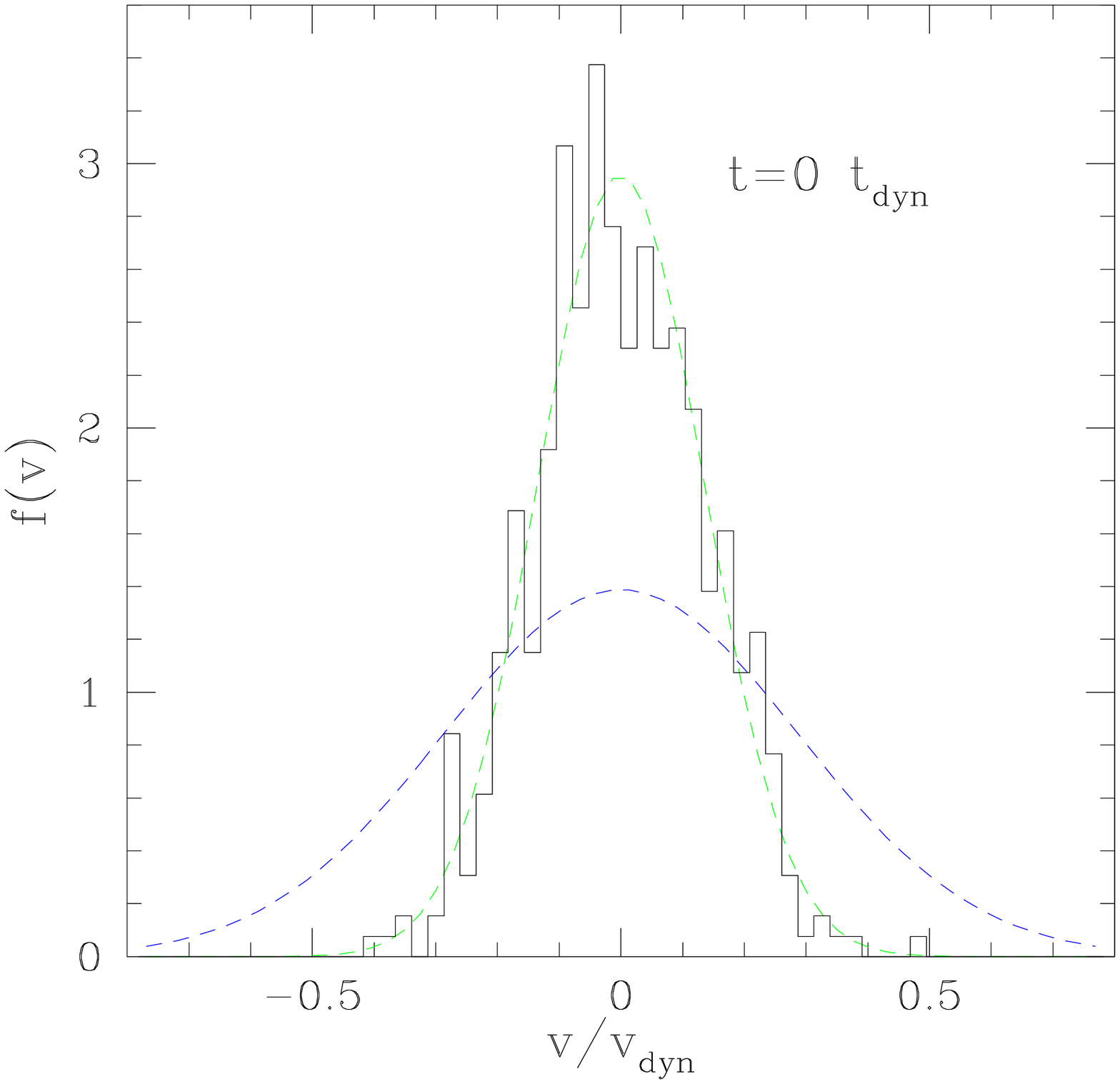}} 
\epsfxsize=4.15 cm \epsfysize=4.5 cm {\epsfbox{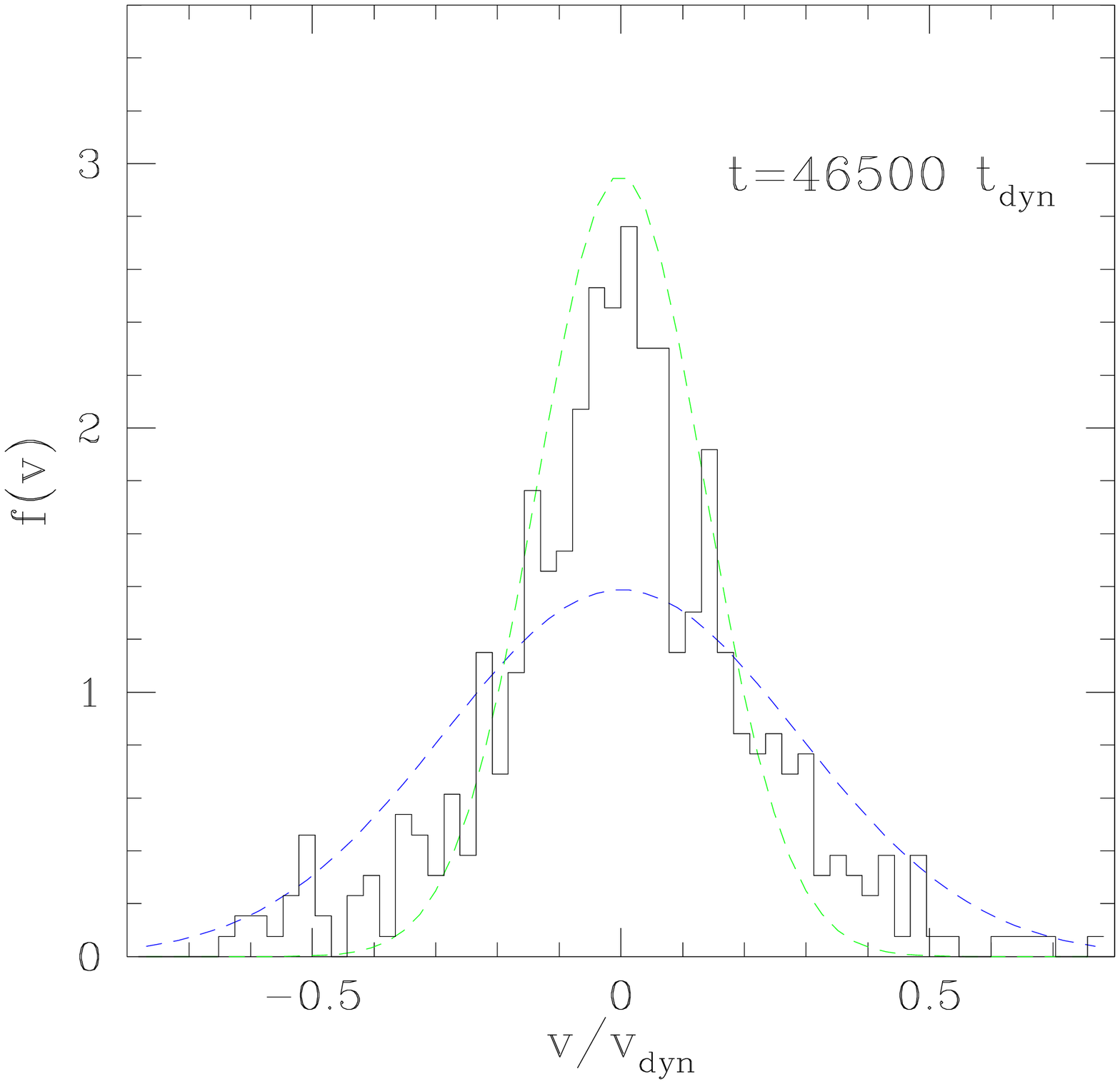}}\\
\epsfxsize=4.15 cm \epsfysize=4.5 cm {\epsfbox{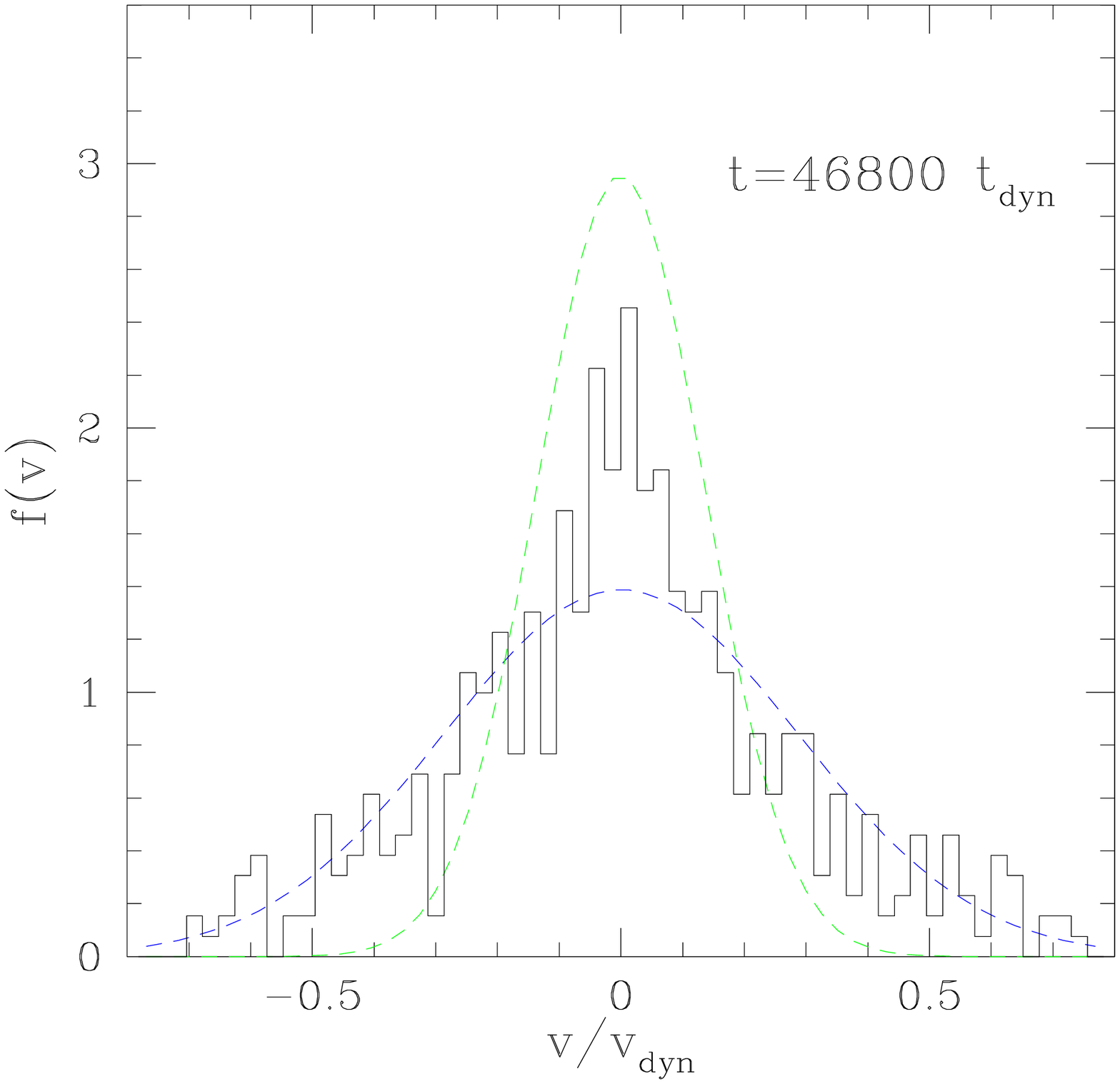}} 
\epsfxsize=4.15 cm \epsfysize=4.5 cm {\epsfbox{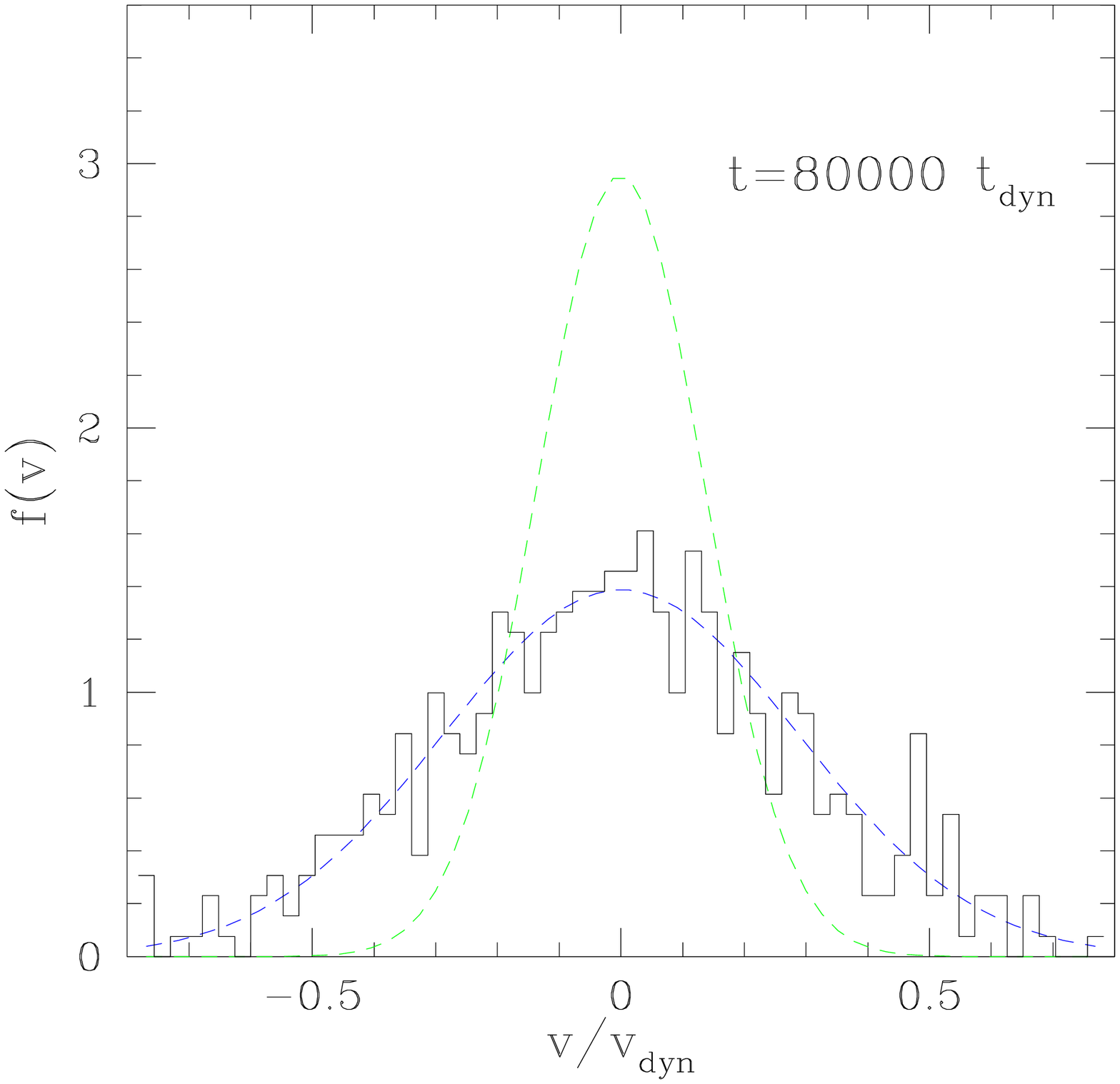}}
\end{center}
\caption{\label{figfv2to1}(Color online) Four snapshots of the velocity 
distribution $f(v)$
at times $t=0, 46500, 46800$ and $80000 t_{\rm dyn}$. The 
histogram corresponds to the $N-$body system whereas the dashed curves
are equilibria $n=2$ (narrow Gaussian) and $n=1$ (broad Gaussian).}
\end{figure}

In a similar fashion, we show in Fig.~\ref{figfv2to1} the velocity distribution
$f(v)$, which is also seen to relax from the equilibrium $n=2$ to state $n=1$. 
We note that the relaxation of the velocity distribution seems to be
more efficient than for the density distribution. Indeed, at 
$t=46500 t_{\rm dyn}$ it is still close to the Gaussian $n=2$
whereas at $t=46800 t_{\rm dyn}$ it is already close to the Gaussian $n=1$.

\begin{figure}
\begin{center}
\epsfxsize=4.15 cm \epsfysize=4.5 cm {\epsfbox{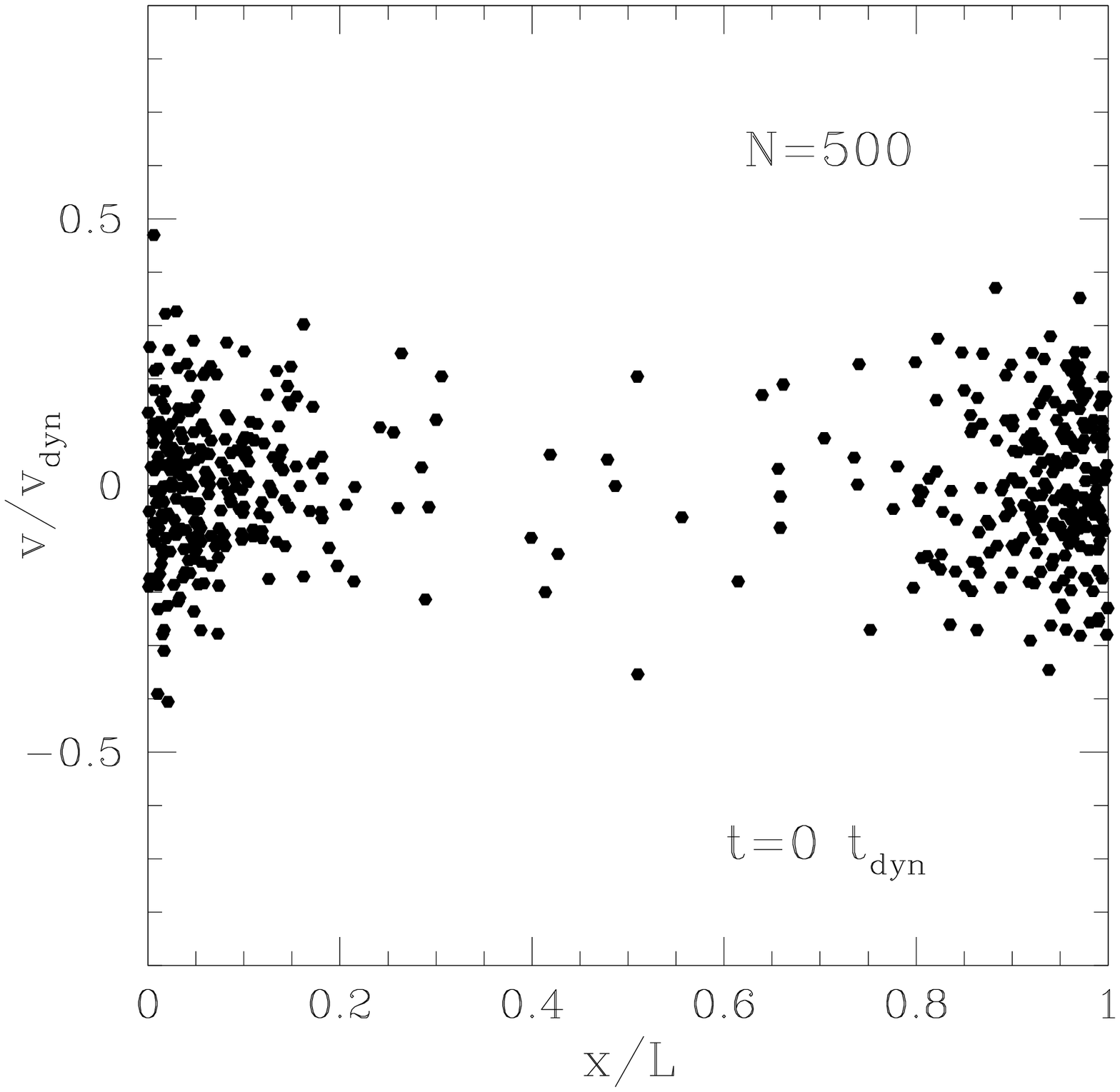}} 
\epsfxsize=4.15 cm \epsfysize=4.5 cm {\epsfbox{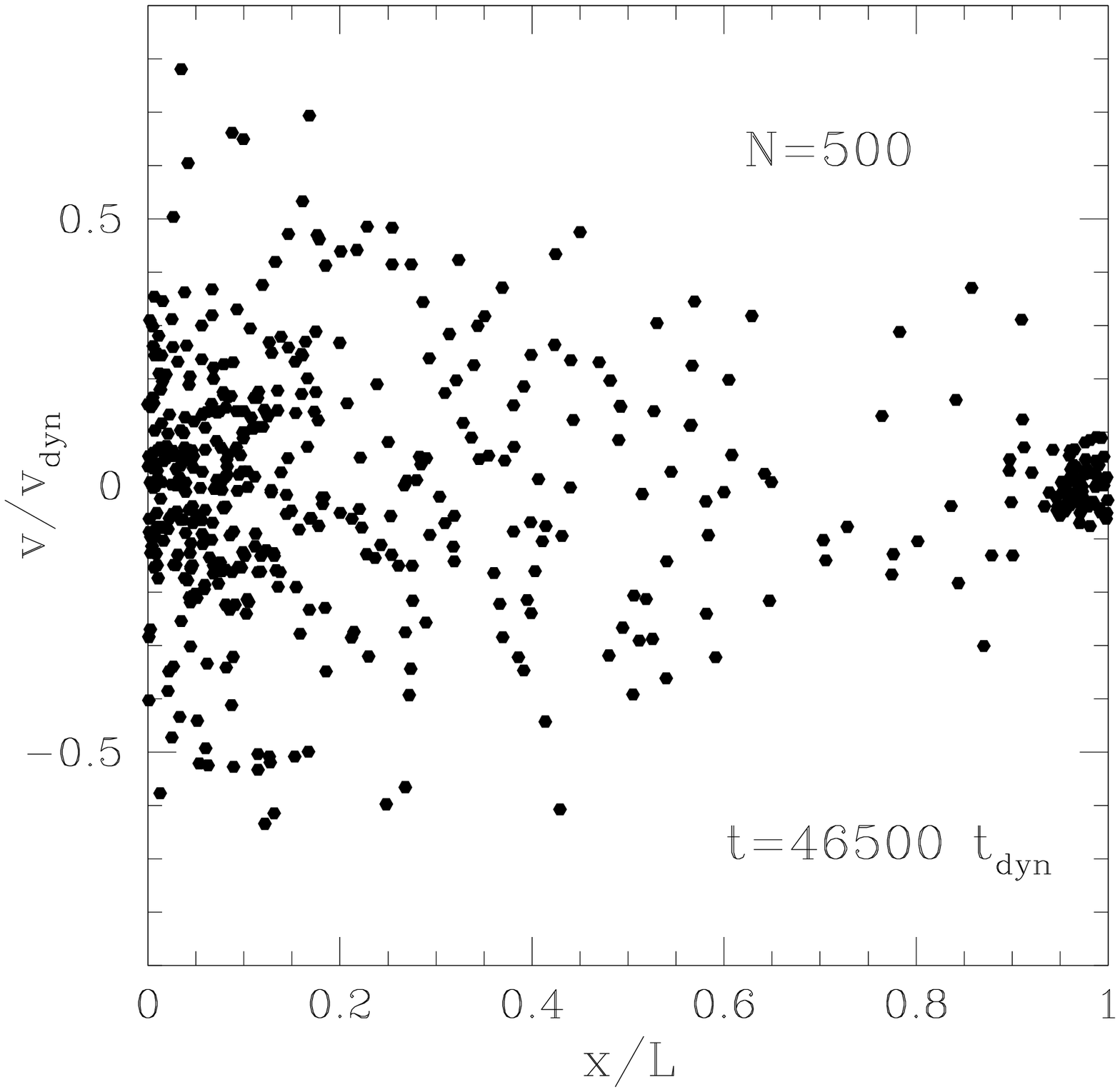}}\\
\epsfxsize=4.15 cm \epsfysize=4.5 cm {\epsfbox{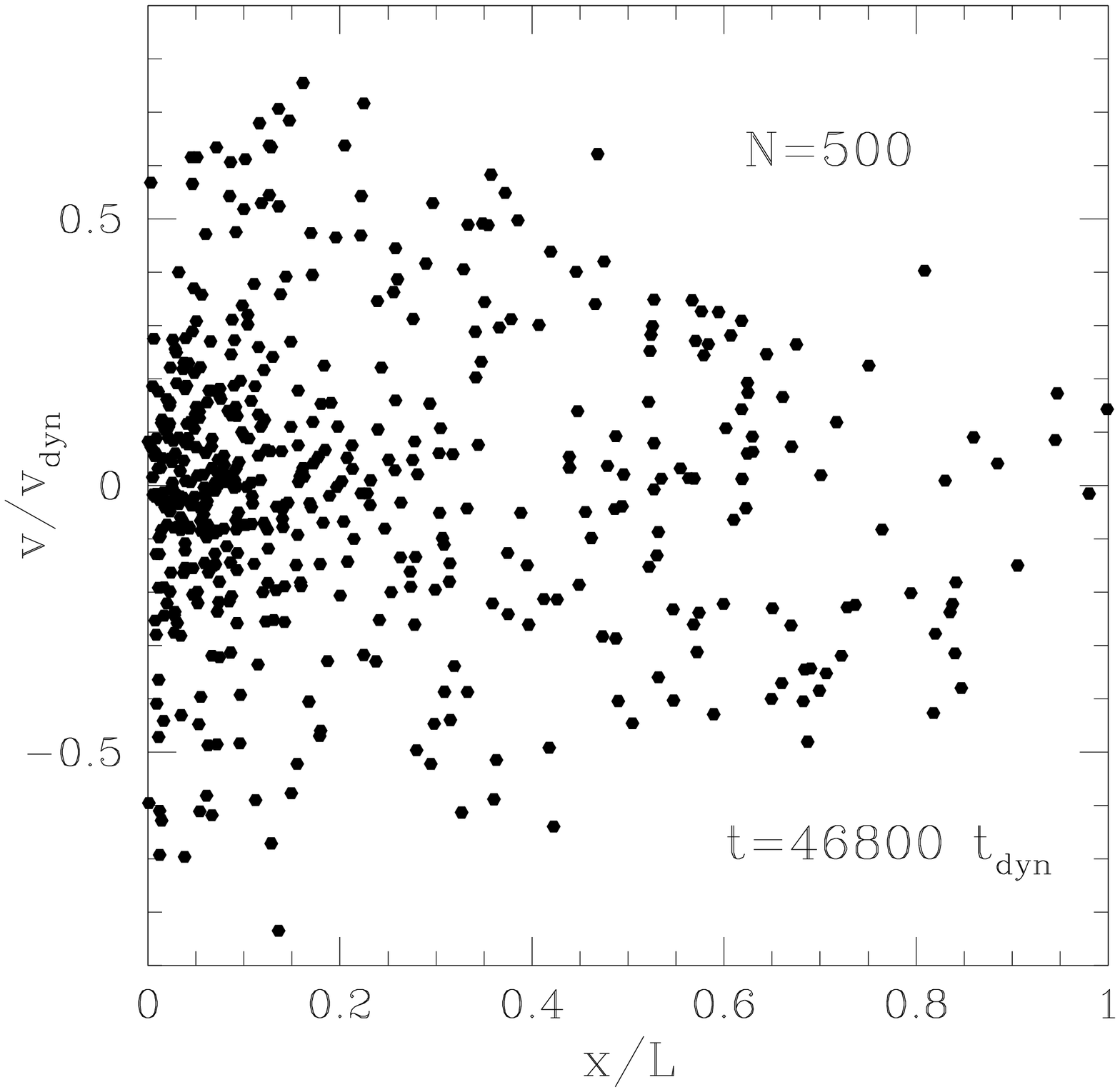}} 
\epsfxsize=4.15 cm \epsfysize=4.5 cm {\epsfbox{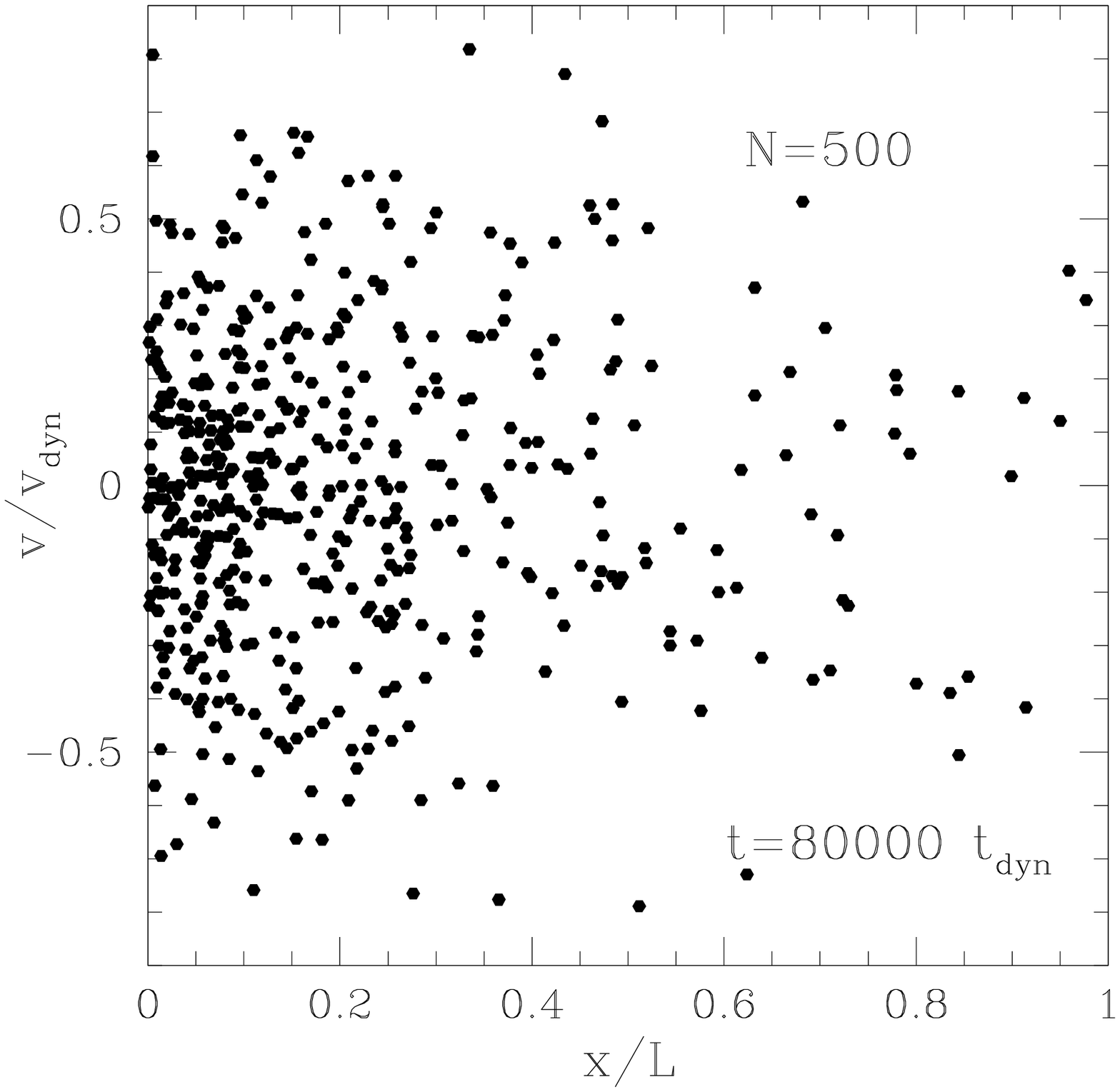}}
\end{center}
\caption{\label{figfxv2to1}(Color online) Four snapshots of the phase-space 
distribution 
$f(x,v)$ at times $t=0, 46500, 46800$ and $80000 t_{\rm dyn}$.}
\end{figure}

We display in Fig.~\ref{figfxv2to1} four snapshots of the phase-space 
distribution $f(x,v)$. It shows again that the transition from $n=2$ to
$n=1$ proceeds in two steps, with a slow diffusion followed by a sudden
disruption of the smallest density peak. In particular, we can see in the
upper right panel of Fig.~\ref{figfxv2to1} that the right peak still exists
as a distinct object at $t=46500 t_{\rm dyn}$ and has disappeared by 
$t=46800 t_{\rm dyn}$.

\begin{figure}
\begin{center}
\epsfxsize=4.15 cm \epsfysize=4.5 cm {\epsfbox{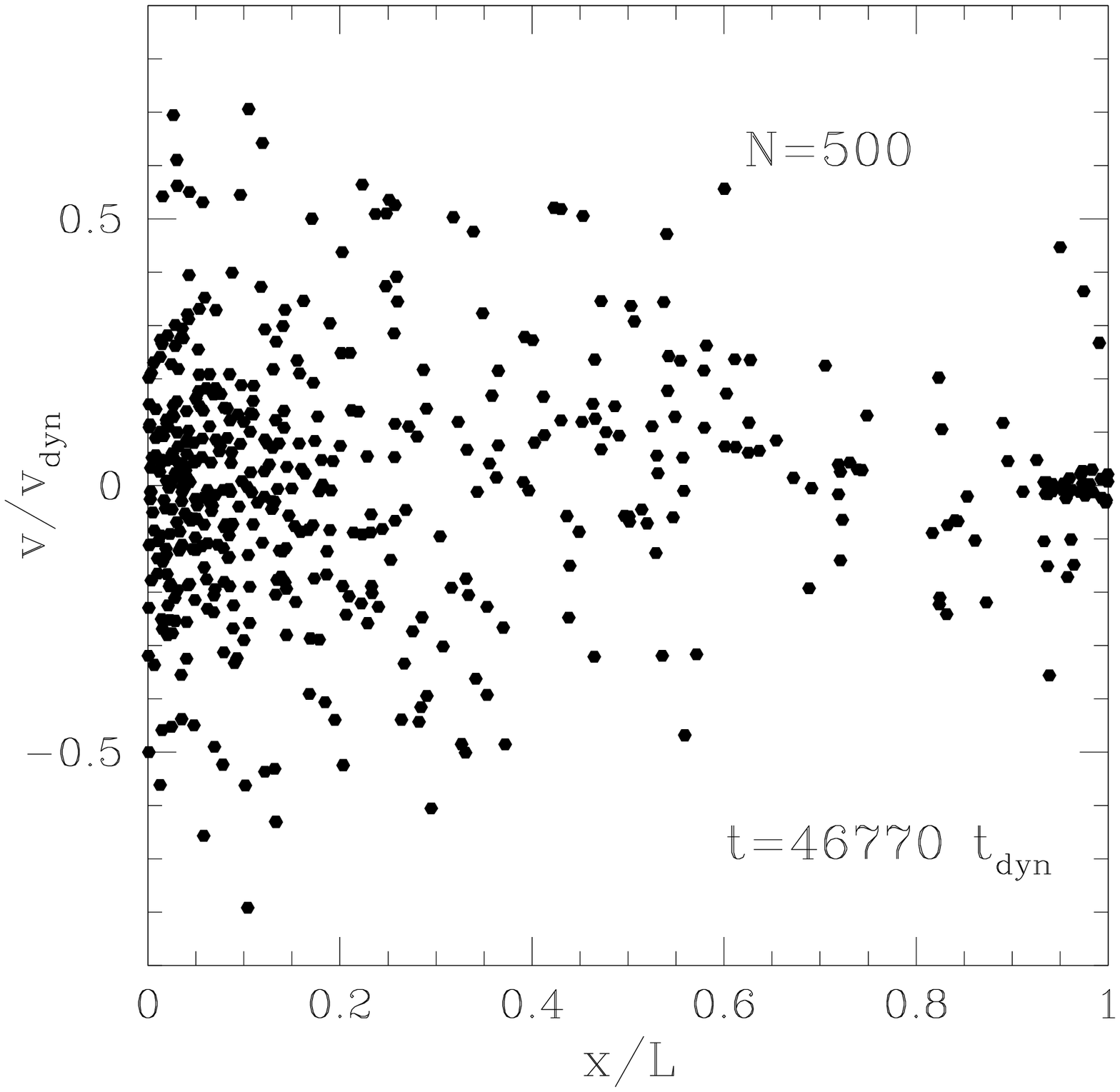}} 
\epsfxsize=4.15 cm \epsfysize=4.5 cm {\epsfbox{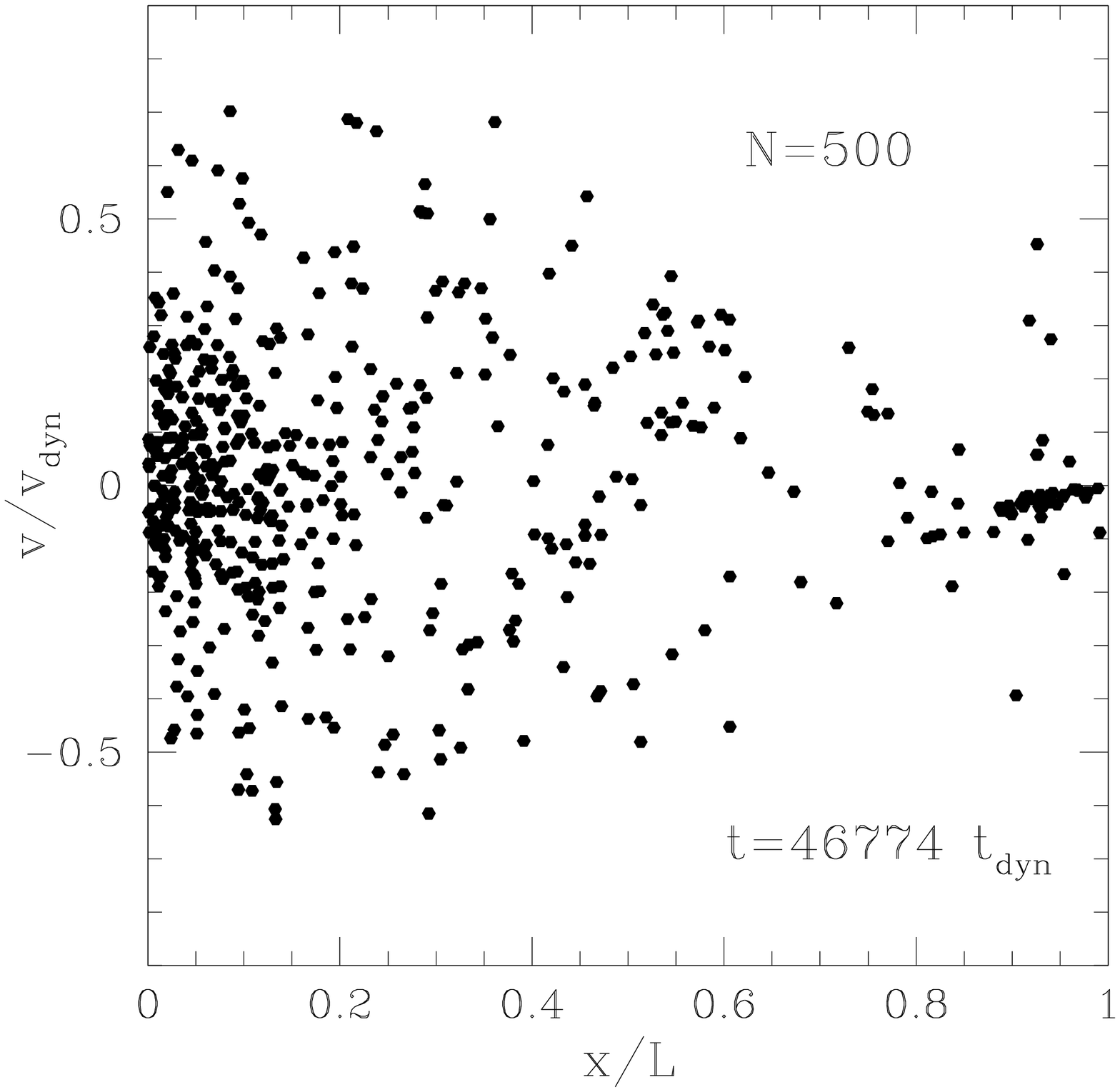}}\\
\epsfxsize=4.15 cm \epsfysize=4.5 cm {\epsfbox{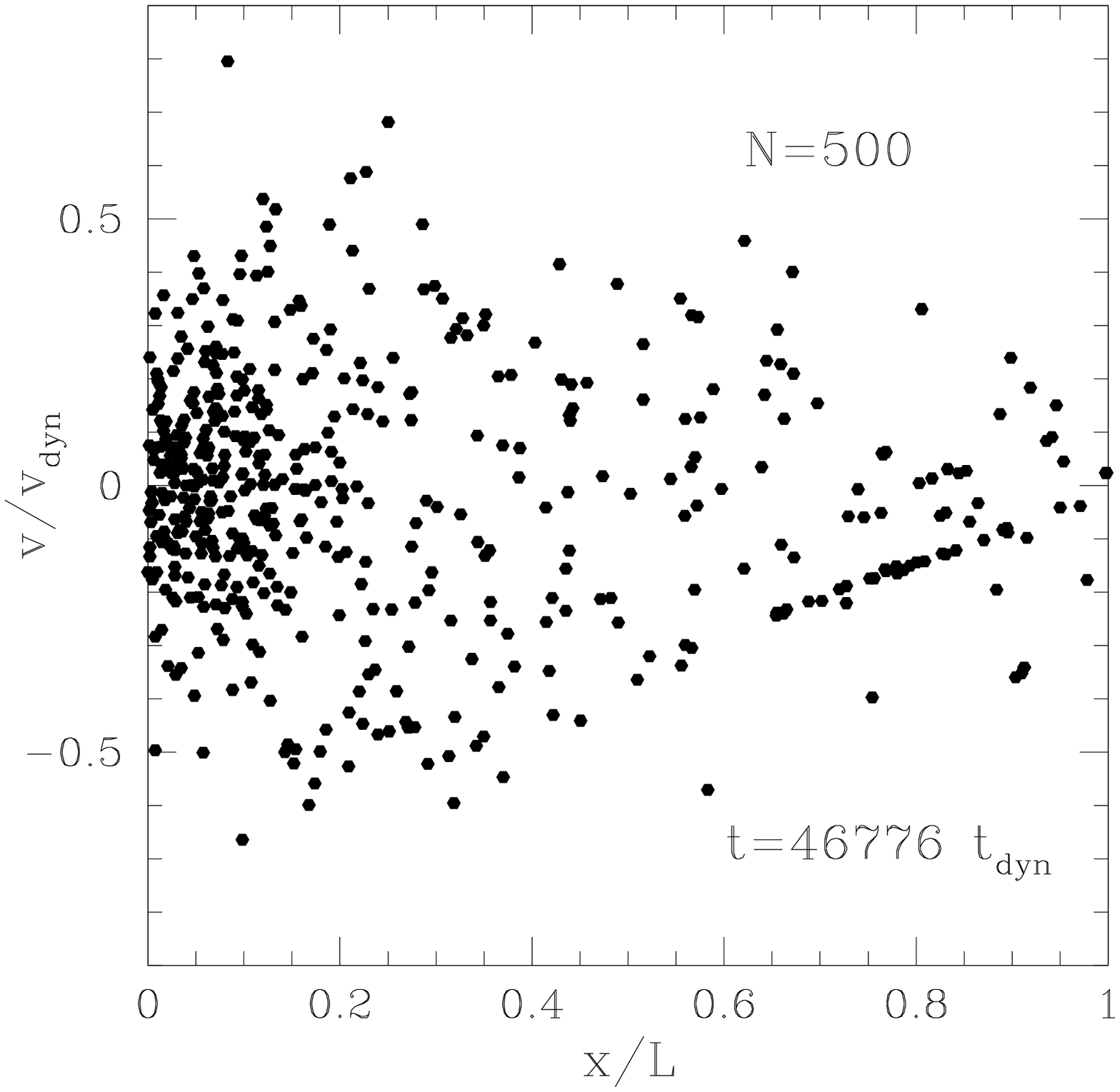}} 
\epsfxsize=4.15 cm \epsfysize=4.5 cm {\epsfbox{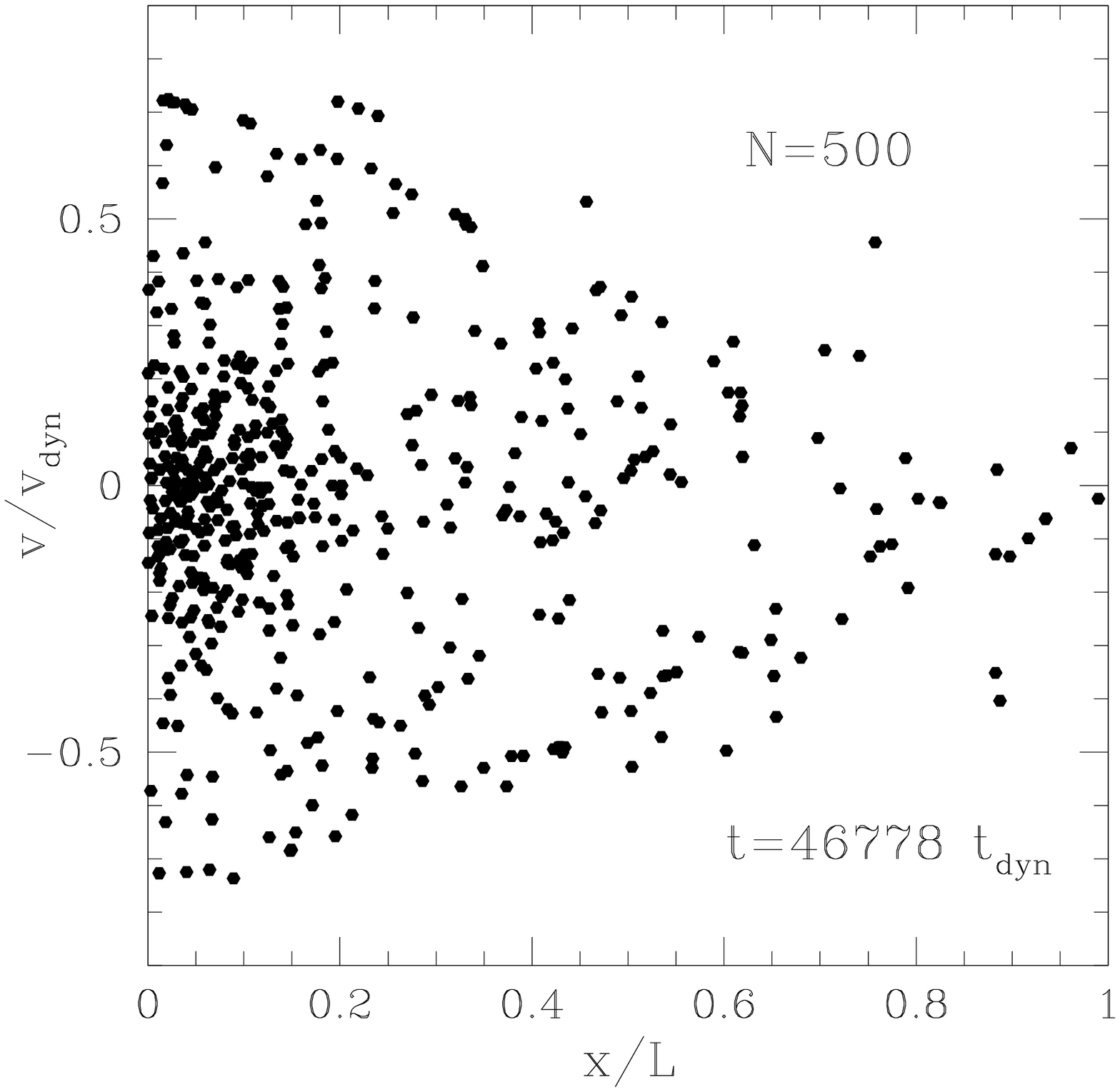}}
\end{center}
\caption{\label{figfxv2to1_disrupt}(Color online) Four snapshots of the 
phase-space 
distribution $f(x,v)$ at times $t=46770, 46774, 46776$ and $46778 t_{\rm dyn}$
around the transition to the one-peak state.}
\end{figure}

Finally, we show in Fig.~\ref{figfxv2to1_disrupt} four snapshots of the 
phase-space distribution around the transition to the one-peak state. 
We can see that what is left of the right peak at $t\sim 46770 t_{\rm dyn}$ 
exhibits a strong deformation at $46774 t_{\rm dyn}$ and is absorbed into 
the left peak at $46776 t_{\rm dyn}$. This describes in details the second 
step of the transition as a collective dynamic instability which merges the 
remains of the smallest peak into the largest one over a few dynamical 
times $t_{\rm dyn}$.

\subsubsection{\label{Diffusion_time}Estimate of the mean transition time
$t_{2\rightarrow\pm 1}$}

We describe in this section how the mean transition times
$t_{2\rightarrow\pm 1}$ shown in Figs.~\ref{figtime2to1}-\ref{figtN2to1}
can be estimated by analytical means. 

The relaxation towards thermal equilibrium of $N-$body systems such as the
OSC model is often studied through the stochastic dynamics of a test particle
in interaction with the rest of the system. 
In such cases, the test particle experiences 
both a systematic drift (such as the dynamical friction suffered by a 
high-velocity star moving through a cloud of low-velocity stars 
\cite{Chandrasekhar1943,Binney1987}) and a diffusion, due to random encounters,
which both grow linearly with time.
Moreover, if the correlation times are small one can use a Markovian
approximation so that the relaxation of the test particle in the thermal
bath is described by a Fokker-Planck equation \cite{vanKampen2004}, which
is fully parameterized by the friction and diffusion coefficients.
One may then describe for instance the relaxation of the velocity distribution
of the test particle and its approach to thermal equilibrium 
\cite{Bouchet2004,Bouchet2005}.

In the case we consider here, the system starts in an inhomogeneous
mean-field equilibrium with two peaks at the boundaries. We model the 
slow loss of matter of the smallest peak, displayed in Fig.~\ref{figEM2to1}, 
as a diffusive process which makes particles located in the smallest potential 
well escape to the other potential well. Since the escape time is shorter for 
the smaller potential well we neglect the flux from particles moving from the
deepest well to the smallest one. This is consistent with the right panel of 
Fig.~\ref{figEM2to1} which shows that the lightest peak is almost steadily
losing mass. Here we shall neglect the evolution with time of the small density
peak due to mass loss and we simply consider the bistable system $n=2$ made of 
the symmetric double potential well $\phi(x)$ which has two minima at the 
walls $x=0,L$ and a maximum at $x=L/2$. Then, we focus on the behavior of a
test mass $m$ orbiting in the left peak and we study the evolution of its
energy $\epsilon$ (whence of its orbit) due to encounters with other discrete 
particles.
Computing the associated friction and diffusion coefficients we obtain the 
average time $\tau$ required to reach the barrier energy $\phi(L/2)-\phi(0)$ 
and escape from the potential well through a Fokker-Planck dynamics.
This allows us to estimate the transition time $t_{2\rightarrow\pm 1}\sim\tau$
to relax to the thermodynamically stable equilibrium $n=1$.

In order to separate the mean-field dynamics from the discrete effects which
give rise to the diffusion of particles we write the Hamiltonian (\ref{HN}) as:
\beq
{\cal H}_N = m({\cal H}_0 + {\cal H}_I)
\label{H0HI}
\eeq
and we defined the mean-field Hamiltonian ${\cal H}_0$ by:
\beq
{\cal H}_0= \sum_{j=1}^N \left[ \frac{v_j^2}{2} + \Phi_0(x_j) + V(x_j) \right]
\label{H0}
\eeq
and the interaction Hamiltonian ${\cal H}_I$ by:
\beq
{\cal H}_I= e^{\omega t} \left[ g m \sum_{j>j'} |x_j-x_{j'}| 
- \sum_j \Phi_0(x_j) \right] .
\label{HI}
\eeq
In Eq.(\ref{H0}) the mean-field gravitational potential $\Phi_0$ is given
by Eq.(\ref{Phi}) where $\rho(x')$ is the mean-field equilibrium distribution
(\ref{Maxwell}). In Eq.(\ref{HI}) we added a factor $e^{\omega t}$ for the 
computation of perturbative eigenmodes and we shall consider the limit 
$\omega\rightarrow 0^+$. Thus, ${\cal H}_0$ describes the mean-field dynamics
whereas ${\cal H}_I$ describes the discrete effects which vanish in the limit
$N \rightarrow \infty$. Therefore, we consider ${\cal H}_I$ as a 
perturbation to ${\cal H}_0$ and we apply a perturbative analysis to the
dynamics of individual particles. Here it is convenient to work with the
action-angle variables $(J,w)$ defined from the Hamiltonian ${\cal H}_0$
which describe the motion of a particle of energy $\epsilon=v^2/2+\phi_0$ 
along its non-perturbed orbit in the equilibrium potential $\phi_0=\Phi_0+V$ 
from position $x_-$ to $x_+$ \cite{Binney1987,Fridman1984}:
\beq
J= \frac{2}{2\pi} \int_{x_-}^{x_+}\d x \sqrt{2(\epsilon-\phi_0(x))} ,
\label{Jdef}
\eeq
and:
\beq 
w= \Omega \int_{x_-}^x \frac{\d x'}{\sqrt{2(\epsilon-\phi_0(x'))}} \;\;\; 
\mbox{with} \;\;\; \Omega = \frac{\d\epsilon}{\d J}  .
\label{wdef}
\eeq
Thus, ${\cal H}_0={\cal H}_0(J_j)$ only depends on the actions $J_j$. On the 
other hand, thanks to the periodicity of $2\pi$ with respect to $w$ of the 
non-perturbed orbits $x(J,w)$ we can write the two-body interaction as:
\beq
g m |x-x'| = \sum_{k,k'=-\infty}^{\infty} \phi_{k,k'}(J,J') e^{i(k w - k'w')},
\label{Phikk'}
\eeq
which defines the Fourier coefficients $\phi_{k,k'}(J,J')$. Besides, from
Eq.(\ref{Phi}) the mean-field potential $\Phi_0(x)$ reads:
\beqa
\Phi_0(x) & = & g \int \d J'\d w' f_0(J') |x-x'| \\
& = & \frac{2\pi}{m} \sum_{k=-\infty}^{\infty} \int\d J' f_0(J') 
\phi_{k,0}(J,J') e^{i k w} ,
\label{Phi0}
\eeqa
where we used the canonical change of variable $\d x\d v= \d J \d w$ and $f_0$
is the mean-field equilibrium distribution (\ref{Maxwell}). This yields:
\beqa
{\cal H}_I &= & e^{\omega t} \left[ \frac{1}{2} \sum_{j,j'}^{j \neq j'} \sum_{k,k'} 
\phi_{k,k'}(J,J') e^{i(k w - k'w')} \right. \nonumber\\ 
& & \left. - \frac{2\pi}{m} \sum_j \sum_k \int\d J' f_0(J') 
\phi_{k,0}(J,J') e^{i k w} \right]
\label{HIJw}
\eeqa	
Then, the equations of motion become:
\beq
\dot{J}_j = - \frac{\pl}{\pl w} ({\cal H}_0+{\cal H}_I) , \;\;\;
\dot{w}_j = \frac{\pl}{\pl J} ({\cal H}_0+{\cal H}_I) .
\label{Jw}
\eeq
We write the action-angle trajectories $\{J(t),w(t)\}$ as the perturbative 
expansions $J=J^{(0)}+J^{(1)}+J^{(2)}+...$ where $J^{(k)}$ is formally of 
order $k$ over ${\cal H}_I$. At zeroth-order we simply have:
\beq
\dot{J}^{(0)}_j = - \frac{\pl {\cal H}_0}{\pl w_j} = 0 , \;\; 
\dot{w}^{(0)}_j = \frac{\pl {\cal H}_0}{\pl J_j} = \Omega(J^{(0)}_j) ,
\label{J0w0}
\eeq
which yields the mean-field equilibrium orbits:
\beq
J^{(0)}_j= \mbox{constant} \;\;\; \mbox{and} \;\;\; w^{(0)}_j=  w^{(0)}_j(0)
+ \Omega(J^{(0)}_j) t .
\label{J0w0t}
\eeq
At first order we obtain:
\beq
\dot{J}^{(1)}_j = - \frac{\pl{\cal H}_I}{\pl w_j} , \;\;\;
\dot{w}^{(1)}_j = \frac{\d\Omega}{\d J} J^{(1)}_j 
+ \frac{\pl{\cal H}_I}{\pl J_j} ,
\label{J1w1}
\eeq
where we can substitute the zeroth-order orbits in the r.h.s. 
Using the property $\phi_{k',k}(J',J)=\phi_{k,k'}(J,J')^*$ obtained from 
Eq.(\ref{Phikk'}) a simple calculation yields (e.g. \cite{Lynden-Bell1972}):
\beq
J^{(1)}_j= -\frac{\pl\chi}{\pl w_j}, \;\;\; w^{(1)}_j= \frac{\pl\chi}{\pl J_j},
\label{J1w1chi}
\eeq
with:
\beqa
\chi & = & e^{\omega t} \left[ \frac{1}{2} \sum_{j,j'}^{j \neq j'} \sum_{k,k'} 
\frac{\phi_{k,k'}}{\omega+i(k\Omega-k'\Omega')} e^{i(k w - k'w')} \right. 
\nonumber\\ 
& & \left. - \frac{2\pi}{m} \sum_j \sum_k \int\d J' f_0(J') 
\frac{\phi_{k,0}}{\omega+i k\Omega} e^{i k w} \right] .
\label{chi}
\eeqa	
At second order we need to follow the first-order orbits in the r.h.s of 
Eq.(\ref{Jw}) and we have for the actions:
\beq
\dot{J}^{(2)}_j = - \sum_{j'} \frac{\pl^2{\cal H}_I}{\pl w_j \pl J_{j'}} 
J^{(1)}_{j'} - \sum_{j'}\frac{\pl^2{\cal H}_I}{\pl w_j\pl w_{j'}} w^{(1)}_{j'}.
\label{J21}
\eeq
Using again $\phi_{k',k}=\phi_{-k,-k'}=\phi_{k,k'}^*$ and taking the average
(with the equilibrium distribution $f_0(J)$) over the actions and angles 
$\{J^{(0)},w^{(0)}(0)\}$ of other particles yields at order $1/N$ 
(note that $\phi_{k,k'} \propto m$ and $m \propto 1/N$):
\beqa
\lefteqn{\!\!\!\!\!\!\! \lag \dot{J}^{(2)} \rag = e^{2\omega t} 
\frac{\pl}{\pl J} \left[ 
\frac{2\pi}{m} \int\d J' f_0(J') \sum_{k,k'} \frac{\omega k^2 |\phi_{k,k'}|^2}
{\omega^2+(k\Omega-k'\Omega')^2} \right. } \nonumber \\ 
&& \!\!\!\!\!\! \left. - \frac{4\pi^2}{N m^2} \int\d J'\d J'' f_0(J') f_0(J'') 
\sum_k \frac{\omega k^2\phi_{k,0}\phi_{k,0}^*}{\omega^2+(k\Omega)^2} \right]
\nonumber \\ 
&& \!\!\!\!\!\! - e^{2\omega t} \frac{2\pi}{m} \int\d J' f_0(J') 
\frac{\pl}{\pl J'} \sum_{k,k'} \frac{\omega k k'|\phi_{k,k'}|^2}
{\omega^2+(k\Omega-k'\Omega')^2}
\label{J22}
\eeqa
Then, using $\lim_{\omega\rightarrow 0^+} \omega/(\omega^2+x^2)= \pi 
\delta_D(x)$ where $\delta_D$ is Dirac's distribution, the limit 
$\omega\rightarrow 0^+$ gives:
\beqa
\lag \dot{J}^{(2)} \rag & = & \frac{2\pi^2}{m} \int\d J' f_0(J') \sum_{k,k'} 
\left( k \frac{\pl}{\pl J} - k' \frac{\pl}{\pl J'}\right) \nonumber \\
&& \times |\phi_{k,k'}|^2 k \delta_D(k\Omega-k'\Omega') ,
\label{J23}
\eeqa
which does not depend on time. From these results we can obtain the mean 
drift and diffusion of the action $J$ of a test particle. First, the average
change $\lag \Delta J\rag=\lag J(t_2)-J(t_1)\rag$ of the action over a 
time-interval $\Delta t=t_2-t_1$ reads at order $1/N$:
\beq
\lag\frac{\Delta J}{\Delta t}\rag = 
\lag\frac{\Delta J^{(1)}+\Delta J^{(2)}}{\Delta t}\rag = \lag\dot{J}^{(2)}\rag
\label{DJ1}
\eeq
since $\lag\dot{J}^{(2)}\rag$ is constant and $\lag J^{(1)}\rag=0$ when we 
average \cite{Lynden-Bell1972} over the angles $w^{(0)}(0)$ as seen from 
Eq.(\ref{J1w1chi}). Next, from Eqs.(\ref{J1w1chi})-(\ref{chi}) the 
mean-square change $\lag(\Delta J)^2\rag$ reads at order $1/N$:
\beqa
\lefteqn{\lag (\Delta J)^2\rag = e^{2\omega t_1} \frac{2\pi}{m}\int\d J' 
f_0(J') \sum_{k,k'} \frac{k^2|\phi_{k,k'}|^2}{\omega^2+(k\Omega-k'\Omega')^2}}
\nonumber \\
&& \times \left( 1+e^{2\omega\Delta t}-2 e^{\omega\Delta t} 
\cos[(k\Omega-k'\Omega')\Delta t] \right) \nonumber \\
&& - e^{2\omega t_1} \frac{4\pi^2}{N m^2} \int\d J'\d J'' f_0(J') f_0(J'') 
\sum_k \frac{k^2\phi_{k,0}\phi_{k,0}^*}{\omega^2+(k\Omega)^2} \nonumber \\
&& \times \left( 1+e^{2\omega\Delta t}-2 e^{\omega\Delta t} 
\cos[k\Omega\Delta t] \right) .
\label{DJ2}
\eeqa
The limit $\omega\rightarrow 0^+$ now gives:
\beqa
\lefteqn{\lag\frac{(\Delta J)^2}{\Delta t}\rag = \frac{2\pi}{m}\int\d J' 
f_0(J') \sum_{k,k'} 2 k^2|\phi_{k,k'}|^2 } \nonumber \\
&& \times \frac{1-\cos[(k\Omega-k'\Omega')\Delta t]}
{\Delta t (k\Omega-k'\Omega')^2}  \nonumber \\
&& - \frac{4\pi^2}{N m^2} \int\d J'\d J'' f_0(J') f_0(J'') 
\sum_k  2 k^2\phi_{k,0}\phi_{k,0}^* \nonumber \\
&& \times \frac{1-\cos[k\Omega\Delta t]}{\Delta t (k\Omega)^2} .
\label{DJ3}
\eeqa
As seen from Eqs.(\ref{J1w1chi})-(\ref{chi}) the action $J$ and angle $w$ of
each particle in a given realization (i.e. without performing any averaging)
are modified by discrete effects over a time-scale which grows as $\sqrt{N}$
and as expected in the limit $N\rightarrow\infty$ we recover the mean-field 
dynamics. Since we consider the limit of large number of particles, which
justifies the perturbative approach (\ref{H0HI}), let us consider in 
Eq.(\ref{DJ3}) time-scales of order $\Delta t \sim \sqrt{N} \rightarrow\infty$.
Using the equality $\lim_{t\rightarrow\infty}(1-\cos tx)/tx^2=\pi\delta_D(x)$
we obtain in this large-$N$ limit:
\beq
\lag\frac{(\Delta J)^2}{\Delta t}\rag_{\infty} = \frac{4\pi^2}{m}\int\d J' 
f_0(J') \sum_{k,k'} k^2|\phi_{k,k'}|^2 \delta_D(k\Omega-k'\Omega').
\label{DJ4}
\eeq
Thus, both the mean drift and diffusion of the action occur through resonances
between particle orbits (see also \cite{Lynden-Bell1972}). Moreover, they both 
grow linearly with time. 
We can note that \cite{Chavanis2005} also obtained a diffusion coefficient
of the form (\ref{DJ3}) which decays as $1/t$ with an oscillatory behavior
for the HMF model in the limit of low temperatures where a density peak 
appears. However, \cite{Chavanis2005} approximated the core orbits
by an harmonic oscillator with an unique frequency $\Omega_0$ so that there was
no integration over $\Omega'$ as in Eqs.(\ref{J23}),(\ref{DJ3}) through the 
dependence $\Omega'(J')$. In order to take the large-time or large-$N$ limit
it is necessary to take into account the distribution of orbital frequencies
$f_0(\Omega)$. Note that our approach also applies to any Hamiltonian system
with two-body interactions, such as the HMF model.
At this point, it is convenient to change variable from the action $J$ to the
energy $\epsilon(J)=v^2/2+\phi_0$ (both are defined from the mean-field
Hamiltonian ${\cal H}_0$). Using the expansion:
\beq
\Delta\epsilon= \Omega\Delta J + \frac{1}{2}\frac{\d\Omega}{\d J} (\Delta J)^2 
+ ..
\label{DE}
\eeq
we obtain at order $1/N$:
\beqa
\lag\frac{\Delta\epsilon}{\Delta t}\rag & = & \frac{4\pi^2}{m} \int\d J' 
f_0(J') \sum_{k,k'=1}^{\infty} \left( k \frac{\pl}{\pl J} 
- k' \frac{\pl}{\pl J'}\right) 
\nonumber \\
&& \times |\phi_{k,k'}|^2 k\Omega \delta_D(k\Omega-k'\Omega') ,
\label{DE1}
\eeqa
and:
\beq
\lag\frac{(\Delta\epsilon)^2}{\Delta t}\rag = \frac{8\pi^2}{m}\!\!\!\!
\int\d J' f_0(J') \sum_{k,k'=1}^{\infty} \!\! |\phi_{k,k'}|^2 k^2\Omega^2 
\delta_D(k\Omega-k'\Omega').
\label{DE2}
\eeq
In Eqs.(\ref{DE1})-(\ref{DE2}) we changed the sums over $k$ and $k'$ from 
$]-\infty,\infty[$ to $[1,\infty[$ which gave a factor $2$. We can check from
Eq.(\ref{DE1}) that the transfer of energy $\Delta\epsilon(J'\rightarrow J)=
- \Delta\epsilon(J\rightarrow J')$ is anti-symmetric over $(J,J')$ so that the
mean change of energy over all particles (with the distribution $f_0(J)$)
vanishes. This is related to the conservation of energy by the Hamiltonian
dynamics. On the other hand $(\Delta\epsilon)^2$ is symmetric.
We now focus on the low temperature regime 
$T\rightarrow 0, \zeta_L\rightarrow\infty$,
where the initial state $n=2$ consists of two narrow density peaks at the 
boundaries $x=0,L$. Then, from the asymptotic behavior (\ref{rhocosh}) and 
the rescaling (\ref{rhon}) we obtain for the properties of the core which
contains most of the mass the scalings:
\beq
\frac{\rho_c}{\rhob} \sim \zeta_L^2 , 
\frac{R_c}{L} \sim \zeta_L^{-2} ,
\frac{\Omega_c}{\Omega_{\rm dyn}} \sim \zeta_L , \frac{v_c}{v_{\rm dyn}} \sim 
\zeta_L^{-1} , \frac{J_c}{J_{\rm dyn}} \sim \zeta_L^{-3} ,
\label{scal1}
\eeq
where $\rho_c,R_c,\Omega_c, v_c$ and $J_c$ are the typical density, orbital 
radius, frequency, velocity and action
of core particles and we defined $\Omega_{\rm dyn}=1/t_{\rm dyn}$, 
$v_{\rm dyn}=L/t_{\rm dyn}$ and $J_{\rm dyn}=L V_{\rm dyn}$, from the dynamical
time $t_{\rm dyn}$ of Eq.(\ref{xieq}). Let us consider the dissipation rate
$\gamma=-\lag\Delta\epsilon/\Delta t\rag$ of a halo particle with an orbit
of order $L/4$ which probes a finite part of the left potential well 
(i.e. not confined to the small core at $x < R_c$). Its typical frequency, 
velocity and action are $\Omega_{\rm dyn}, v_{\rm dyn}$ and 
$J_{\rm dyn}$. Since most of the mass (except for an exponentially small
fraction, see Eqs.(\ref{rhocosh})-(\ref{rhovoid})) is in the core the 
dissipation rate obtained from Eq.(\ref{DE1}) is dominated by core particles
$J'$. Therefore, we have $k/J\sim \zeta_L$ and $k'/J'\sim \zeta_L^3$ hence
expression (\ref{DE1}) is governed by its second term. Next, using:
\beq
\frac{\d f_0}{\d J'} = \Omega' \frac{\d f_0}{\d\epsilon'} = - \beta \Omega' f_0
\label{df0}
\eeq
for the equilibrium distribution (\ref{Maxwell}) and integrating by parts we
obtain:
\beq
\gamma = \frac{4\pi^2\beta}{m} \int\d J' f_0(J') \sum_{k,k'=1}^{\infty} 
|\phi_{k,k'}|^2 k^2\Omega^2 \delta_D(k\Omega-k'\Omega') .
\label{gamma1}
\eeq
As could be expected, we find that the energy drift $\lag\Delta\epsilon\rag$ 
corresponds to a damping term ($\gamma$ is positive) of order $1/N$. Thus, 
the high-energy halo particle loses its energy to lower-energy core particles. 
This is similar to the dynamical friction suffered by a high-speed star 
moving through an homogeneous background of lower velocity stars 
\cite{Binney1987,Chandrasekhar1943}.
Moreover, the diffusion coefficient $D=\lag(\Delta\epsilon)^2/\Delta t\rag$
obtained in Eq.(\ref{DE2}) and the dissipation coefficient $\gamma$ are 
related by:
\beq
D = \frac{2\gamma}{\beta} = 2\gamma T .
\label{Dgamma}
\eeq
Thus we recover the usual Einstein relation (\ref{Dgamma}) for a halo
particle. Next, from the definition (\ref{Phikk'}) we have:
\beq
\phi_{k,k'} = g m \int_0^{\pi} \frac{\d w \d w'}{\pi^2} |x-x'| \cos(kw) 
\cos(k'w') .
\label{Phikk'cos}
\eeq
If we approximate each orbit by its first harmonic $x=R\sin(w/2)$ we obtain:
\beq
\phi_{k,k'} \simeq - \frac{gm}{2\pi} \frac{R'^2}{R} \delta_{k',1} \;\;\; 
\mbox{for} \;\;\; k\geq1, k'\geq 1, R' \ll R.
\label{Phikk'approx}
\eeq
Taking into account the exact orbital trajectory would remove the Kronecker
factor $\delta_{k',1}$ (higher $k'$ would contribute) but would not change the
scaling $gmR'^2/R$. Next, from Eq.(\ref{HIJw}) we see that the action and angle
of each core particle in a given realization (i.e. without performing any 
averaging) are modified by discrete effects over a time-scale $\Delta t_c$
of order:
\beq
\Delta t_c \sim \frac{J}{\dot J} \sim \frac{\sqrt{N}}{\Omega_c} 
\sim \frac{\sqrt{N}}{\zeta_L} t_{\rm dyn} ,
\label{tc}
\eeq
whereas for halo particles we obtain:
\beq
\Delta t_h \sim \sqrt{N}\zeta_L^3 t_{\rm dyn} \gg \Delta t_c ,
\label{th}
\eeq
where we used the scalings (\ref{scal1}) and (\ref{Phikk'approx}). Therefore,
we see that the core particles are significantly perturbed over a time-scale
which is much smaller than for the halo particles. Then, the resonances between
halo and core particles will be detuned over times of order $\Delta t_c$
during which the trajectory of the halo particle has only suffered small
deviations. This means that the core particles act as an external noise
characterized by small time-scales with respect to the halo particle. This
justifies a Markovian approximation for the behavior of the halo particle
and allows us to write a Fokker-Planck equation \cite{vanKampen2004} for the 
evolution of the probability distribution $P(\epsilon)$ of its energy: 
\beq
\frac{\pl P}{\pl t} = \frac{\pl}{\pl\epsilon}(\gamma P) + \frac{1}{2} \frac{\pl^2}{\pl\epsilon^2}(DP) .
\label{FokkerPlanck}
\eeq
The orbit-averaged Fokker-Planck equation (\ref{FokkerPlanck}) clearly 
separates the slow changes in phase-space coordinates due to encounters from 
the fast orbital motion in the mean-field potential \cite{Binney1987}.
Indeed, the dissipation and diffusion coefficients scale as $1/N$ from
Eqs.(\ref{DE1})-(\ref{DE2}). It is interesting to note that 
for the OSC model (\ref{HN}) particles which do not cross the test mass do
not contribute. This can be checked from Eq.(\ref{Phikk'cos}) by noting that
in such cases the absolute value can be dropped which yields $\phi_{k,k'}=0$
for $k\neq 0$ and $k'\neq 0$. This is related to the fact that in 1-D gravity 
the force is merely proportional to the relative number of particles to the 
left and to the right of the test particle, independently of distances, as 
seen from Eqs.(\ref{HN}) or (\ref{xi}). Therefore, only particle crossings can 
lead to fluctuations of the force seen by the test mass. 
The stationary solution $P_{\rm st}$ of the Fokker-Planck equation 
(\ref{FokkerPlanck}) is:
\beq
P_{\rm st}(\epsilon) \propto \frac{1}{D} e^{-(2\gamma/D) \epsilon} .
\label{Pst}
\eeq
We can note that this distribution agrees with the statistical equilibrium
state ``$n=2$'' described by the Maxwellian (\ref{Maxwell}) thanks to 
Einstein's relation (\ref{Dgamma}). On the other hand, we must point out that
the relaxation of core particles cannot be described by the Fokker-Planck 
equation (\ref{FokkerPlanck}) because there is no separation of time-scales
as in Eqs.(\ref{tc})-(\ref{th}) which prevents the use of the Markovian
approximation. This also explains why the Einstein relation (\ref{Dgamma})
does not hold for core particles (since we cannot any longer neglect
the first term of Eq.(\ref{DE1})). From Eq.(\ref{FokkerPlanck}) we can obtain
the mean time $\tau$ it takes for a particle starting at energy $\epsilon_-$ 
in the potential well $\phi_0$ to reach the maximum potential energy 
$\epsilon_+$. It reads \cite{vanKampen2004}:
\beq
\tau = \int_{\epsilon_-}^{\epsilon_+} \d\epsilon \; e^{F(\epsilon)} 
\int_{\epsilon_-}^{\epsilon} \d\epsilon' \frac{2}{D(\epsilon')} 
e^{-F(\epsilon')} ,
\label{tau1}
\eeq
with:
\beq
F(\epsilon) = \int_{\epsilon_-}^{\epsilon} \d\epsilon' 
\frac{2\gamma(\epsilon')}{D(\epsilon')} = \beta (\epsilon-\epsilon_-) .
\label{F1}
\eeq
Eqs.(\ref{tau1})-(\ref{F1}) clearly show that the escape time is 
dominated by the time spent at large values of $\epsilon$, that is at large 
radii of order $L/2$ far from the small core. This yields:
\beq
\tau \sim \frac{2}{\beta^2 D} \left[ e^{\beta(\epsilon_+-\epsilon_-)} 
- 1 - \beta(\epsilon_+-\epsilon_-) \right] .
\label{tau2}
\eeq
On the other hand, the diffusion coefficient (\ref{DE2}) can be simplified
by first integrating over $J'$:
\beq
D = \left. \frac{8\pi^2}{m}\sum_{k,k'=1}^{\infty}\left|\frac{\d J'}{\d\Omega'}
\right| f_0(J') |\phi_{k,k'}|^2 k' \Omega'^2 \right|_{\Omega'=k\Omega/k'} .
\label{D1}
\eeq
Since $\Omega/\Omega'\sim 1/\zeta_L \ll 1$ the sum over $k$ can now be 
approximated by an integral over $J'$ which yields:
\beq
D \simeq \left. \frac{8\pi^2}{m}\sum_{k'=1}^{\infty} \int \d J' f_0(J') 
|\phi_{k,k'}|^2 \frac{k'^2\Omega'^2}{\Omega} \right|_{k=k'\Omega'/\Omega} .
\label{D2}
\eeq
From the scalings (\ref{scal1}) and (\ref{Phikk'approx}) we obtain:
\beq
D \sim \frac{1}{N\zeta_L^2} \frac{T^2}{t_{\rm dyn}} .
\label{D3}
\eeq
Taking for the energy gap $\Delta\epsilon=\epsilon_+-\epsilon_-$ the potential 
barrier $\Delta\Phi=\Delta\phi$ between the bottom of the potential well at 
$x=0$ and the maximum at $x=L/2$ we obtain:
\beqa
\tau & \sim & t_{\rm dyn} N \zeta_L^2 
\left[ e^{\beta\Delta\varphi} -1 -\beta\Delta\varphi \right] \nonumber \\
& \sim & t_{\rm dyn} N \zeta_L^2 \left[ \frac{\rho_{\rm max}}{\rho_{\rm min}} 
- 1 - \ln \frac{\rho_{\rm max}}{\rho_{\rm min}} \right] ,
\label{tau3}
\eeqa
where we used Eq.(\ref{offset}). Thus, at low temperatures we obtain from 
Eqs.(\ref{rhocosh})-(\ref{rhovoid}) the asymptotic behavior:
\beq
\tau \sim  t_{\rm dyn} N \zeta_L^4 e^{\zeta_L^2/8} 
\sim t_{\rm dyn} N \left(\frac{T_{c2}}{T}\right)^2 e^{\pi^2 T_{c2}/2T} ,
\label{tau4}
\eeq
where we used the rescaling (\ref{rhon}). From Eq.(\ref{tau3}) we can estimate 
the mass flux from one peak to the other one as:
\beq	
\frac{\d M}{\d t} = \frac{M}{2 t_M} \;\; \mbox{with} \;\; 
\frac{t_M}{t_{\rm dyn}}= 0.2 \frac{N \zeta_L^2}{2} 
\left[ \frac{\rho_{\rm max}}{\rho_{\rm min}} 
- 1 - \ln \frac{\rho_{\rm max}}{\rho_{\rm min}} \right] .
\label{dMdt}
\eeq
Here the factors $M/2$ and $N/2$ express the fact that only half of the total
mass is within each peak whereas the factor $0.2$ of order unity has been
chosen so as to match the slope of the mass transfer at early times displayed
in right panel of Fig.~\ref{figEM2to1}, which shows the evolution of a typical
system with a transition time 
$t_{2\rightarrow\pm 1}\simeq 4.7\times 10^4 t_{\rm dyn}$ equal to the average
value obtained from 200 numerical simulations (solid line in right panel of 
Fig.~\ref{figtime2to1}). Then, in order to take into account the acceleration
of the mass transfer close to the transition we simply write the mean
transition time $t_{2\rightarrow\pm 1}$ as:
\beq
t_{2\rightarrow\pm 1} = \frac{\Delta M}{M/2} t_M \;\; \mbox{with} \;\; 
\Delta M = (M_L(n=1)-M/2)/4 ,
\label{t21}
\eeq
where $M_L(n=1)$ is the final mass located at $x<L/2$ in the equilibrium state
$n=1$ whereas the factor $4$ accounts for the late steepening of the
mass transfer estimated from Fig.~\ref{figEM2to1}. The transition time 
(\ref{t21}) is shown by the dashed-lines in 
Figs.~\ref{figtime2to1}-\ref{figtN2to1}. We can verify that it agrees with
numerical simulations and recovers the steep increase obtained at low total
energies, as the exponential of the inverse temperature (see Eq.(\ref{tau4})). 
The departure at higher energies close to the transition $T_{c2}$ is expected
since the approach described above was performed in the low-temperature
limit $\zeta_L \gg 1$ where the hierarchy (\ref{th}) holds. Close to $T_{c2}$
one cannot distinguish halo and core particles and one cannot use a 
Fokker-Planck equation as (\ref{FokkerPlanck}).
Besides, Eq.(\ref{tau4}) shows that the relaxation time increases linearly
with the number of particles $N$, in agreement with the numerical results
shown in Fig.~\ref{figtN2to1}. This is different from the relaxation of
homogeneous states obtained in the HMF model \cite{Bouchet2005} where the 
relaxation time was seen to grow faster than $N$ (numerical simulations 
gave $\sim N^{1.7}$ \cite{Yamaguchi2004,Zanette2003}).
However, we must note that Eq.(\ref{tau4}) was derived in the low-energy 
limit where the system is strongly inhomogeneous. Therefore, the physical
process involved in the relaxation (the escape of particles from the smallest
potential well) is rather different.
On the other hand, \cite{Miller1996} also found a relaxation time 
proportional to $N$ for the 1-D gravitational system (which is identical 
to the OSC model without the reflecting walls and the external potential $V$).
However, using the Fokker-Planck equation derived in \cite{Yawn1995} for the 
diffusion in $(v,a)$ space (velocity $v$ and acceleration $a$) from
a Markovian approximation, \cite{Miller1996} obtained a Fokker-Planck 
equation in energy space as in Eq.(\ref{FokkerPlanck}) but without the
friction term. Indeed, as seen above the latter arises from the correlations 
between particles which accumulate over time but these were not considered 
in \cite{Yawn1995} which computed the transport coefficients in $(v,a)$ space
in the limit of infinitesimal time-steps and next assumed a Markovian
evolution. This procedure cannot be applied here since correlations do not
decrease exponentially with time (see for instance the cosine dependence in 
Eqs.(\ref{chi}),(\ref{DJ3})) and the dynamics cannot be described as
Markovian over time-steps of the order of a few orbital times (which
amounts to erase all correlations every few orbital times).
Moreover, the dissipation term in Eq.(\ref{FokkerPlanck}) is clearly required
by physical consistency to recover the stationary distribution (\ref{Pst})
and Einstein's relation. 

We can note that halo and core particles are rather strongly coupled in the 
sense that the dissipation and diffusion coefficients only decrease as a power 
of the ratio of orbital frequencies $\Omega'/\Omega$ (through the factors
$(R'/R)^2$ in Eq.(\ref{Phikk'approx}) and $(\Omega'/\Omega)^2$ in 
Eq.(\ref{D2})) which gives rise to the inverse power of $\zeta_L$ in
Eq.(\ref{D3}). This is significantly different from the coupling in 3D gravity 
between a small dense core and unbounded scattering particles. In that 
case the smooth gravitational interaction (for non-zero impact parameter) 
leads to a coupling and a dissipation which decrease exponentially with the 
ratio of frequencies \cite{Ispolatov2004}. In the case we consider here
the large transition time observed at low temperature is not due to 
exponentially small couplings between halo and core particles. It is merely
due to the usual Arrhenius factor $e^{-\Delta\epsilon/T}$ associated with
the diffusion through a finite potential barrier. 
Finally, we can note that our results also apply to any Hamiltonian system
with a two-body interaction which shows similar equilibria states.
We only need to use the relevant scalings over the temperature $T$ (in our
case written in terms of $\zeta_L$).

\section{\label{Conclusion}Conclusion}

We have studied in this article the relaxation of a 1-D gravitational
system (OSC model) which was originally derived from the formation of 
large-scale structures in cosmology \cite{Aurell2001,Fanelli2002,Valageas2006}.
We have checked that the homogeneous equilibrium state $n=0$ becomes unstable
below the critical energy $E_{c1}$ and exhibits a relaxation to the one-peak
state $n=\pm 1$ over a few dynamical times. This is consistent with the fact 
that the homogeneous state is unstable both from a thermodynamical analysis and
from a dynamical mean-field analysis. Therefore, the linear instability
of the Vlasov dynamics leads to a violent relaxation to the stable equilibrium
$n=\pm 1$ (or to a nearby one-peak state) which develops through a collective
dynamical instability.

On the other hand, close to the transition $E_{c1}$ we have found that for 
moderate numbers of particles ($N=50$) the fluctuations due to finite $N$
effects are large enough to prevent the system from converging towards a
stable equilibrium. Indeed, the system keeps wandering over left and right
peak states since it can easily jump from one equilibrium to another one.
For larger numbers of particles ($N=1000$) the fluctuations due to discrete
effects are much smaller than the distance (in terms of energy levels or
mass ratios) between homogeneous, left and right peak states (unless we
go closer to $E_{c1}$). Then, starting from the homogeneous unstable 
equilibrium the system wanders again for a long time ($\sim 5000 t_{\rm dyn}$) 
over left and right peak states but it eventually manages to settle in
a stable left or right peak equilibrium.

Finally, at low energies below $E_{c2}$ we have noticed that in some cases
the homogeneous state does not relax directly to a left or right peak
configuration but first converges over a few dynamical times towards a
two-peak state close to equilibrium $n=2$. Then, the system undergoes a
slow collisional relaxation towards a one peak state $n=\pm 1$.

Next, we have investigated the relaxation to thermodynamical equilibrium of
the two-peak equilibrium $n=2$. As expected, we have found that since this
is a stable equilibrium of the mean-field Vlasov dynamics the relaxation
involves a slow diffusion process over a time-scale $t_{2\rightarrow\pm 1}$
which diverges with $N$. Moreover, after the smallest density peak has lost
most of its mass the two-peak configuration (with a high mass ratio) becomes
dynamically unstable and the system converges to a one-peak state over a few
dynamical times. We have estimated analytically the mean transition time
$t_{2\rightarrow\pm 1}$ by
describing the slow diffusion of particle energies due to finite $N$ effects
with a Fokker-Planck equation. We have found that the friction and diffusion
coefficients of halo particles satisfy Einstein's relation and we have obtained
a mean transition time of the form $t_{2\rightarrow\pm 1} \sim N e^{1/T}$
which is proportional to the number of particles and grows at low temperatures.
We have checked that this prediction agrees reasonably well with our numerical
simulations. The relaxation involves an efficient coupling between the halo
particles which extend close to the barrier at $L/2$ and the core particles
buried in the density peaks thanks to efficient resonances at high harmonics. 
Thus, although halo particles have a much smaller orbital frequency than core 
particles the dissipation and diffusion due to the build-up of correlations 
and encounters do not vanish exponentially with the ratio of orbital 
frequencies.

Therefore, we have found that the relaxation of the OSC model proceeds in
a fashion similar to some other long-range systems. It first involves a violent
relaxation phase governed by dynamical instabilities, where the system
converges to stable solutions of the mean-field Vlasov dynamics over a few
dynamical times. This is followed by a second much slower collisional 
relaxation phase where the system goes through a series of quasi-stationary
states of the Vlasov dynamics until it reaches thermal equilibrium.
Moreover, we have found that this slow evolution can be followed
by another violent relaxation step as the series of quasi-stationary states
may lead to an unstable configuration (the least stable eigenvalue increases
along the series and eventually becomes positive) which quickly relaxes to 
a new quasi-stationary solution. Here we note that the slow relaxation of
the system as it goes through the series of quasi-stationary states is due
to dynamical constraints and not to metastability as for the cases discussed 
for instance in \cite{Chavanis2005a} where the system is trapped in local
entropy maxima. On the other hand, the route to thermal equilibrium clearly
depends on the initial conditions. Therefore, the relaxation time exhibits
a strong dependence on the initial conditions, as shown by the comparison
of $t_{0\rightarrow\pm 1}$ and $t_{2\rightarrow\pm 1}$ studied in this
article.

\bibliography{a2}

\end{document}